\documentclass[aps,prb,twocolumn,raggedbottom,floatfix,showpacs]{revtex4}
\usepackage{amsmath}
\usepackage{amssymb}
\usepackage{graphicx}
\usepackage{bm}

\newcommand{\comment}[1]{}

\newcommand{\w}{\omega}

\newcommand{\Ss}{\Sigma_{\sigma}}

\newcommand{\Gs}{G_{\sigma}}

\newcommand{\pd}{\phantom\dagger}

\newcommand{\tm}{\tilde{\mu}}

\newcommand{\nimp}{n_{\mathrm{imp}}}
\newcommand{\nimps}{n_{\mathrm{imp},\sigma}}
\newcommand{\nimpas}{n_{\mathrm{imp},A\sigma}}
\newcommand{\mimp}{m_{\mathrm{imp}}}
\newcommand{\mimpa}{m_{\mathrm{imp},A}}
\newcommand{\rhoimps}{\Delta\rho_{\mathrm{imp},\sigma}(\w, h)}

\begin{document}

\title{Common non-Fermi liquid phases in quantum impurity physics}
\author{David E. Logan}
\author{Adam P. Tucker}
\author{Martin R. Galpin}
\affiliation{Oxford University, Chemistry Department, Physical \& Theoretical Chemistry, 
South Parks Road, Oxford, OX1~3QZ, UK.}

\begin{abstract}
We study correlated quantum impurity models which undergo a local quantum phase transition (QPT) from a strong 
coupling, Fermi liquid phase to a non-Fermi liquid phase with a globally doubly degenerate ground state. Our 
aim is to establish what can be shown exactly about such `local moment' (LM) phases; of which the permanent 
(zero-field) local magnetization is a hallmark, and an order parameter for the QPT. A description of the zero-field 
LM phase is shown to require two distinct self-energies, which reflect the broken symmetry nature of the phase and 
together determine the single self-energy of standard field theory. Distinct Friedel sum rules for \emph{each} phase 
are obtained, via a Luttinger theorem embodied in the vanishing of appropriate Luttinger integrals. By contrast, the 
standard Luttinger integral is non-zero in the LM phase, but found to have universal magnitude. 
A range of spin susceptibilites are also considered; including that corresponding to the local order parameter, 
whose exact form is shown to be RPA-like, and to diverge as the QPT is approached. Particular attention is given to 
the pseudogap Anderson model, including the basic physical picture of the transition, the low-energy behavior of 
single-particle dynamics, the quantum critical point itself, and the rather subtle effect of an applied local field.
A two-level impurity model which undergoes a QPT (`singlet-triplet') to an underscreened LM phase is also considered, 
for which we derive on general grounds some key results for the zero-bias conductance in both phases. 
\end{abstract}

\pacs{71.27.+a, 71.10.Hf, 72.15.Qm}

\maketitle

\section{Introduction}
\label{sec:intro}

Since its inception more than half a century ago,~\cite{anderson} the Anderson impurity model (AIM) --
a single, correlated level coupled to a metallic conduction band -- has played a central role in understanding strongly correlated electron systems,~\cite{hewsonbook} with a resurgence of interest in recent years arising 
from the advent of quantum dot devices.~\cite{kondorevival}
Its essential physics in the regime where the impurity/dot is in essence singly occupied, is that of the Kondo effect: the impurity spin degree of freedom is completely quenched on coupling to
the metallic conduction band, and a strong coupling (SC), many-body singlet ground state arises.

\begin{figure}
\includegraphics{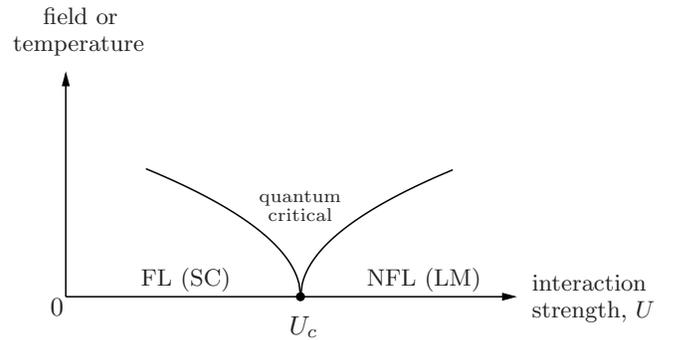}
\caption{\label{fig:fig0} 
A zero-field QPT ($T=0$) occurs at a critical interaction $U_{c}$.
It separates a FL (or strong coupling) phase for $U<U_{c}$, which is adiabatically connected to
the non-interacting limit; from a NFL, local moment phase for $U>U_{c}$. Solid lines denote 
low-energy scales characteristic of each phase, which vanish as the QPT is approached and set the
scale for crossover to quantum critical behavior at finite $T$ or field. 
}
\end{figure}

Yet the metallic AIM is atypical in one important sense. 
The system is a Fermi liquid (FL) for \emph{any} value of the interaction strength:
the model lacks a local (or boundary) quantum phase transition (QPT) to a phase in which the local 
spin degree of freedom is incompletely quenched. Local QPTs occur of course
at $T=0$, and in the absence of a field that would otherwise destroy them.
The familiar situation is sketched in Fig.\ \ref{fig:fig0}, with generic interaction strength (`$U$') 
as the abscissa and the QPT occurring at a critical $U_{c}$. The transition separates two distinct phases. 
One is perturbatively connected to the non-interacting limit, and in that general sense 
is thus a Fermi liquid. Separated from it by the QPT, the other is not then perturbatively connected 
to the non-interacting limit. As such, it is a non-Fermi liquid (NFL) phase.

Continuous quantum phase transitions between FL and NFL phases are in fact quite typical
in quantum impurity physics. For single-level models, examples include the pseudogap AIM~\cite{WithoffFradkin,ChenJayap1995,BPH1997,GB-Ingersent,MTG_SPAIM,MTGEPJB,nrglmacomp,Vojta_Bulla2001,Zitzler2002,Ingersent-Si2002,Kircan_Vojta2003,MTG_APAIM,MTGSPAIMepl,Vojta_Fritz2004,Fritz_Vojta2004,mattgarethPAIM,Lee2005,Fritz_Florens_Vojta2006,Vojta_Fritz_Bulla2010,MTGKirchner2011,Pixley2012}
(where the conduction band density of states has a soft-gap at the Fermi level), as well as
the gapped AIM;~\cite{fnrefgappedAIM} while many examples arise in multi-level and multi-impurity models, 
including e.g.\ single-channel two-level impurity systems which undergo a QPT to an underscreened spin-1 phase,~\cite{fnreftwo-levelgen,CJW2009,CJWfield} and impurity models for double~\cite{fnrefdqdmrg} and triple~\cite{fnrefTQDgen,Jarrold2013} quantum dot devices. In all these cases the NFL phase 
is `common' in the sense that it occupies a finite fraction of the model parameter space
(i.e.\ does not require fine-tuning of parameters to be realized~\cite{fnfinetune}). 
The associated QPTs are diverse in character, ranging from a quantum critical point with a fixed 
point (FP) distinct from that characteristic of either the FL or NFL phases, through a critical end-point
of a line of FPs characteristic of one or other phase (Kosterlitz-Thouless transitions), to a simple first-order level-crossing transition. And the NFL phases are commonly (globally) doubly degenerate states, characterized as such by a degenerate  $SU(2)$ spin-like degree of freedom. The latter is typically a `real' spin,
whence we refer to them as local moment (LM) phases; although it can arise also from underlying charge degrees 
of freedom.~\cite{fnrefdqdmrg}

These degenerate LM phases are the primary focus of the present paper, and are 
certainly non-trivial -- the local spin degree of freedom is not `free', but 
incompletely quenched by coupling to the conduction band. Part of our motivation arises
from the Local Moment Approach (LMA),~\cite{LET,MTGLMA_asym,nigelscalspec} 
where the notion of local moments enters centrally from the outset, and which
provides a rather successful description of the 
pseudogap,~\cite{MTG_SPAIM,nrglmacomp,MTG_APAIM,MTGSPAIMepl,mattgarethPAIM}
gapped~\cite{fnrefgappedAIM} and metallic~\cite{LET,MTGLMA_asym,nigelscalspec,fnrefLMAgeneralAIM} 
AIMs; as well as correlated lattice-fermion models~\cite{fnrefLMAPAM} 
within the framework of dynamical mean-field theory.~\cite{MetznerVollhardt1989,dmftgeorgeskotliar}
Yet the LMA is of course approximate (and in the first instance local moments enter explicitly at 
mean-field level). Our aim here by contrast is to show what can be deduced exactly about LM 
(as well as SC) phases, unfettered by approximations.

 For any impurity model the Hamiltonian has the form 
$H=H_{\mathrm{imp}}+H_{\mathrm{CB}}+H_{\mathrm{hyb}}$,
where $H_{\mathrm{imp}}$ refers to the impurity itself, $H_{\mathrm{CB}}$ to the
conduction band, and the hybridization term $H_{\mathrm{hyb}}$ couples 
the impurity and conduction band degrees of freedom.
For most of the paper we consider explicitly the case of a single-level impurity 
(and for $T=0$ unless specified otherwise). This is in part for notational simplicity, 
since much of the following can be generalized to multi-level impurities (which we 
consider in sec.\ \ref{section:Luttintgen}).
The free impurity Hamiltonian, with level energy $\epsilon$ and local Coulomb interaction $U$, is then
$H_{\mathrm{imp}} = \sum_{\sigma} (\epsilon -\sigma h) d^{\dagger}_{\sigma}d^{\pd}_{\sigma}+U\hat{n}_{\uparrow}\hat{n}_{\downarrow}$
(with $\sigma = \pm$ for $\uparrow$$/$$\downarrow$-spins), where
$\hat{n}_{\sigma}=d^{\dagger}_{\sigma}d^{\pd}_{\sigma}$ is the local $\sigma$-spin number operator
and $h=\tfrac{1}{2}g\mu_{B}B$ denotes a magnetic field. The conduction band 
$H_{\mathrm{CB}}= \sum_{\mathbf{k},\sigma}(\epsilon_{\mathbf{k}}^{\pd}-\sigma h)c^{\dagger}_{\mathbf{k}\sigma}c^{\pd}_{\mathbf{k}\sigma}$, where for generality we include 
a field acting on the band states;
and the hybridization term is
$H_{\mathrm{hyb}}=\sum_{\mathbf{k},\sigma}(V_{\mathbf{k}}^{\pd}d^{\dagger}_{\sigma}c^{\pd}_{\mathbf{k}\sigma}+ \mathrm{h.c.})$.\cite{fn-2}


\subsection{Overview}
\label{subsec:overview}

As the paper is quite broad ranging, we first give an overview of it.
Sec.\ \ref{section:LMgeneral} begins by considering the $T=0$ local (impurity)
magnetization $m(h)$; in particular its $h$-dependence in the LM phase 
(sec.\ \ref{subsection:secIILMphase}) and the non-vanishing permanent local 
moment $m(h=0+)$ that is both a characteristic signature of the phase
and a natural order parameter for the QPT.

The behavior of $m(h)$ has strong implications for the structure of single-particle propagators 
in the LM phase, particularly at zero-field (sec.\ \ref{subsection:secIIh=0}).
Here, reflecting the spin-degeneracy of the ground state, we show that a description
of the zero-field propagator requires consideration of the two distinct self-energies
which reflect the broken symmetry nature of the LM phase; and which together determine
the conventional single self-energy of standard field theory.
This `two-self-energy' description also underlies the LMA~\cite{LET,MTGLMA_asym,nigelscalspec} 
(where it is used to describe both LM and SC phases), but here it is exact.

In sec.\ \ref{section:sumrules} we obtain the Friedel-Luttinger sum rules which relate a scattering 
phase shift to the corresponding `excess' charge (sec.\ \ref{subsection:excessprops}) and a Luttinger 
integral. Luttinger's theorem holds in the SC, FL phase, i.e.\ the Luttinger integral vanishes
(`universally', throughout the  phase).
In sec.\ \ref{subsection:Luttintgen} we show that for $U>U_{c}$ in the LM phase, the
corresponding Luttinger integral appropriate to the two-self-energies likewise vanishes;
whence a Friedel sum rule arises for \emph{each} phase, with important consequences considered in 
secs.\ \ref{section:PAIMh=0} \emph{ff}.
By contrast, sec.\ \ref{subsection:Luttinth=0} considers again the zero-field LM phase,
with a phase shift defined in terms of the conventional single self-energy. Here we
show that while a Friedel-Luttinger sum rule again arises, the associated Luttinger
integral, $I_{L}$, cannot be argued to vanish.

Following brief discussion (sec.\ \ref{section:atomlimh=0}) of the simple atomic limit
-- an uncoupled, correlated level -- we turn in secs.\ \ref{section:PAIMh=0}-\ref{section:finitefield}
to the pseudogap AIM (PAIM),~\cite{WithoffFradkin,ChenJayap1995,BPH1997,GB-Ingersent,MTG_SPAIM,MTGEPJB,nrglmacomp,Vojta_Bulla2001,Zitzler2002,Ingersent-Si2002,Kircan_Vojta2003,MTG_APAIM,MTGSPAIMepl,Vojta_Fritz2004,Fritz_Vojta2004,mattgarethPAIM,Lee2005,Fritz_Florens_Vojta2006,Vojta_Fritz_Bulla2010,MTGKirchner2011,Pixley2012} 
where the SC and LM phases are separated by a quantum critical point (QCP).~\cite{GB-Ingersent}
Beginning with zero-field, the implications of the respective Friedel sum rules for SC and LM phases 
are considered in turn (secs.\ \ref{subsection:PAIMh=0sc},\ref{subsection:PAIMh=0lm}).
The generic physical picture of the SC to LM transition is thereby shown to be that, 
immediately on entering the LM phase, the entire system acquires (or loses) a single additional 
electron, which is fully spin-polarized for even an infinitesimal field; 
but that precisely at the QCP there is no spin-density on the impurity, which by contrast is on 
the verge of acquiring a permanent local moment. 

The conventional Luttinger integral in the LM phase is then considered (sec.\ \ref{subsection:ILprime}), 
and its magnitude shown to be universal but non-vanishing, $|I_{L}| =\tfrac{\pi}{2}$.
In sec.\ \ref{subsection:SSElowomega} the asymptotic low-energy LM phase single self-energy (and hence single-particle spectrum) is obtained, with the resultant low-$\w$ behavior seen to be symptomatic of the NFL nature of the phase. The corresponding low-$\w$ asymptotics for the two-self-energies inherent to the broken-symmetry LM phase are then deduced (sec.\ \ref{section:SSElowomegaLW}), by self-consistently adapting  
Luttinger's original analysis,~\cite{Luttingerf1961} based on the underlying all-orders skeleton expansion for the two-self-energies.

Sec.\ \ref{section:QCP} considers the universal scaling of single-particle dynamics
close to the transition, where the low-energy scales characteristic of each phase vanish 
(e.g.\ the Kondo scale for the SC phase); and, relatedly, the interacting QCP itself (sec.\ \ref{subsection:QCPperse}), including spectral signatures of both the symmetric and asymmetric QCPs~\cite{GB-Ingersent} (for which we supplement analytical considerations with numerical renormalization 
group calculations). The effects of a non-zero local field, $h$, are considered in 
sec.\ \ref{section:finitefield}. While the pristine local QPT between SC and LM phases is 
destroyed for any finite $h$, we show that the situation here is subtle, and
physically rather rich, due to an underlying \emph{bulk} level-crossing `transition' that is quite
distinct from the local QCP.

Moving away from the PAIM \emph{per se}, sec.\ \ref{section:susceptibilities} 
considers a range of static spin susceptibilities which probe the transition between LM and 
SC phases. The susceptibility corresponding to the local order parameter
is first considered, \emph{viz} the $T=0$ local (impurity) susceptibility in response to a local 
field $h$, as $h \rightarrow 0$. Its exact functional form is obtained, and shown both to be `RPA-like' 
(sec.\ \ref{section:localchi}) and to diverge as the QPT is approached and the permanent local moment 
vanishes. Results are then obtained for the $h=0$ local susceptibility as $T\rightarrow 0$ (sec.\ \ref{subsection:localsuscfiniteT}), followed (sec.\ \ref{subsection:impsusceps}) by the
corresponding local and global spin susceptibilities in response to a globally applied uniform field;
all of which are naturally Curie-like in form, but with coefficients vanishing with different
powers of the order parameter as the QPT is approached.

Finally, in sec.\ \ref{section:Luttintgen} we turn briefly to multilevel impurity systems;
notably to a two-level impurity (or quantum dot) coupled to metallic leads in 1-channel 
fashion,~\cite{fnreftwo-levelgen,CJW2009,CJWfield} which, due to a Hund's rule 
coupling, undergoes a QPT from a SC phase to an underscreened LM phase. Here we 
derive and understand on general grounds some key results, hitherto
inferred numerically,~\cite{CJW2009} for the $T=0$ zero-bias conductance and the conventional
Luttinger integral in the LM phase, with $|I_{L}| =\tfrac{\pi}{2}$ again shown to arise.


\section{Local Moments and local propagators}
\label{section:LMgeneral}

In considering LM and SC phases, and the key differences between them, our natural initial focus is the 
$T=0$ local magnetization, $m(h)$. The field $h$ may be applied either locally to the impurity, or globally 
(i.e.\ acting also on conduction band states); at present we do not need to specify which, but will do so when necessary. 

 We begin with generalities applicable to both phases. The local $m(h)$, and local charge $n(h)$, are defined by
\begin{equation}
\label{eq:mn}
m(h)~=~ \sum _{\sigma}~\sigma n_{\sigma}(h)
~~~~~~ n(h)~=~ \sum _{\sigma}~ n_{\sigma}(h)
\end{equation}
where $n_{\sigma}(h)=\langle\hat{n}_{\sigma}\rangle$ is given by
\begin{equation}
\label{eq:nsig}
n_{\sigma}(h)= \int^{E_{\mathrm{F}} = 0}_{-\infty}d\w~ D_{\sigma}(\w, h) ~~ 
=~-\tfrac{1}{\pi} \mathrm{Im} \int^{0}_{-\infty}d\w~ G_{\sigma}(\w, h)
\end{equation}
in terms of the local spectral density 
$D_{\sigma}(\w, h)=-\tfrac{1}{\pi}\mathrm{Im}\Gs(\w,h)$;
and where $\Gs(\w,h) \equiv \Gs^{r}(\w, h)$ is the retarded impurity Green function
given by~\cite{hewsonbook} 
\begin{equation}
\label{eq:Gs}
\Gs (\w, h)= \left[\w^{+} -\epsilon +\sigma h -\Gamma_{\sigma}(\w, h) -\Ss(\w, h) \right]^{-1}
\end{equation}
($\w^{+} =\w +i\eta$ and $\eta = 0+$), with $\Ss(\w, h)$ the local interaction self-energy.
$\Gamma_{\sigma}(\w, h)$ denotes the usual one-electron impurity/conduction band 
hybridization;~\cite{hewsonbook} if the field $h$ is applied purely locally, it is independent of 
$h$ (and $\sigma$), $\Gamma_{\sigma}(\w, h) \equiv \Gamma (\w)$ 
($=\sum_{\mathbf{k}}V_{\mathbf{k}}^{2}[\w^{+} -\epsilon_{\mathbf{k}}]^{-1}$).
With the field sign-convention chosen, $\epsilon_{\sigma}=\epsilon -\sigma h$, 
$\uparrow$-spins are favored over $\downarrow$ and hence $\mathrm{sgn}(m(h))= \mathrm{sgn}(h)$.
Since $H$ is invariant under $\sigma \leftrightarrow -\sigma$ and $h \leftrightarrow -h$, it follows that
\begin{equation}
\label{eq:gsig}
\Gs(\w, h)~=~G_{-\sigma}(\w, -h)~.
\end{equation}
From eqs.\ (\ref{eq:mn},\ref{eq:nsig}) the magnetization
\begin{equation}
\label{eq:modd}
m(h)~=~ -m(-h)
\end{equation}
is thus naturally odd in $h$, while $n(h) =n(-h)$ is even.


\subsection{Local moment phase}
\label{subsection:secIILMphase}

\begin{figure}
\includegraphics[height=9.5cm, width=7cm]{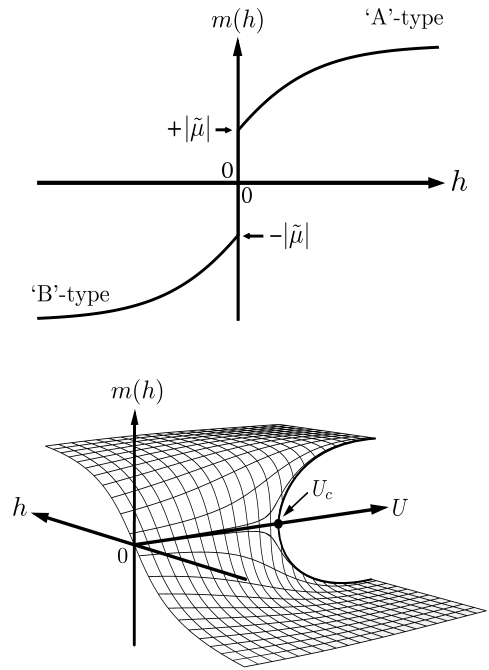}
\caption{\label{fig:fig1} \emph{Upper:} schematic of $T=0$ local magnetization, $m$, in a LM phase: 
$m(h)=-m(-h)$ \emph{vs} $h$ for fixed $U>U_{c}$. As $h\rightarrow 0\pm$, $m(h) \rightarrow \pm |\tm|$, with $|\tm|$ the magnitude of the permanent local moment. \emph{Lower:} qualitative behavior of $m$ as a function of $U$ and $h$, with the critical $U_{c}$ indicated.}
\end{figure}

The above holds whether the phase is SC or LM. What distinguishes the two is of course 
the low-field behavior of the local magnetization. In a SC phase, $m(h)$ vanishes continuously as $h\rightarrow 0$. In a LM phase by contrast, the system possesses a permanent (zero-field) local moment, with magnitude denoted by $|\tm|$. Hence as illustrated schematically in Fig. \ref{fig:fig1}, application of a field $h$ in some arbitrary direction leads to a  non-vanishing magnetization along that direction, which we refer to as an `A-type' LM state for $h>0$ and `B-type' for $h<0$; such that as $h\rightarrow 0\pm$,
\begin{equation}
\label{eq:mlimits}
m(h\rightarrow 0\pm)~=~ \pm|\tm| 
\end{equation}

\noindent The local magnetization is thus discontinuous across $h=0$, which is one hallmark of a LM phase. 
For $h=0$ by contrast, 
\begin{equation}
\label{eq:mh=0}
m(h=0)~=~ 0  
\end{equation}
reflecting simply the fact that the $h=0$ ground state is locally doubly degenerate (with a permanent 
moment that is `equally probably up or down').~\cite{fn0} 

An obvious order parameter for the zero-field QPT between LM and SC phases is thus 
the local moment $|\tm| = m(h=0+)$, since it is non-zero in the LM phase $U>U_{c}$ but vanishes in the SC phase;
and we make the natural assumption that it vanishes continuously as $U \rightarrow U_{c}+$ from the LM phase. Two brief points should be noted here. \\
\noindent (a) The local moment $|\tm|$ is of course a familiar order parameter within a static mean-field approximation (unrestricted Hartree-Fock); which is however well known to overestimate the tendency to local moment formation, and is hence liable to predict spuriously a QPT to a LM phase (as it does e.g.\ for the metallic 
AIM~\cite{anderson}). We emphasize that the present work has nothing to do with mean-field theory.
Nevertheless, where a QPT between SC and LM phases genuinely occurs, an appropriate order parameter for it is certainly the $h\rightarrow 0+$ local magnetiztation -- the local moment $|\tm|$. This fact has arguably been somewhat overlooked in the literature on local quantum phase transitions (exceptions include 
e.g.\ [\onlinecite{Ingersent-Si2002,Pixley2012}]), possibly due to its association with naive
mean-field theory. We will also demonstrate it explicitly in sec.~\ref{subsection:PAIMh=0lm} for the 
pseudogap AIM, from numerical renormalization group (NRG) calculations (see Fig.\ \ref{fig:fig2}).\\
\noindent (b) As mentioned above the magnetization $m(h)$ is non-zero for any $h\neq 0$, 
whether for $U>U_{c}$ or $U<U_{c}$. The pristine QPT is thus strictly destroyed
on application of a finite field: physically, the ground state for $h\neq 0$
in the  $U>U_{c}$  LM phase is no longer doubly degenerate, the field `picking out' 
one or other of the A and B components according to whether $h\gtrless 0$.


\subsubsection{Structure of propagators}
\label{subsubsection:propagators}

The above behavior in the LM phase has important implications for the structure of the local propagators 
which, via eqs.\ (\ref{eq:mn},\ref{eq:nsig}), determine the magnetization. For $h>0$, 
$\Gs(\w,h) \equiv G_{A\sigma}(\w,|h|)$, where the broken symmetry propagator $G_{A\sigma}(\w,|h|)$
reflects the LM state corresponding to $m(h) >0$ for all $h > 0$, and with $m(0+) =+|\tm|$. For 
$h<0$ by contrast, $\Gs(\w,h) \equiv G_{B\sigma}(\w,-|h|)$ refers to $m(h) <0$ for all $h < 0$, with 
$m(0-) =-|\tm|$; such that, from the invariance of $H$ under
$\sigma \leftrightarrow -\sigma$ and $h\leftrightarrow -h$,~\cite{fnABnotation} 
\begin{equation}
\label{eq:ABrel}
G_{A\sigma}(\w, |h|)~=~ G_{B-\sigma}(\w, -|h|)~.
\end{equation}
In otherwords,
\begin{equation}
\label{eq:g2g}
\Gs(\w, h)=\theta(h)~ G_{A\sigma}(\w, |h|)+ \theta(-h)~G_{B\sigma}(\w, -|h|)
\end{equation}
with $\theta(x)$ the unit step function.
In parallel to the discontinuity in the magnetization across $h=0$, the propagator $\Gs(\w,h)$ in 
the LM phase is likewise discontinuous across $h=0$; 
i.e.\ $\Gs(\w, h=0+)=G_{A\sigma}(\w, 0)$ and $\Gs(\w, h=0-)=G_{B\sigma}(\w, 0)$,
where $G_{A\sigma}(\w, 0)$ and $G_{B\sigma}(\w, 0) =G_{A-\sigma}(\w, 0)$ (eq.\ \ref{eq:ABrel}) do not 
coincide $\forall ~\w$ in the LM phase -- as implied by the very existence of a permanent moment
(see eqs.\ (\ref{eq:mAzero},\ref{eq:maitospec})). 

Precisely at $h=0$ by contrast, eq. \ref{eq:g2g} gives
\begin{subequations}
\label{eq:gsigh=0}
\begin{align}
\Gs(\w, h=0)~=&~ \tfrac{1}{2} \left[G_{A\sigma}(\w, 0)~+~ G_{B\sigma}(\w, 0)\right]
\\
=&~\tfrac{1}{2} \left[G_{A-\sigma}(\w, 0)~+~ G_{B-\sigma}(\w, 0)\right]
\\
=&~G_{-\sigma}(\w,0)
\end{align}
\end{subequations}
reflecting the fact that in the absence of a field the ground state is doubly degenerate. Moreover, 
since $G_{A\sigma}(\w, 0)=G_{B-\sigma}(\w, 0)$, eq.\ \ref{eq:gsigh=0} shows that $\Gs(\w,0)$ is indeed 
rotationally invariant ($\sigma$-independent), as it must be at zero-field.

Since $\Gs(\w,h)$ is discontinuous across $h=0$, so too is the spectrum 
$D_{\sigma}(\w, h)=-\frac{1}{\pi}\mathrm{Im}\Gs(\w,h)$. In particular,
\begin{equation}
D_{\sigma}(\w, h=0+)~=~D_{A\sigma}(\w, 0) 
\nonumber
\end{equation}
while for $h=0$ by contrast, $D_{\sigma}(\w, 0)$ is given (eq.\ref{eq:gsigh=0}) by
\begin{equation}
\label{eq:lincolncity}
D_{\sigma}(\w, h=0)~=~\tfrac{1}{2} \left[D_{A\sigma}(\w, 0)~+~ D_{B\sigma}(\w, 0)\right]
\end{equation}
with $D_{B\sigma}(\w, 0)=D_{A-\sigma}(\w, 0) \neq D_{A\sigma}(\w,0)$. Hence,
on switching on even an infinitesimal field, the $\w$-dependence of the single-particle spectrum must 
change abruptly. Such behavior is indeed found in NRG studies of LM phases in a range of different problems, including e.g.\ the underscreened spin-1 phase arising in two-level quantum dots,~\cite{CJW2009,CJWfield}
where it is evident in an abrupt redistribution of spectral weight in the Hubbard bands of $D_{\sigma}$
upon application of even the tiniest magnetic field.~\cite{CJWfield}\\

 Now return to eq.\ \ref{eq:nsig} for $n_{\sigma}(h)$. From eq.\ \ref{eq:g2g}, it is of form
\begin{equation}
\label{eq:nsigh}
n_{\sigma}(h)~=~\theta(h)~n_{A\sigma}^{\pd}(|h|)~+~\theta(-h)~n_{B\sigma}^{\pd}(-|h|)
\end{equation}
with
$n_{A\sigma}^{\pd}(|h|)=~-\frac{1}{\pi}\mathrm{Im} \int^{0}_{-\infty}d\w ~ G_{A\sigma}(\w, |h|)$
(and similarly for $n_{B\sigma}^{\pd}(-|h|)$), such that
\begin{equation}
\label{eq:noddy0}
n^{\pd}_{B-\sigma}(-|h|)~=~n^{\pd}_{A\sigma}(|h|)
\end{equation}
from eq.\ \ref{eq:ABrel}. 
The local charge $n(h) =\sum_{\sigma} n_{\sigma}(h)$ follows using eqs.\ (\ref{eq:nsigh},\ref{eq:noddy0}) as
$n(h)=[\theta(h) +\theta(-h)]~\sum_{\sigma} n^{\pd}_{A\sigma} (|h|)$ $=\sum_{\sigma} n^{\pd}_{A\sigma}(|h|)$,
with the resultant $n(h)= n(-h)$ thus continuous  across $h=0$.
We also define the obvious magnetization
\begin{equation}
\label{eq:noddy1}
m^{\pd}_{A}(|h|)~=~\sum_{\sigma}~\sigma n^{\pd}_{A\sigma}(|h|) 
\end{equation}
such that $m^{\pd}_{A}(|h|)>0$ for all $h \geq 0$ and $U>U_{c}$,  with
\begin{equation}
\label{eq:mAzero}
m^{\pd}_{A}(0) ~=~ |\tm|
\end{equation}
the permanent moment; and likewise $m^{\pd}_{B}(-|h|)=\sum_{\sigma}~\sigma n^{\pd}_{B\sigma}(-|h|)$, 
such that $m^{\pd}_{B}(-|h|)= - m^{\pd}_{A}(|h|)$ from eq.\ \ref{eq:noddy0}.
Hence from eqs.\ (\ref{eq:mn},\ref{eq:nsigh},\ref{eq:noddy1}), the local magnetization
$m(h)=\theta(h)~m^{\pd}_{A}(|h|)+\theta(-h)~m^{\pd}_{B}(-|h|)$ is given simply by
\begin{equation}
\label{eq:basicma}
m(h)~=~\left[\theta(h) ~-~\theta(-h) \right]~m^{\pd}_{A}(|h|)
\end{equation}
This form naturally recovers the symmetries of eqs.\ (\ref{eq:modd}-\ref{eq:mh=0}), 
and just embodies formally the behavior sketched in Fig. \ref{fig:fig1} (upper).
It shows in particular that $m(h)$ in the LM phase is entirely determined (for any field $h$) by
\begin{equation}
\label{eq:maitospec}
m^{\pd}_{A}(|h|)~=~-\tfrac{1}{\pi}~ \mathrm{Im} \sum_{\sigma}~\sigma \int^{E_{\mathrm{F}}=0}_{-\infty}d\w ~ G_{A\sigma}(\w, |h|)
\end{equation}
which is calculable from the spectral densities 
$D_{A\sigma}(\w, |h|) = -\tfrac{1}{\pi}\mathrm{Im}G_{A\sigma}(\w, |h|)$ of
the broken symmetry propagators.

In fact, as generally employed in the following (and for any field $h\gtrless 0$), B-type propagators 
can be eliminated from further consideration, and only the A-type propagators need be considered: 
from eqs.\ (\ref{eq:g2g},\ref{eq:ABrel}), $\Gs(\w, h)$ can be written as
\begin{equation}
\label{eq:g2g1}
\Gs(\w, h)=\theta(h)~ G_{A\sigma}(\w, |h|)~+~ \theta(-h)~G_{A-\sigma}(\w, |h|)
\end{equation} 
and hence for $n_{\sigma}(h)$ (from eqs.\ (\ref{eq:nsigh},\ref{eq:noddy0}))
\begin{equation}
\label{eq:genericnsigma}
n_{\sigma}^{\pd}(h)~=~\theta(h)~n_{A\sigma}^{\pd}(|h|)~+~\theta(-h)~n_{A-\sigma}^{\pd}(|h|)~.
\end{equation}


\subsubsection{Self-energies}
\label{subsubsection:s-estructure}
The propagator $G_{A\sigma}(\w,|h|)$ is given by the Dyson equation in terms of its 
self-energy $\Sigma_{A\sigma}(\w,|h|)$:
\begin{subequations}
\label{eq:Aprop_se}
\begin{align}
G_{A\sigma}(&\w, |h|)=\left[\w^{+} -\epsilon + \sigma h -\Gamma_{\sigma}(\w,h) -\Sigma_{A\sigma}(\w,|h|)
\right]^{-1}
\\ \nonumber \\
=& \left[\w^{+}-\epsilon + \sigma h  -\Gamma_{\sigma}(\w,h) -\Sigma\left[\{G_{A\sigma^{\prime}}\}\right]~\right]^{-1}
\label{eq:Aprop_seb}
\end{align}
\end{subequations}
Here $\Sigma [\{G_{A\sigma^{\prime}}\}]$ indicates that $\Sigma_{A\sigma}$ is a functional of
(the time-ordered) $G_{A\uparrow}$ and $G_{A\downarrow}$, obtained  as a functional derivative of 
the Luttinger-Ward functional (as exploited below) and given diagrammatically by the skeleton expansion.~\cite{LW1960} As such, for any $h \geq 0$, the Dyson equation \ref{eq:Aprop_seb} amounts  
to a self-consistency condition which, given $\Sigma_{A\sigma} =\Sigma [\{G_{A\sigma^{\prime}}\}]$, 
in principle determines the $\{G_{A\sigma}(\w,|h|)\}$ (and in turn the $\{\Sigma_{A\sigma}(\w,|h|)\}$).

 Now consider any field $h \neq0$, say $h>0$ for specificity, for which eq.\ \ref{eq:g2g} gives $\Gs(\w, h) = G_{A\sigma}(\w, |h|)$. Since the propagators $\Gs(\w, h)$ and $G_{A\sigma}(\w, |h|)$ are coincident, so too (trivially) are the associated self-energies (eqs.~\ref{eq:Gs},\ref{eq:Aprop_se}), i.e.\
$\Ss (\w, h) \equiv \Sigma_{A\sigma}(\w, |h|)$, or equivalently
$\Sigma [\{G_{\sigma^{\prime}}\}]\equiv \Sigma [\{G_{A\sigma^{\prime}}\}]$.
There is nothing particularly subtle here: that $\Gs(\w, h) = G_{A\sigma}(\w, |h|)$ 
simply reflects the fact that for $h> 0$ (and $h\neq 0$ generally) the ground state of the system in the 
LM phase is no longer doubly degenerate, the field `picking out' as the ground state the appropriate spin 
component (as illustrated in Fig.\ \ref{fig:fig1}). Although the transition is strictly destroyed on 
application of a field, it is nevertheless natural to retain the `A' label in $G_{A\sigma}(\w, |h|)$ and $\Sigma_{A\sigma}(\w, |h|)$ for $U>U_{c}$; in recognition of the fact that $\Gs(\w, h)$ as a function of 
$\w$ is discontinuous across $h=0$, and to indicate that as $h$ vanishes the system possesses the permanent 
local moment that is symptomatic of the LM phase.


\subsection{Zero field}
\label{subsection:secIIh=0}

But now let us consider $h=0$, of particular interest since here the transition at $U=U_{c}$
between SC and LM phases is pristine. This case is quite subtle. From eq.\ \ref{eq:g2g1},
the (spin-independent) \emph{zero-field} propagator is 
$\Gs(\w,0)= \tfrac{1}{2} \left[G_{A\sigma}(\w, 0)+ G_{A-\sigma}(\w, 0)\right]$.
This is a \emph{two-self-energy} (TSE) description of $\Gs(\w,0)$, since $\Gs(\w,0)$ is thereby 
specified in terms of the two distinct self-energies $\Sigma_{A\uparrow}(\w,0)$ and $\Sigma_{A\downarrow}(\w,0)$. $\Gs(\w,0)$ is however given by eq.\ \ref{eq:Gs} in terms of the \emph{single} self-energy 
$\Sigma(\w) \equiv \Ss(\w,0)$ $(=\Sigma_{-\sigma}(\w,0))$, whence direct comparison between 
eq.\ \ref{eq:Gs} and eq.\ \ref{eq:Aprop_se} for $\tfrac{1}{2} \left[G_{A\sigma}(\w, 0)+ G_{A-\sigma}(\w, 0)\right]$ specifies the exact relation between $\Sigma(\w)$ and the two self-energies $\Sigma_{A\uparrow}(\w,0)$, $\Sigma_{A\downarrow}(\w,0)$ characteristic of the LM phase:
\begin{equation}
\label{eq:SSE_TSE}
\begin{split}
\Sigma (\w)~&=~\tfrac{1}{2} \left[\Sigma_{A\uparrow}(\w,0)+\Sigma_{A\downarrow}(\w,0)
\right]
\\
& +~\frac{\left[\tfrac{1}{2} \left(\Sigma_{A\uparrow}(\w,0)-\Sigma_{A\downarrow}(\w,0) \right)\right]^{2}}
{G_{0}^{-1}(\w,0) - \tfrac{1}{2} \left[\Sigma_{A\uparrow}(\w,0)+\Sigma_{A\downarrow}(\w,0) \right]}
\end{split}
\end{equation}
with
\begin{equation}
G_{0}^{\pd}(\w, 0)~=~ \left[ \w^{+} -\epsilon - \Gamma(\w)\right]^{-1}
\end{equation}
the non-interacting propagator (and $\Gamma(\w) = \Gamma_{\sigma}(\w, 0)$). The ramifications of this will be considered in secs.~\ref{section:sumrules} \emph{ff}, but here we make three initial comments. \\
\noindent (i) We emphasize that it is the self-energies $\Sigma_{A\sigma}$ which are directly calculable from 
 many-body perturbation theory in the LM phase, e.g.\ by the implicit self-consistency equation~\ref{eq:Aprop_seb} 
as mentioned above. In particular (see sec. \ref{section:sumrules}), it is \emph{these} self-energies to which 
classic methods of many-body theory~\cite{LW1960,Luttingerf1960FermiSurf,Luttingerf1961} apply (as will be
exploited in subsequent sections). In this sense $\Sigma(\w) = \Ss(\w,0)$ -- the conventional single self-energy 
in the LM phase for $h=0$ -- is a derivative quantity, being calculable via eq. \ref{eq:SSE_TSE} from a knowledge 
of the $\Sigma_{A\sigma}(\w,0)$.\\
\noindent (ii) It is nonetheless the single self-energy $\Ss(\w, 0)$ of eq.\ \ref{eq:Gs} that is traditionally considered (and referred to as `the' self-energy), even in a LM phase. This single self-energy 
(for any $h$) is also directly calculable using NRG,~\cite{Wilson75,kww1,kww2,NRGrmp} 
via the ratio~\cite{bullahewprus} $\Ss(\w, h)=F_{\sigma}(\w,h)/\Gs(\w,h)$;
where (in standard notation) the correlation function
$F_{\sigma}(\w,h)=\langle\langle [d_{\sigma}, H_{I}];d_{\sigma}^{\dagger}\rangle\rangle$,
with $H_{I}$  the interaction part of the Hamiltonian ($H_{I}=U\hat{n}_{\uparrow}\hat{n}_{\downarrow}$ 
for a single-level AIM). Both $\Gs(\w,h)$ and $F_{\sigma}(\w,h)$ are directly computable from NRG,~\cite{bullahewprus} enabling $\Ss(\w,h)$ to be obtained, and thus in particular the zero-field 
$\Sigma(\w) \equiv \Ss(\w,0)$.\\
(iii) We point out that the self-energies $\Sigma_{A\sigma}(\w, 0)$ are \emph{also} directly calculable 
from NRG -- they can be obtained by just the same method, but in the limit $h\rightarrow 0+$. We have thus 
used the full density matrix (FDM) generalisation~\cite{PetersPruschkeAnders2006,WeichselbaumDelft2007} 
of NRG to calculate separately both $\Ss(\w, 0)$ and the $\Sigma_{A\sigma}(\w, 0)$, for the LM phase of the 
pseudogap AIM; and have thereby confirmed that the relation between them, eq.~\ref{eq:SSE_TSE},
is indeed satisfied by the numerics. 

While our emphasis here has naturally been on the LM phase, we would also point out that all equations, 
from eqs.\ (\ref{eq:mlimits} - \ref{eq:Aprop_se}), hold equally for the SC phase $U<U_{c}$ on simply 
dropping the A or B labels (and of course setting $|\tm| =0$). In this case, the relevant equations 
reduce either to trivial identities, or to one or other of 
eqs.\ (\ref{eq:Gs},\ref{eq:gsig},\ref{eq:modd},\ref{eq:mh=0}).
The minor point here is that the notation used for the LM phase reduces very simply to that appropriate 
in the SC phase. Note also for $h=0$ in particular that, on dropping the A-labels, precisely the same self-consistency condition as for the LM phase (eq.\ \ref{eq:Aprop_seb}) determines the propagators $\Gs$ 
in the SC phase; but simply with different self-consistent \emph{solutions} thereto, according to whether 
$U<U_{c}$ or $>U_{c}$.

One aspect of the preceding discussion merits further brief comment. Conventional diagrammatic field theory 
for the zero-temperature, $t$-ordered  propagators assumes the global ground state $|\Psi_{0}\rangle$ of the full $H$ to be non-degenerate.~\cite{fwbook, agdbook} In the presence of a field, however small, that is indeed the case whether $U>U_{c}$ or $<U_{c}$. But in the absence of a field, while the global ground state is non-degenerate in the SC phase $U<U_{c}$, it is doubly degenerate in the LM phase for $U>U_{c}$. It is essentially for this reason that, to gain a tangible handle on the zero-field LM phase, it is necessary to consider the general case of 
$h\neq 0$; with the degenerate zero-field LM phase obtained from the limits $h\rightarrow 0\pm$, as described above.


\subsection{`Excess' charge and magnetization}
\label{subsection:excessprops}

We have so far considered the local (impurity) $n_{\sigma}(h)$. Of well known importance in any quantum 
impurity problem~\cite{hewsonbook} are of course so-called `excess' quantities, namely the \emph{difference} 
in a particular property calculated with and without the impurity present. Central among these is 
$\nimps(h)$, the difference in the number of $\sigma$-spin electrons in the entire system, with and without 
the impurity:
$\nimps(h) =\langle \sum_{\mathbf{k}} \hat{n}^{\pd}_{\mathbf{k}\sigma} +\hat{n}_{\sigma}\rangle -\langle \sum_{\mathbf{k}} \hat{n}^{\pd}_{\mathbf{k}\sigma}\rangle_{0}^{\pd}$, where 
$\langle \cdot\cdot\cdot \rangle_{0}^{\pd}$ denotes an average in the absence of the impurity (and
$\hat{n}^{\pd}_{\mathbf{k}\sigma} =c_{\mathbf{k}\sigma}^{\dagger}c_{\mathbf{k}\sigma}^{\pd}$,
$\hat{n}_{\sigma}^{\pd} =d_{\sigma}^{\dagger}d_{\sigma}^{\pd}$). 
The importance of $\nimps$ arises in large part because the Friedel(-Luttinger) sum rule applies to it, 
as shown in sec.\ \ref{section:sumrules} \emph{ff}. Using standard equation of motion methods,~\cite{Zubarev}
$\nimps(h)$ is easily shown to be~\cite{hewsonbook}
\begin{equation}
\label{eq:nimpsig}
\nimps(h) ~=~ \int^{0}_{-\infty}d\w ~ \rhoimps
\end{equation}
where the excess density of states  $\rhoimps$ is given by
\begin{equation}
\label{eq:rhoimp}
\rhoimps = -\tfrac{1}{\pi}\mathrm{Im}\left(\Gs(\w, h) \left[ 1- \frac{\partial \Gamma_{\sigma}(\w, h)}{\partial\w}
\right] \right)
\end{equation}
in terms of the local propagator $\Gs(\w,h)$. The corresponding excess magnetization and charge are 
then given obviously (\emph{cf} eq.\ \ref{eq:mn}) by
\begin{equation}
\label{eq:mimpnimp}
\mimp(h)~=~\sum_{\sigma} ~\sigma \nimps (h) ~~~~\nimp(h)~=~\sum_{\sigma} ~\nimps(h)~.
\end{equation}

Now focus on the LM phase (all relevant formulae for the SC phase again follow from those appropriate to the LM phase, simply by dropping the A or B labels). Since the local propagator enters eq.\ \ref{eq:rhoimp}, it takes the same form as eq.\ \ref{eq:g2g} for $\Gs$, namely
$\Delta\rho^{\pd}_{\mathrm{imp},\sigma}(\w, h)=\theta (h)\Delta\rho^{\pd}_{\mathrm{imp},A\sigma}(\w, |h|)
+\theta(-h)\Delta\rho^{\pd}_{\mathrm{imp},B\sigma}(\w, -|h|)$;
where (\emph{cf} eq.\ \ref{eq:ABrel})
$\Delta\rho^{\pd}_{\mathrm{imp},A\sigma}(\w, |h|)=\Delta\rho^{\pd}_{\mathrm{imp},B-\sigma}(\w, -|h|)$
such that only $\Delta\rho^{\pd}_{\mathrm{imp},A\sigma}(\w, |h|)$ need ever be considered. The basic `excess' quantities thus have exactly the same form as for their purely local counterparts 
(eqs.\ \ref{eq:genericnsigma}, \ref{eq:basicma}, \ref{eq:noddy1},\ref{eq:maitospec}), namely
\begin{equation}
\label{eq:rhoimpAB}
\nimps (h) =\theta (h) ~ \nimpas (|h|)~+~\theta(-h)~n^{\pd}_{\mathrm{imp},A-\sigma}(|h|)
\end{equation}
and     
\begin{equation}
\label{eq:mimpLM}
\mimp (h) =\left [\theta(h)~-~\theta(-h)\right]~\mimpa(|h|)
\end{equation}
with $\mimpa (|h|) =\sum_{\sigma}\sigma \nimpas (|h|)$ given by
\begin{equation}
\label{eq:mimpAlm}
\mimpa (|h|) =\sum_{\sigma}~\sigma ~\int^{0}_{-\infty}d\w ~ 
\Delta\rho^{\pd}_{\mathrm{imp}, A\sigma}(\w, |h|)~.
\end{equation}


\section{Sum Rules and Luttinger integrals}
\label{section:sumrules}

In this section, for both $U<U_{c}$ and $>U_{c}$, we consider the Friedel-Luttinger sum 
rules~\cite{CJW2009} (eqs.\ (\ref{eq:flsrsc},\ref{eq:flsrlm})) which relate a static scattering 
phase shift to the corresponding excess charge and a Luttinger integral. The arguments given below 
hold for any field, including the $h=0$ case of particular interest. The relevant Luttinger integrals 
are then shown in sec.\ \ref{subsection:Luttintgen} to vanish in \emph{both} the SC and LM phases; leading 
to recovery of a Friedel sum rule~\cite{hewsonbook,Langreth1966} for \emph{each} phase, and thereby relating 
the excess charges $\nimps$ and $\nimpas$ to the so-called renormalized levels for the appropriate phase. 
In sec.\ \ref{subsection:Luttinth=0} we consider the LM phase at zero-field, and the phase shift defined 
in terms of the conventional single self-energy $\Sigma(\w) \equiv \Ss(\w, h=0)$. Here we show that while 
a Friedel-Luttinger sum rule again arises (eq.\ \ref{eq:pssse}), the associated Luttinger integral cannot 
be argued to vanish.

We consider first $U<U_{c}$, and hence the SC phase. The static phase shift 
$\delta_{\sigma}^{\pd}(h)$  $(=\delta_{-\sigma}^{\pd}(-h))$ is defined by
\begin{equation}
\label{eq:pssc}
\delta_{\sigma}^{\pd}(h) = \mathrm{arg}\left[\Gs(\w, h)\right]\Bigr{|}_{\w =-\infty}^{\w = 0} =
\mathrm{Im} \ln \Gs(0, h)+ \pi
\end{equation}
(with $\mathrm{arg}[\Gs(-\infty, h)] =-\pi$ from eq.\ \ref{eq:Gs}). Writing equivalently 
$\delta^{\pd}_{\sigma}(h) = \mathrm{Im}\int^{0}_{-\infty} d\w \frac{\partial}{\partial \w} \ln \Gs(\w,h)$, and using eq.\ \ref{eq:Gs} for $\Gs (\w, h)$, a short calculation then gives
\begin{equation}
\label{eq:flsrsc}
\delta_{\sigma}^{\pd}(h)~=~  \pi \nimps(h) + ~I_{L_{\sigma}}(h). 
\end{equation}
This Friedel-Luttinger sum rule relates the phase shift to the excess charge $\nimps(h)$ 
(eq.\ \ref{eq:nimpsig}) and the Luttinger integral $I_{L_{\sigma}}(h)$ $( =I_{L_{-\sigma}}(-h))$ 
given by:
\begin{equation}
\label{eq:LISC}
I_{L_{\sigma}}(h)~=~\mathrm{Im} \int^{0}_{-\infty} d\w ~\frac{\partial \Sigma_{\sigma}(\w,h)}{\partial \w}~
G_{\sigma}(\w, h) 
\end{equation}
We remind the reader that the functions here are all retarded, such that e.g. 
$\Ss(\w, h) = \Ss^{R}(\w, h) -i\Ss^{I}(\w,h)$; with the $t$-ordered self-energy 
$\Ss^{t}(\w,h)= \Ss^{R}(\w, h) -i\mathrm{sgn}(\w)\Ss^{I}(\w,h)$ (or alternatively its 
Matsubara counterpart) given as a functional derivative of the Luttinger-Ward functional, 
\emph{viz} $\Ss^{t}(\w, h)=\delta \Phi_{LW}/\delta \Gs^{t}(\w,h)$ with
$\Phi_{LW}^{\pd} \equiv \Phi_{LW}^{\pd}[\{\Gs^{t} \}]$ a functional of 
the $\{\Gs^{t}(\w, h)\}$.

From  eq.\ \ref{eq:pssc} the phase shift is expressible solely in terms of $\mathrm{arg}[\Gs(\w , h)]$
at the Fermi level $\w =0$. A simple calculation using eq.\ \ref{eq:Gs} then gives
\begin{equation}
\label{eq:pssc*}
\delta_{\sigma}^{\pd}(h)~=~ \frac{\pi}{2}~-~\mathrm{tan}^{-1}\left[
\frac{\epsilon^{*}_{\sigma}(h)}{\eta +\Gamma^{I}_{\sigma}(0,h) +\Ss^{I}(0,h)}
\right]
\end{equation}
(where the $\mathrm{arctan} \in [-\frac{\pi}{2}, +\frac{\pi}{2}]$). This relates the phase shift to
the renormalized level $\epsilon^{*}_{\sigma}(h)$ $(=\epsilon^{*}_{-\sigma}(-h))$ given by
\begin{equation}
\label{eq:rlsc}
\epsilon_{\sigma}^{*}(h)~=~\epsilon -\sigma h +\Gamma^{R}_{\sigma}(0,h) + \Ss^{R}(0,h),
\end{equation} 
which embodies the interaction-induced renormalization of the bare level energy $\epsilon$.

The above results are considered further below.  They are of course well known for the SC phase,~\cite{hewsonbook} which is perturbatively connected (in $U$) to the non-interacting limit and is thus a Fermi liquid. For $U>U_{c}$ by contrast, 
the zero-field LM phase is separated from the SC phase by a quantum phase transition 
at $U_{c}$. It is not therefore perturbatively connected to the non-interacting limit, and as such is a non-Fermi liquid. Importantly, however, directly analogous results to those given above carry over 
\emph{mutatis mutandis} for $U>U_{c}$, but now for the phase shift 
$\delta_{A\sigma}^{\pd}(|h|)=\mathrm{arg} G_{A\sigma}(0, |h|)+ \pi$ defined in terms of the broken symmetry propagators [and with $\delta_{A\sigma}^{\pd}(|h|)=\delta_{B-\sigma}^{\pd}(-|h|)$ such that only $h=|h|\geq 0$ 
and hence the A-type phase shifts need be considered]; specifically
\begin{equation}
\label{eq:flsrlm}
\delta_{A\sigma}^{\pd}(|h|)~=~  \pi \nimpas(|h|) + ~I_{L_{A\sigma}}(|h|)  
\end{equation}
where the Luttinger integral $I_{L_{A\sigma}}(|h|)$ $(= I_{L_{B-\sigma}}(-|h|))$ is now
\begin{equation}
\label{eq:LILM}
I_{L_{A\sigma}}(|h|)~=~\mathrm{Im} \int^{0}_{-\infty} d\w ~\frac{\partial \Sigma_{A\sigma}(\w,|h|)}{\partial \w}~
G_{A\sigma}(\w, |h|) ~.
\end{equation}
Note that the self-energy is again given from
$\Sigma_{A\sigma}^{t} (\w, |h|)=\delta \Phi_{\mathrm{LW}}/\delta G_{A\sigma}^{t}(\w,h)$, 
with $\Phi_{\mathrm{LW}}^{\pd} \equiv \Phi_{\mathrm{LW}}^{\pd}[\{G_{A\sigma}^{t} \}]$ precisely the same 
functional of the $\{G_{A\sigma}^{t} \}$ for $U>U_{c}$ as it is of the $\{ \Gs^{t} \}$ in the SC phase. 
Likewise the phase shift is given by 
\begin{equation}
\label{eq:pslm*}
\delta_{A\sigma}^{\pd}(|h|)= \frac{\pi}{2}-\mathrm{tan}^{-1}\left[
\frac{\epsilon^{*}_{A\sigma}(|h|)}{\eta +\Gamma^{I}_{\sigma}(0,|h|) +\Sigma_{A\sigma}^{I}(0,|h|)}
\right]
\end{equation}
in terms of the corresponding renormalized level $\epsilon_{A\sigma}^{*}(|h|)$ $(=\epsilon_{B-\sigma}^{*}(-|h|))$:
\begin{equation}
\label{eq:rllm}
\epsilon_{A\sigma}^{*}(|h|)~=~\epsilon -\sigma |h| +\Gamma^{R}_{\sigma}(0,|h|) + \Sigma_{A\sigma}^{R}(0,|h|)
\end{equation}
As expected (sec.\ \ref{section:LMgeneral}), results for $\delta_{\sigma}^{\pd}(h)$ thus follow directly from those for $\delta_{A\sigma}^{\pd}(|h|)$, simply on dropping the A-label.

We also add here that the static phase shifts are simply related to the local single-particle spectrum at 
the Fermi level $\w =0$, by
\begin{equation}
\label{eq:dosps1}
\pi \left[\Gamma^{I}_{\sigma}(0, |h|) + \Sigma^{I}_{A\sigma}(0, |h|)\right] D_{A\sigma}(0, |h|))~=~
\mathrm{sin}^{2}\left(\delta^{\pd}_{A\sigma}(|h|) \right)
\end{equation}
(and likewise for $U<U_{c}$ on dropping the A-labels).


\subsection{Luttinger integrals and Friedel sum rules}
\label{subsection:Luttintgen}

The propagators and self-energies are analytic functions of frequency  everywhere except on the real axis. From this, using standard methods of complex analysis, it is straightforward (if lengthy) to show that the Luttinger integral is given by 
\begin{equation}
\label{eq:LIdetail1}
\begin{split}
 I_{L_{A\sigma}}(|h|)= ~& \frac{1}{2i}   \int^{+\infty}_{-\infty} d\w ~\Sigma_{A\sigma}^{t}(\w, |h|) \frac{\partial G_{A\sigma}^{t}(\w, |h|)}{\partial \w}
\\
&~-~ \Sigma_{A\sigma}^{I}(0, |h|)~ G_{A\sigma}^{R}(0, |h|)
\end{split}
\end{equation}
and likewise for $I_{L_{\sigma}}(h)$ on dropping the A-label. It is the integral on the right hand side of
eq.\ \ref{eq:LIdetail1} that appears in Luttinger and Ward's seminal
work,\cite{LW1960} and which is readily shown to vanish (for any field $h$ and spin $\sigma$) by a standard argument we briefly recap: recalling that the self-energy is a functional derivative of $\Phi_{LW}^{\pd}$ (as above), one considers a variation $\delta\Phi_{LW}^{\pd}$ in which the frequencies of all propagators of a given spin type, $\sigma$, in any given diagram are shifted uniformly from $\w$ to $\w +\delta\w^{\prime}$; i.e.\ 
$\delta G_{A\sigma}^{t}(\w, |h|) =G_{A\sigma}^{t}(\w +\delta\w^{\prime}, |h|) - G_{A\sigma}^{t}(\w, |h|) \equiv \delta\w^{\prime} (\partial G_{A\sigma}^{t}(\w, |h|)/\partial\w)$, 
for which
$\delta\Phi_{LW}^{\pd}=\int^{\infty}_{-\infty}d\w ~(\delta \Phi_{LW}/\delta G_{A\sigma}^{t}(\w, |h|))~\delta G_{A\sigma}^{t}(\w, |h|)$. 
But by virtue of the fact that both frequency and spin are conserved at each vertex in 
any closed linked diagram contributing to $\Phi_{LW}^{\pd}$,~\cite{LW1960,Luttingerf1961} it follows simply 
that $\delta\Phi_{LW}^{\pd} =0$ for \emph{any} $\delta\w^{\prime}$. Hence 
$0= \int^{\infty}_{-\infty}d\w ~(\delta \Phi_{LW}/\delta G_{A\sigma}^{t}(\w, |h|))~
(\partial G_{A\sigma}^{t}(\w, |h|)/\partial\w) $
$=\int^{\infty}_{-\infty}d\w ~\Sigma^{t}_{A\sigma}(\w, |h|)~
(\partial G_{A\sigma}^{t}(\w, |h|)/\partial\w)$, as required.

Similarly, considering order by order the skeleton expansion diagrams for the self-energy,
its imaginary part at the Fermi level can also be shown to vanish, following the analysis 
(`phase space arguments') given originally by Luttinger,~\cite{Luttingerf1961} i.e.\
\begin{equation}
\label{eq:ses=0}
\Sigma_{A\sigma}^{I}(0, |h|)~=~0 ~=~\Ss^{I}(0, h)
\end{equation}
$\Ss^{I}(0, h) =0$ is of course the standard result for the Fermi liquid, SC phase, but we emphasize that the argument for $\Sigma_{A\sigma}^{I}(0, |h|)=0$ is just the same (reflecting the fact that, in skeleton form, $\Sigma_{A\sigma}^{t}(\w, |h|)$ is the same functional of the $\{G_{A\sigma}^{t}(\w, |h|)\}$ that $\Ss^{t}(\w, h)$ is of the $\{\Gs^{t}(\w, h)\}$ in the SC phase). Further details will be given in 
sec.\ \ref{section:SSElowomegaLW}, since the same arguments allow the  low-$\w$ asymptotic behavior of 
the self-energies to be obtained.

 Using eq.\ \ref{eq:ses=0}, Luttinger's theorem thus  holds for \emph{both} the SC and LM phases, i.e.\
the Luttinger integrals vanish 
\begin{equation}
\label{eq:LIs=0}
I_{L_{\sigma}}(h)~=~ 0~=~I_{L_{A\sigma}}(|h|)   ~~~(~~=~I_{L_{B-\sigma}}(-|h|) ),
\end{equation}
which we reiterate holds for all fields, including $h=0$.
With this eq.\ \ref{eq:flsrlm}, as well as its familiar counterpart eq.\ \ref{eq:flsrsc} for the SC phase, reduces to a Friedel sum rule~\cite{hewsonbook,Langreth1966} relating the phase shift to the excess charge; which in turn is related to the renormalized levels by eqs.\ \ref{eq:pslm*},\ref{eq:pssc*}, \emph{viz}:
\begin{subequations}
\label{eq:nimpsboth}
\begin{align}
\nimpas (|h|) ~=&~\frac{1}{2}~-~\frac{1}{\pi}~\mathrm{tan}^{-1}\left[
\frac{\epsilon^{*}_{A\sigma}(|h|)}{\eta +\Gamma^{I}_{\sigma}(0,|h|)}
\right]
\\ 
\nimps (h) ~ =&~\frac{1}{2}~-~\frac{1}{\pi}~\mathrm{tan}^{-1}\left[
\frac{\epsilon^{*}_{\sigma}(h)}{\eta +\Gamma^{I}_{\sigma}(0,h)}
\right]
\end{align}
\end{subequations}
As will be shown in secs.\ \ref{section:PAIMh=0} \emph{ff}, these equations enable us to make a number 
of exact deductions about the nature of the SC and LM phases for the pseudogap (and also gapped) Anderson 
impurity model. First, however, we revisit the LM phase at zero-field, and the phase shift expressed in 
terms of the conventional single self-energy.


\subsection{The zero-field LM phase and the conventional single self-energy}
\label{subsection:Luttinth=0}

The above results hold for any field, including $h=0$, and for both $U<U_{c}$ and $U>U_{c}$.
But now let us look from another angle at $h=0$, where the QPT between the LM and SC phase is pristine.
Here, as explained in sec.\ \ref{subsection:secIIh=0}, the conventional single self-energy 
$\Sigma(\w)\equiv \Ss(\w, 0)$ in the LM phase for $U>U_{c}$ does not coincide with the 
$\Sigma_{A\sigma}(\w, 0)$, but is instead related to the two-self-energies by eq.\ \ref{eq:SSE_TSE}. 
The propagator $\Gs(\w, 0)$ (which is spin-independent for $h=0$) is nevertheless  the same object regardless 
of whether we choose to express it as eq.\ \ref{eq:gsigh=0} in terms of the two self-energy description, or 
as eq.\ \ref{eq:Gs} in terms of the single self-energy. Accordingly, in the LM phase we can repeat just the same analysis for the static phase shift in terms of the single self-energy, that was given above 
(eqs.\ \ref{eq:flsrsc}-\ref{eq:rlsc}) for the SC phase. For $U>U_{c}$ the phase shift $\delta$ (independent 
of spin $\sigma$ for $h=0$) is again defined by eq.\ \ref{eq:pssc} ($\delta = \mathrm{arg}[\Gs (0,0)]+\pi$). Repeating the simple calculation leading to eq.\ \ref{eq:flsrsc}, using eq.\ \ref{eq:Gs} for $\Gs(\w, 0)$ in terms of the single self-energy $\Sigma(\w)\equiv \Ss(\w, 0)$, gives 
\begin{equation}
\label{eq:pssse}
\delta = \pi \nimps (0)~+~I_{L} ~~~~~ =~\tfrac{\pi}{2} \nimp(0)~+~I_{L}
\end{equation}
with
\begin{equation}
\label{eq:LIsse}
I_{L}~=~\mathrm{Im}\int^{0}_{-\infty}d\w ~\frac{\partial \Sigma (\w)}{\partial \w}~\Gs(\w, 0)
\end{equation}
a Luttinger integral expressed in terms of the single self-energy. 
Similarly, proceeding just as in sec.\ \ref{section:sumrules} above, $\delta$ is related to a ($\sigma$-independent) renormalized level $\epsilon^{*}$ by
\begin{equation}
\label{eq:pslmsse}
\delta~=~ \frac{\pi}{2}~-~\mathrm{tan}^{-1}\left[
\frac{\epsilon^{*}}{\eta +\Gamma^{I}_{\sigma}(0,0) +\Sigma^{I}(0)}
\right]
\end{equation}
($\Gamma_{\sigma}(\w, 0) \equiv \Gamma(\w, 0)$ is naturally independent of spin),
with $\epsilon^{*}$ given by:
\begin{equation}
\label{eq:rllmsse}
\epsilon^{*}~=~\epsilon +\Gamma^{R}_{\sigma}(0,0) + \Sigma^{R}(0)
\end{equation}

The single-particle spectrum at the Fermi level is likewise related to the phase shift $\delta$ by
(\emph{cf} eq.\ \ref{eq:dosps1})
\begin{equation}
\label{eq:specsseps}
\pi \left[\Gamma^{I}_{\sigma}(0,0) + \Sigma^{I}(0)\right] D_{\sigma}(0,0))~=~
\mathrm{sin}^{2}(\delta).
\end{equation}
This applies to the generic case where $\delta (\w)=\mathrm{arg}[\Gs (\w,0)]$ is continuous in $\w$ 
across the Fermi level. If by contrast it is discontinuous across $\w =0$, of form
$\delta (\w) \overset{\w \rightarrow 0\pm}\sim \delta \pm\Delta$, then 
the generalization of eq.\ \ref{eq:specsseps} is readily shown to be
\begin{equation}
\label{eq:dosps1a}
\underset{\w \rightarrow 0\pm}{\mathrm{lim}}
\left( \pi [\Gamma^{I}_{\sigma}(\w,0) + \Sigma^{I}(\w)] D_{\sigma}(\w,0))\right)=
\mathrm{sin}^{2}(\delta \pm \Delta).
\end{equation}
In practice, this case is relevant only to the p-h symmetric limit of the pseudogap AIM,
as considered in sec.\ \ref{subsection:SSElowomega}.\\

Eqs.\ (\ref{eq:pssse}-\ref{eq:specsseps}) are simply the direct analogues of their counterparts in the SC phase, given above and expressed in terms of the single self-energy. However the arguments given in 
sec.\ \ref{subsection:Luttintgen} for the vanishing of the Luttinger integrals $I_{L_{\sigma}}$ and $I_{L_{A\sigma}}$ hinge on the fact that in each case the relevant self-energy in skeleton form is a functional derivative of $\Phi_{LW}^{\pd}$ with respect to the appropriate propagator ($G_{A\sigma}$ or $\Gs$).
The zero-field single self-energy in the LM phase is not by contrast expressible as such a functional derivative, and the Luttinger integral $I_{L}$ cannot therefore be argued to vanish.  It must thus be determined in some other way. We consider this in sec.\ \ref{subsection:ILprime} for the pseudogap AIM, in sec.\ \ref{section:atomlimh=0} for the simple case of a single correlated level, and in  sec.\ \ref{section:Luttintgen} for 
the underscreened spin-1 phase of a two-level impurity model;~\cite{CJW2009}
finding in all cases that $I_{L}$ is generically non-vanishing, and that $|I_{L}|$ has a universal value characteristic of the LM phase.


\section{Simple example: atomic limit}
\label{section:atomlimh=0}

We turn briefly to an almost trivial problem: the zero-field atomic limit of an AIM, i.e.\ a single correlated level with $H= \sum_{\sigma} \epsilon\hat{n}_{\sigma}+U\hat{n}_{\uparrow}\hat{n}_{\downarrow}$.
Simple though it is, and devoid of a QPT worth the name, it nonetheless provides the simplest 
illustration of parts of the preceding discussion; for which reason we include it.

The exact Green function here follows straightforwardly from equation of motion methods,~\cite{hewsonbook,Zubarev} is given by
\begin{equation}
\label{eq:galeom}
\Gs (\w, 0)~=~ \frac{1-n_{-\sigma}(0)}{\w^{+} -\epsilon}~+~\frac{n_{-\sigma}(0)}{\w^{+} -\epsilon-U}
\end{equation}
and is of course independent of $\sigma$. The ground state occupancy 
$n_{\sigma}(0) =\langle\hat{n}_{\sigma}\rangle$ is easily determined, and two distinct 
regimes arise:~\cite{hewsonbook} \\
\noindent (a) For $\epsilon <-U$,  the gound state is doubly occupied with
$n_{\uparrow}(0) =1=n_{\downarrow}(0)$ [for $\epsilon >0$ there is of course a hole-analogue with 
$n_{\uparrow}(0) =0=n_{\downarrow}(0)$, which we omit from explicit consideration]. 
This is the `Fermi liquid' regime.\\
\noindent (b) For $-U < \epsilon < 0$, the ground state is by contrast singly occupied, with
$n_{\uparrow}(0) =\tfrac{1}{2}=~n_{\downarrow}(0)$. This is the LM regime, accessed 
for given $\epsilon <0$ by increasing $U$ through `$U_{c}$'$=-\epsilon$, where a trivial ground 
state level-crossing between doubly- and singly-occupied regimes occurs.

 In the doubly-occupied regime (a), the propagator follows from
eq.\ \ref{eq:galeom} as $G_{\sigma}(\w,0) = [\w^{+} -\epsilon -U]^{-1}$; i.e. is of form
(eq.\ \ref{eq:Gs}) $G_{\sigma}(\w,0) = [\w^{+} -\epsilon -\Ss(\w,0)]^{-1}$ with a self-energy
$\Ss(\w,0) =U$. Three points should be noted. 
(i) This self-energy corresponds simply to first-order self-consistent perturbation theory, i.e.\ to
$\Ss(\w,0) =U n_{-\sigma}(0)$.~\cite{fn8a}
(ii) Since $\langle\hat{n}_{-\sigma}\rangle =1$ for all $U<-\epsilon$ --- and thus in particular for $U=0$ --- $\Ss(\w,0) =U$ corresponds equivalently to `bare' perturbation theory in $U$ about the non-interacting limit 
(the Hartree bubble, $U\langle\hat{n}_{-\sigma}\rangle_{U=0}^{\phantom\dagger})$; reflecting the fact that 
this regime is perturbatively connected to the non-interacting limit $U=0$, and as such is a Fermi liquid 
(albeit a trivial one). 
(iii) Since $\Ss(\w,0) =U$ is $\w$-independent, it follows that the Luttinger integral 
(eq.~\ref{eq:LISC}) trivially vanishes, $I_{L_{\sigma}}(h=0) =0$; as the general arguments of
sec.~\ref{subsection:Luttintgen} indeed require (eq.\ \ref{eq:LIs=0}).

 For the singly-occupied LM regime (b), eq.\ \ref{eq:galeom} (with $n_{-\sigma}(0)=\tfrac{1}{2}$) instead gives
\begin{equation}
\label{eq:Gsallm}
\Gs (\w, 0)~=~ \frac{1}{2}\left[\frac{1}{\w^{+} -\epsilon}~+~\frac{1}{\w^{+} -\epsilon-U}
\right]~.
\end{equation}
This is indeed precisely of form eq.\ \ref{eq:gsigh=0}, with 
\begin{subequations}
\label{eq:altsepropsh=0}
\begin{align}
G_{A\uparrow}(\w, 0)~=~G_{B\downarrow}(\w, 0)~=&~
\left[\w^{+}-\epsilon\right]^{-1}
\label{eq:altsepropsh=0a}
\\
G_{A\downarrow}(\w, 0)~=~G_{B\uparrow}(\w, 0)~=&
~\left[\w^{+}-\epsilon -U\right]^{-1}~.
\label{eq:altsepropsh=0b}
\end{align}
\end{subequations}
The zero-field `A' and `B'-type states correspond respectively to
$h\rightarrow 0\pm$ (Fig.\ \ref{fig:fig1}), i.e.\ $n_{A\sigma}(0) =n_{\sigma}(h=0+)$
and $n_{B\sigma}(0) =n_{\sigma}(h=0-)$ (eq.\ \ref{eq:nsigh}), with $n_{A\sigma}(0)=n_{B-\sigma}(0)$ by symmetry.
Physically, the situation here is simple: the A-type state corresponds to an $\uparrow$-spin occupied ground
state, with $n_{A\uparrow}(0) =1,~n_{A\downarrow}(0)=0$ and hence a fully saturated local moment 
$m_{A}(0) =n_{A\uparrow}(0)-n_{A\downarrow}(0)=1$.
[Likewise the B-type state is  $\downarrow$-spin occupied,  such that (eq.\ \ref{eq:nsigh})
$n_{\sigma}(0)= \tfrac{1}{2}[n_{A\sigma}(0)+n_{B\sigma}(0)]$
$=\tfrac{1}{2}[n_{A\sigma}(0)+n_{A-\sigma}(0)]$
indeed gives $n_{\sigma}(0) =\tfrac{1}{2}$ for both $\sigma =\uparrow,\downarrow$.]

Focusing as  usual on the A-propagators, $G_{A\uparrow}(\w, 0)$ in eq.~\ref{eq:altsepropsh=0a} thus corresponds 
to removing the $\uparrow$-electron from the $\uparrow$-occupied level, and $G_{A\downarrow}(\w, 0)$ to adding 
a $\downarrow$-electron. Each propagator is of form 
$G_{A\sigma}(\w,0) =[\w^{+}-\epsilon -\Sigma_{A\sigma}(\w,0)]^{-1}$
(eq.\ \ref{eq:Aprop_se}), with self-energies $\Sigma_{A\downarrow}(\w,0) =U$ and 
$\Sigma_{A\uparrow}(\w,0) =0$. As for the spin-independent single self-energy in the doubly-occupied 
regime, the $\Sigma_{A\sigma}(\w,0)$ likewise correspond to first-order self-consistent perturbation theory,
i.e.\ to $\Sigma_{A\sigma}(\w,0) =U n_{A-\sigma}(0)$ (but of course with different self-consistent solutions 
than regime (a)). Unlike regime (a), however, the self-energies $\Sigma_{A\sigma}(\w,0)$ obviously do 
\emph{not} correspond to bare perturbation theory in $U$ about the non-interacting limit (since $\langle\hat{n}_{-\sigma}\rangle_{U=0} =1$); i.e.\ the LM regime is not perturbatively connected to 
the non-interacting limit. But, since the $\Sigma_{A\sigma}(\w,0)$ are $\w$-independent, it follows 
that the Luttinger integrals  eq.\ \ref{eq:LILM} appropriate to \emph{these} self-energies also
trivially vanish, $I_{L_{A\sigma}}(0) =0$; again as required from the general arguments of
sec.\ \ref{subsection:Luttintgen} (eq.\ \ref{eq:LIs=0}).

 While the preceding comments refer to the two-self-energies $\Sigma_{A\sigma}(\w,0)$, the conventional single self-energy in the LM regime, $\Sigma (\w) \equiv \Ss(\w,0)$, follows directly from them via eq.\ \ref{eq:SSE_TSE};
\emph{viz}
\begin{subequations}
\label{eq:alSSE}
\begin{align}
(\Ss(\w,0) \equiv )~~~~
\Sigma (\w)~=&~\tfrac{1}{2}U ~+~
\frac{\tfrac{1}{4}U^{2}}
{\w^{+}-\epsilon -\frac{U}{2}}
\label{eq:alSSEa}
\\
~=&~\tfrac{1}{2}U ~+~\tfrac{1}{4}U^{2}g_{0}^{\pd}(\w)\frac{1}{1-\frac{U}{2}g_{0}^{\pd}(\w)}
\label{eq:alSSEb}
\end{align}
\end{subequations}
with $g_{0}^{\pd}(\w)=[\w^{+}-\epsilon]^{-1}$ the non-interacting propagator.
This is entirely different from the $\Sigma_{A\sigma}(\w,0)$, being both  $\w$-dependent and 
containing in general all orders in the interaction $U$; and with the absence of perturbative 
continuity to the non-interacting limit evident  in the ${\cal{O}}(U)$ contribution of $U/2$, which 
is \emph{not} equal to the $U\langle\hat{n}_{-\sigma}\rangle_{U=0}$ of leading order perturbation theory 
(as $\langle\hat{n}_{-\sigma}\rangle_{U=0}=1$ for all $\epsilon <0$).
The sole exception is the p-h symmetric point $\epsilon =-U/2$, where $\epsilon$ is slaved to $U$. Here, 
no ground state level-crossing occurs on increasing $U$ from $0$, with 
$\langle\hat{n}_{\sigma}\rangle =\tfrac{1}{2}$ for all $U\geq 0$. In this case $\Sigma (\w)$ 
(eq.\ \ref{eq:alSSEa}) terminates at the ${\cal{O}}(U^{2})$ term and second-order perturbation theory in $U$  
is exact.~\cite{hewsonbook} 

Finally, since the single self-energy $\Sigma (\w)$ for the LM phase is known fully (together with 
$\Gs(\w,0)$, eq.\ \ref{eq:Gsallm})), the corresponding zero-field Luttinger integral $I_{L}$ 
(eq.\ \ref{eq:LIsse}) for the atomic limit can thus be evaluated explicitly.~\cite{Rosch_Lutt}
The integrals are elementary, and with $x =\epsilon +\tfrac{U}{2}$ (such that $x=0$ corresponds to the p-h symmetric point $\epsilon =-U/2$), the result for $I_{L} \equiv I_{L}(x)$ is 
$I_{L}(x) =\tfrac{\pi}{2}[\theta(x)-\theta(-x)]$: the Luttinger integral is thus indeed generically non-zero 
in the LM phase, with $|I_{L}| =\tfrac{\pi}{2}$ for any $x\neq 0$ in this regime. Further discussion of this 
result is given in sec.\ \ref{subsection:ILprime}, since here it arises merely in the trivial 
context of the atomic limit.


\section{Pseudogap Anderson model: zero field}
\label{section:PAIMh=0}

 As a paradigm model exhibiting a quantum phase transition between a SC and a LM phase, we now consider the pseudogap Anderson model (PAIM),~\cite{WithoffFradkin,ChenJayap1995,BPH1997,GB-Ingersent,MTG_SPAIM,MTGEPJB,nrglmacomp,Vojta_Bulla2001,Zitzler2002,Ingersent-Si2002,Kircan_Vojta2003,MTG_APAIM,MTGSPAIMepl,Vojta_Fritz2004,Fritz_Vojta2004,mattgarethPAIM,Lee2005,Fritz_Florens_Vojta2006,Vojta_Fritz_Bulla2010,MTGKirchner2011,Pixley2012}
where in the absence of a magnetic field the host-impurity hybridization vanishes in power-law fashion as the Fermi level is approached, $\Gamma^{I}(\w) \propto |\w|^{r}$ (with $r=0$ corresponding to the normal metallic AIM).
The two phases are known to be separated by a quantum critical point (QCP).~\cite{GB-Ingersent}
For simplicity we consider the wide-band limit of the model (where the host bandwidth 
$D\rightarrow \infty$), and as such $r$ in the range $0\leq r< 1$.~\cite{MTGEPJB} We shall consider both the particle-hole asymmetric and symmetric cases of the model, which in the SC phase are known to be qualitatively distinct~\cite{GB-Ingersent}
(the stable RG fixed points corresponding respectively to the asymmetric and symmetric SC FPs).

 As mentioned in sec.\ \ref{section:LMgeneral}, a magnetic field may be applied either locally to the impurity, or globally (acting also on the conduction band states). We will consider both cases in the following, but with primary interest in the more subtle local field case. For a global field, the (retarded) hybridization function 
$\Gamma_{\sigma}(\w, h)=\Gamma_{\sigma}^{R}(\w, h) -i \Gamma_{\sigma}^{I}(\w, h)$ is given by
$\Gamma_{\sigma}^{I}(\w, h)=\Gamma_{0}|\w +\sigma h|^{r}$, with
$\Gamma_{\sigma}^{R}(\w, h)=  -\mathrm{sgn}(\w +\sigma h) \Gamma_{0} \beta(r)|\w +\sigma h|^{r}$,
 where $\beta(r) = \mathrm{tan}(\tfrac{\pi}{2}r)$. For a locally applied field, by contrast, the hybridization 
is independent of $h$, and thus given by 
\begin{equation}
\Gamma (\w) = \Gamma_{\sigma}(\w, 0)= -[\mathrm{sgn}(\w)\beta(r) +i]~\Gamma_{0}|\w|^{r}.
\end{equation}

One further symmetry can usefully be exploited, namely that arising from the particle-hole (p-h) transformation 
\begin{equation}
\label{eq:phst}
d_{\sigma}^{\dagger} \leftrightarrow d_{-\sigma}^{\pd} ~~~~~~
c_{\mathbf{k}\sigma}^{\dagger} \leftrightarrow -c_{-\mathbf{k}-\sigma}^{\pd}
\end{equation}
in which particles/holes and spins are simultaneously exchanged. Labelling temporarily the dependence 
of the Hamiltonian upon $x =\epsilon +\tfrac{U}{2}$, it is readily shown that under this canonical 
transformation, $H(x) \equiv H(-x)$; and that the propagators in turn satisfy 
\begin{equation}
\label{eq:propsymmpht}
G_{A\sigma}(\w; x, |h|)~=~ -\left[
G_{A-\sigma}(-\w; -x, |h|)
\right]^{*}
\end{equation}
(or $G_{A\sigma}^{t}(\w; x, |h|)= - G_{A-\sigma}^{t}(-\w; -x, |h|)$ for the $t$-ordered propagators), 
and likewise for the $G_{\sigma}(\w; x, h)$ appropriate to $U<U_{c}$. Hence, only one or other of
$x =\epsilon +\tfrac{U}{2} \leq 0$ or $\geq 0$ need be considered explicitly. 

In the remainder of this section we consider the zero-field case. Our primary focus is the LM phase, but we begin with  brief consideration of the SC phase.


\subsection{SC phase, $\mathbf{U<U_{c}}$}
\label{subsection:PAIMh=0sc}

Since the zero-field hybridization vanishes at the Fermi level, eq. \ref{eq:nimpsboth}b gives
(with $\eta=0+$ as ever)
\begin{equation}
\label{eq:nimph=0sc}
\nimps (0) ~ =~\frac{1}{2}~-~\frac{1}{\pi}~\mathrm{tan}^{-1}\left[
\frac{\epsilon^{*}(0)}{\eta}
\right]
\end{equation}
where  $\nimps (0) = \tfrac{1}{2}\nimp(0)$, such that the zero-field excess magnetization 
(eq.\ \ref{eq:mimpnimp}) naturally vanishes; and from eq.\ \ref{eq:rlsc} the renormalized level 
$\epsilon^{*}(0) \equiv \epsilon^{*}_{\sigma}(h=0)$ (likewise independent of 
$\sigma$ for $h=0$) is  $\epsilon^{*}(0)=\epsilon + \Ss^{R}(0,0)$. There are then just three possibilities:\\
\noindent (a) If the renormalized level lies below the Fermi level, $\epsilon^{*}(0) <0$, then $\nimp(0) =2$ uniquely; while  \\
\noindent (b) if $\epsilon^{*}(0) >0$, then $\nimp(0) =0$; and \\
\noindent (c) if $\epsilon^{*}(0) =0$ then  $\nimp(0) =1$. \\
This behavior is physically natural -- one expects intuitively that the change in number of electrons due to addition of the impurity should be integral -- and for the SC phase these results are 
known.~\cite{MTG_APAIM} Note further that under the p-h transformation eq.\ \ref{eq:phst},
where $x \rightarrow -x$, it is readily shown that $\nimp (0) \rightarrow 2-\nimp(0)$. Cases (a) and (b) above 
thus correspond to the generic p-h asymmetric model (with $\nimp (0) =2$ or $0$ characteristic of the asymmetric 
SC FP). Case (c) by contrast, $\nimp(0)=1$, corresponds \emph{uniquely} to the p-h symmetric point 
$\epsilon =-U/2$, and is characteristic of the symmetric SC FP.


\subsection{LM phase, $\mathbf{U>U_{c}}$}
\label{subsection:PAIMh=0lm}

Now we turn to the LM phase, where eq. \ref{eq:nimpsboth}a gives
\begin{equation}
\label{eq:nimpAh=0lm}
\nimpas (0) ~=~\frac{1}{2}~-~\frac{1}{\pi}~\mathrm{tan}^{-1}\left[
\frac{\epsilon^{*}_{A\sigma}(0)}{\eta}
\right]
\end{equation}
with the renormalized level 
$\epsilon_{A\sigma}^{*}(0)=\epsilon  + \Sigma_{A\sigma}^{R}(0,0)$ (eq. \ref{eq:rllm}). 
The quantities to be determined here are the excess charge, $\nimp(0)=\sum_{\sigma}\nimpas (0)$
(eqs.\ (\ref{eq:mimpnimp},\ref{eq:rhoimpAB})), and the excess magnetization 
$\mimpa (0) =\sum_{\sigma}~\sigma \nimpas (0)$ (eq.\ \ref{eq:mimpAlm}); 
where $\mimpa (0)$  is the `excess' analogue of the permanent \emph{local} moment $m_{A}^{\pd}(0) =|\tm| >0$ 
(eq.\ \ref{eq:mAzero}), and is likewise non-vanishing in the LM phase.
Since $\mathrm{sgn}(\mimpa(0)) =\mathrm{sgn}(m_{A}(0))$ -- i.e.\ the excess and local moments lie 
in the same direction -- eq.\ \ref{eq:nimpAh=0lm} is satisfied only if
\begin{equation}
\label{eq:levelsupdownlm}
\epsilon^{*}_{A\uparrow}(0) ~<~0 ~~~~~~~~\epsilon^{*}_{A\downarrow}(0) ~>~0 
\end{equation}
corresponding to
\begin{equation}
\label{eq:nassupdownlm}
n_{\mathrm{imp},A\uparrow}(0) ~=~1 ~~~~~~~~n_{\mathrm{imp},A\downarrow}(0) ~=~0 
\end{equation}
and hence 
\begin{equation}
\nimp (0)~=~ 1 ~~~~~~~~ \mimpa(0)~=~1
\end{equation}
From  $\mimp (h)=[\theta(h)-\theta(-h)]~\mimpa(|h|)$ (eq.\ \ref{eq:mimpLM}), 
the zero-field magnetization itself obviously vanishes, $\mimp (0) =0$; but as $h \rightarrow 0\pm$, 
$|\mimp (0\pm)| =\mimpa(0)=1$, which we refer to in obvious terms as a `fully saturated' excess moment. 

\begin{figure}
\includegraphics[height=6cm, width=8.5cm]{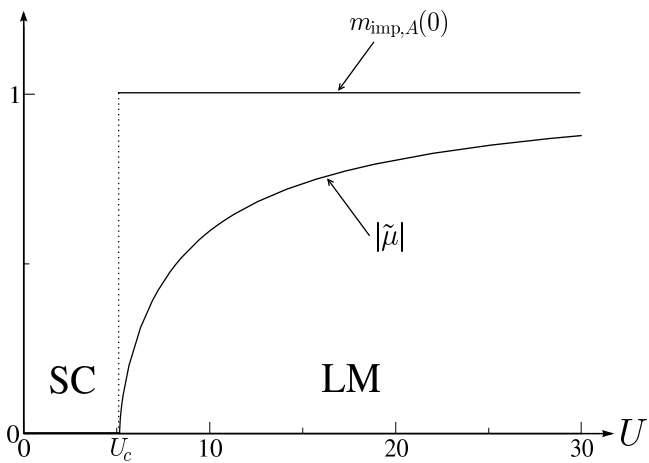}
\caption{\label{fig:fig2} 
Evolution of the zero-field, permanent local moment $|\tm|$ ($= m_{A}(0) =m(0+)$, 
eqs.\ (\ref{eq:mAzero},\ref{eq:basicma}b)), and the excess magnetization $\mimpa(0)$ ($=\mimp (0+$)), 
on crossing the transition at $U=U_{c}$ to the LM phase. $|\tm|$ vanishes continuously as $U\rightarrow U_{c}+$, while $\mimpa(0)$ generically changes discontinuously to full saturation on crossing the transition.
($|\tm|$ shown here has been obtained from NRG for $r=0.4$, see text).
}
\end{figure}

Two important points should be noted here. 
First, and in marked contrast to the SC phase where $\nimp(0) =1$ arises only at p-h symmetry, 
$\nimp(0) =1$ occurs \emph{throughout} the LM phase, regardless of whether or not the system is 
p-h symmetric. This is seen to be a consequence of a Luttinger theorem in terms of the two-self-energy 
description, i.e. $I_{L_{A\sigma}} =0$ (which led directly to eq.\ \ref{eq:nimpAh=0lm}). 
 
Second, note then that on crossing the QPT at $U=U_{c}$, the excess charge generically changes 
discontinuously, with $\delta\nimp(0) = \nimp(U=U_{c}+)-\nimp(U=U_{c}-) =\pm 1$.
Since the total charge in the \emph{absence} of the impurity (the free conduction band)
obviously changes continuously as $U_{c}$ is crossed, this indicates that an
additional electron is acquired (or lost) by the \emph{entire} system on entering the LM phase. 
Hence, on crossing to the LM phase at $U=U_{c}$, the entire system abruptly acquires an additional 
fully saturated moment, $\mimpa(0)=1$. But the \emph{local} impurity moment $m_{A}(0) =|\tm|$ -- the 
local order parameter for the transition -- by contrast evolves continuously from zero on increasing 
$U$ above $U_{c}$ (saturating to unity only as $U \rightarrow \infty$).
\emph{The generic physical picture of the transition to the LM phase is thus that an additional fully saturated moment `condenses' in the entire system, but that at $U=U_{c}$ (the QCP) it has no weight on the impurity, 
which by contrast is on the verge of acquiring a permanent local moment}. To our knowledge, this is a 
new physical perspective of the transition. The p-h symmetric point is of course special in the sense that 
$\delta\nimp(0) =0$. But here again the excess moment in the LM phase is fully saturated for all $U>U_{c}$, while the local impurity moment $m_{A}(0) =|\tm|$ vanishes as $U\rightarrow U_{c}+$; the same physical picture thus emerging.
 
 The situation just described is illustrated in Fig.\ \ref{fig:fig2}. Since the local moment $|\tm|$ 
$= m_{A}(0) =m(0+)$ (eqs.\ (\ref{eq:mAzero},\ref{eq:basicma}b))) 
is simply the local magnetization as $h\rightarrow 0+$, it can be calculated numerically using the FDM-NRG.~\cite{PetersPruschkeAnders2006,WeichselbaumDelft2007,NRGrmp} 
 The specific results shown in Fig.\ \ref{fig:fig2} have in fact been obtained in this way, for $r=0.4$ 
and fixed $x=\epsilon + \tfrac{U}{2} = -0.4$ (with $[\Gamma_{0}]^{1/(1-r)} \equiv 1$ as the unit of energy).
Note also that the situation here is of course quite different physically from the atomic limit
(sec.\ \ref{section:atomlimh=0}) -- in the latter case, the trivial level-crossing
at `$U_{c}$'$=-\epsilon$ leads to condensation of a fully saturated moment on the 
\emph{impurity} itself ($|\tm| = 1$).

\begin{figure}
\includegraphics[height=6cm, width=8.5cm]{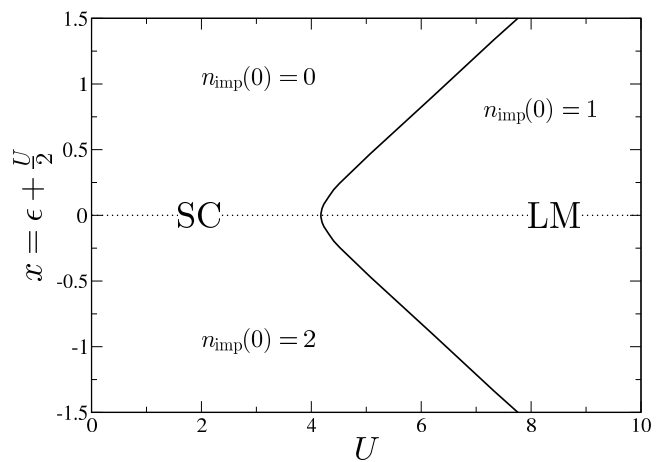}
\caption{\label{fig:phasediagram} 
Phase diagram for the PAIM with $r=0.4$, determined numerically via the
FDM-NRG.~\cite{PetersPruschkeAnders2006,WeichselbaumDelft2007,NRGrmp} 
Shown in the $(U, x)$-plane, with $x=\epsilon+\tfrac{U}{2}$ (and $[\Gamma_{0}]^{1/(1-r)} \equiv 1$ 
as the unit of energy). SC and LM phases are separated by a (solid) line of quantum critical points 
(see sec.\ \ref{section:QCP}); the dotted line merely shows the p-h symmetric line, $x=0$. Characteristic $\nimp(0)$'s for the SC and LM phases (secs.\ \ref{subsection:PAIMh=0sc},\ref{subsection:PAIMh=0lm}) are 
indicated.
}
\end{figure}

For illustration and later reference, Fig.\ \ref{fig:phasediagram} shows a representative phase diagram 
for the PAIM, determined via FDM-NRG. SC and LM phases are separated by a line of quantum critical points 
(sec.\ \ref{section:QCP}) in the $(U, x)$-plane; and the characteristic $\nimp(0)$'s for 
SC and LM phases, as determined above, are indicated.

\subsection{The Luttinger integral $\mathbf{I_{L}}$}
\label{subsection:ILprime}

From a knowledge of the renormalized levels in the zero-field LM phase, 
$\epsilon^{*}_{A\sigma}(0)=\epsilon +\Sigma_{A\sigma}^{R}(0,0)$, 
we can in turn determine the renormalized level $\epsilon^{*}$ (eq. \ref{eq:rllmsse}) appropriate to 
a conventional single self-energy description of the propagator (eq.\ \ref{eq:Gs}).
From eq.\ \ref{eq:rllmsse}, $\epsilon^{*}= \epsilon +\Sigma^{R}(0)$; with the single self-energy 
$\Sigma(\w)=\Sigma^{R}(\w)-i\Sigma^{I}(\w)$ itself determined by eq.\ \ref{eq:SSE_TSE} from
the two-self-energies $\{\Sigma_{A\sigma}(\w, 0)\}$. This we now consider, focusing initially 
on the generic p-h asymmetric case. From eq.\ \ref{eq:ses=0} the two self-energies are purely real at 
the Fermi level, $\Sigma_{A\sigma}^{I}(0,0) =0$. From eq.\ \ref{eq:SSE_TSE}, two important results then 
follow. First, that $\Sigma^{I}(0)=0$, so the single-self-energy likewise vanishes at the Fermi level in 
the general p-h asymmetric case (we determine its asymptotic behavior as $\w \rightarrow 0$ in 
sec.\ \ref{subsection:SSElowomega} below). Second, that
$\epsilon^{*} = \tfrac{1}{2}(\epsilon^{*}_{A\uparrow}+\epsilon^{*}_{A\downarrow})
-(\tfrac{1}{2} [\epsilon^{*}_{A\uparrow} -\epsilon^{*}_{A\downarrow}])^{2}/(\tfrac{1}{2}
[\epsilon^{*}_{A\uparrow}+\epsilon^{*}_{A\downarrow}])$, i.e.\
\begin{equation}
\label{eq:levelitolevels}
\epsilon^{*}~=~ \frac{2\epsilon^{*}_{A\uparrow}\epsilon^{*}_{A\downarrow}}
{\epsilon^{*}_{A\uparrow}+\epsilon^{*}_{A\downarrow}}   ~~~ \longleftrightarrow 
~~~
\frac{1}{\epsilon^{*}}~=~
\frac{1}{2} \left[
\frac{1}{\epsilon_{A\uparrow}^{*}} ~+~\frac{1}{\epsilon_{A\downarrow}^{*}}
\right]
\end{equation}
where $\epsilon_{A\sigma}^{*} \equiv \epsilon_{A\sigma}^{*}(h=0)$.~\cite{fn9a}

The preceding results enable the Luttinger integral $I_{L}$ (eq.\ \ref{eq:LIsse}) to be determined.
The phase shift $\delta$ is given from eq.\ \ref{eq:pssse} by $\delta = \tfrac{\pi}{2}\nimp(0) +I_{L}$; 
but we have shown above that throughout the LM phase $\nimp(0) =1$, whence:
\begin{equation}
\label{eq:ILdet1}
\delta ~=~ \frac{\pi}{2}~+~I_{L}
\end{equation}
And since $\Sigma^{I}(0) =0$ has been shown above, eq.\ \ref{eq:pslmsse} gives
\begin{equation}
\label{eq:ILdet2}
\delta ~=~ \frac{\pi}{2} ~-~ \mathrm{tan}^{-1}\left[\frac{\epsilon^{*}}{\eta}\right]~,
\end{equation}
whence $I_{L} = -\mathrm{tan}^{-1}(\epsilon^{*}/\eta)$.
But throughout the LM phase $\epsilon_{A\uparrow}^{*}<0$ and $\epsilon_{A\downarrow}^{*}>0$
(eq.\ \ref{eq:levelsupdownlm}); so eq.\ \ref{eq:levelitolevels} yields
$\epsilon^{*}>0$ if $|\epsilon_{A\downarrow}^{*}| < |\epsilon_{A\uparrow}^{*}|$, and
$\epsilon^{*}<0$ if $|\epsilon_{A\downarrow}^{*}| > |\epsilon_{A\uparrow}^{*}|$.
Hence the desired result for the Luttinger integral:
\begin{subequations}
\label{eq:ILdetresult}
\begin{align}
I_{L} ~=~&\mathrm{Im}\int^{0}_{-\infty}d\w ~\frac{\partial \Sigma (\w)}{\partial \w}~\Gs(\w, 0)
\nonumber
\\
~~=&~-\tfrac{\pi}{2} ~~:~|\epsilon_{A\downarrow}^{*}| < |\epsilon_{A\uparrow}^{*}|
\\
=&~  +\tfrac{\pi}{2} ~~:~|\epsilon_{A\downarrow}^{*}| > |\epsilon_{A\uparrow}^{*}|
\end{align}
\end{subequations}
The sign change here is expected, for under the the p-h transformation eq.\ \ref{eq:phst},
it is readily shown that $I_{L} \equiv I_{L}(x)$ ($x=\epsilon+U/2$ as usual) satisfies
\begin{equation}
\label{eq:ILpht}
I_{L}(x)~=~ -I_{L}(-x)
\end{equation}
and that $\epsilon_{A\uparrow}^{*}(x) =- \epsilon_{A\downarrow}^{*}(-x)$ such that
$|\epsilon_{A\uparrow}^{*}| =|\epsilon_{A\downarrow}^{*}|$ at the p-h symmetric point $\epsilon =-U/2$. 
Precisely at this point, eq.\ \ref{eq:ILpht} implies $I_{L}(x=0) =0$, as indeed can be confirmed by direct consideration of this case. 

 Several comments should be made here.\\
\noindent (a) $I_{L}$ is the direct analogue, in the LM phase, of the standard Luttinger integral 
eq.\ \ref{eq:LISC} appropriate to the SC phase (each being expressed in terms of the conventional single self-energy). In the SC, Fermi liquid phase the Luttinger integral has a value (of zero) that is independent of underlying bare model parameters. As such it is an intrinsic hallmark of the FL phase. The above results show that the magnitude of the Luttinger integral $I_{L}$ is likewise intrinsic to the zero-field LM phase, with $|I_{L}| = \tfrac{\pi}{2}$ arising generically for all $x\neq 0$. \\
\noindent (b)  The result eq.\ \ref{eq:ILdetresult} for $I_{L}$ is not specific to the PAIM.
The arguments given above apply equally to the gapped AIM (where $\Gamma(\w=0) =0$, as for the PAIM). 
As seen in sec.\ \ref{section:atomlimh=0}, it arises too for the free atomic limit of the model, 
in the LM regime where the impurity is singly-occupied. $|I_{L}| =\tfrac{\pi}{2}$ is also known (from NRG calculations) to arise for the underscreened spin-1 phase of a two-level quantum dot~\cite{CJW2009}
-- and we shall give a general argument to demonstrate this in sec.\ \ref{section:Luttintgen} --
as well as for multi-dot models known to contain LM phases.~\cite{Jarrold2013}\\
\noindent (c) We have also checked eq.\ \ref{eq:ILdetresult} numerically, as an integral over all frequency given the zero-field propagator and single self-energy calculated directly from NRG.

 Finally, we reiterate that the considerations above apply exclusively to $h=0$. For any non-zero field the QPT between the SC and LM phases is strictly destroyed, the ground state of the system is always singly degenerate, and all Luttinger integrals vanish (as in eq.\ \ref{eq:LIs=0}).


\subsection{Low-frequency behavior of $\mathbf{\Sigma(\w)}$: LM phase}
\label{subsection:SSElowomega}

In considering the LM phase, we introduce an energy scale $\w_{*}$ defined by~\cite{MTG_APAIM}
\begin{equation}
\label{eq:w*def}
\Gamma_{0}~\w_{*}^{r} ~=~ 
\tfrac{1}{2}|\epsilon_{A\uparrow}^{*}-\epsilon_{A\downarrow}^{*}|
\end{equation}
(the $\tfrac{1}{2}$ is merely for convenience). On the natural assumption that the  
$\epsilon_{A\sigma}^{*}\equiv \epsilon_{A\sigma}^{*}(0)$ are continuous in $U$,
$\epsilon_{A\uparrow}^{*}-\epsilon_{A\downarrow}^{*}$ will vanish as the transition is 
approached from the LM phase (since the zero-field renormalized level is independent of 
spin in the SC phase). As elaborated below, $\w_{*}$ is the low-energy scale characteristic of 
the LM phase (the counterpart of the Kondo scale $\w_{K}$ characterizing the SC phase). 
We make two initial points here: \\
\noindent (i) Since $\w_{*}$ vanishes as the transition is approached, physical properties 
(including single-particle dynamics) should exhibit scaling in terms of it, i.e.\ will be universal 
functions of $\w/\w_{*}$ (as indeed shown below). \\
\noindent (ii) Since $\epsilon_{A\uparrow}^{*}<0$ and $\epsilon_{A\downarrow}^{*}>0$ throughout the 
LM phase (eq.\ \ref{eq:levelsupdownlm}), 
$|\epsilon_{A\uparrow}^{*}-\epsilon_{A\downarrow}^{*}| = |\epsilon_{A\uparrow}^{*}|+
|\epsilon_{A\downarrow}^{*}|$. The vanishing of $\w_{*}$ as $U\rightarrow U_{c}+$ thus implies that the renormalized levels $\epsilon_{A\sigma}^{*}$ \emph{separately} vanish as the transition is approached.
This holds generally,  whether the system is p-h symmetric or asymmetric. We have also confirmed it 
numerically, from NRG calculations (see Fig.\ \ref{fig:fig6}).

 As already seen, it is necessary to distinguish between the p-h asymmetric and symmetric cases in the 
zero-field LM phase. This is conveniently embodied in the following ratio of renormalized levels:
\begin{equation}
\label{eq:gammadef}
\gamma = \frac{\tfrac{1}{2}(\epsilon_{A\uparrow}^{*}+\epsilon_{A\downarrow}^{*})}
{\Gamma_{0}\w_{*}^{r}}
~=~\frac{\epsilon_{A\uparrow}^{*}+\epsilon_{A\downarrow}^{*}}
{|\epsilon_{A\uparrow}^{*}-\epsilon_{A\downarrow}^{*}|}
~=~ \frac{|\epsilon_{A\downarrow}^{*}|-|\epsilon_{A\uparrow}^{*}|}
{|\epsilon_{A\downarrow}^{*}|+|\epsilon_{A\uparrow}^{*}|}
\end{equation}
As noted above, it follows from the p-h transformation eq.\ \ref{eq:phst} that 
$\epsilon_{A\uparrow}^{*}(x) =- \epsilon_{A\downarrow}^{*}(-x)$.
$\gamma$ thus vanishes throughout the LM phase at p-h symmetry, $x=\epsilon+\tfrac{U}{2}=0$ 
(where $\epsilon_{A\uparrow}^{*} =-\epsilon_{A\downarrow}^{*}$). Generically, however, it is non-zero 
away from the p-h symmetric point; and with $\gamma(x) =-\gamma (-x)$. Note further that $\gamma$ is 
strictly bounded, $\gamma \in (-1,+1)$; and thus tends to a finite limit when the
$\epsilon^{*}_{A\sigma}$ vanish as $U\rightarrow U_{c}+$ and the QCP is approached. 

Now we turn to the low-$\w$ behavior of the single self-energy in the zero-field LM phase, 
$\Sigma (\w)=\Sigma^{R}(\w)-i\Sigma^{I}(\w) $, as may be obtained from eq.\ \ref{eq:SSE_TSE} given 
a knowledge of the $\{\Sigma_{A\sigma}(\w, 0)\}$. We consider first the generic asymmetric case.


\subsubsection{Particle-hole asymmetric case}
\label{subsection:asymmSSElowomega}

As shown in sec.\ \ref{subsection:ILprime}, 
$\Sigma^{I}(\w=0)=0 $; and the renormalized level $\epsilon^{*} = \epsilon +\Sigma^{R}(0)$ is given 
by eq.\ \ref{eq:levelitolevels}, which may be written equivalently as
\begin{equation}
\label{eq:estargamma}
\frac{1}{\Gamma_{0}\w_{*}^{r}} ~\epsilon^{*}
~ = ~ \gamma ~-~ \frac{1}{\gamma}
\end{equation}
in terms of $\w_{*}$ and $\gamma$ introduced above. Note that since $\gamma$ remains finite as 
$U \rightarrow U_{c}+$, eq.\ \ref{eq:estargamma} shows that the renormalized level  vanishes as
$\epsilon^{*} \propto \omega_{*}^{r}$ as the transition is approached and $\w_{*}$ vanishes.

To obtain the leading low-$\w$ behavior of $\Sigma^{I}(\w)$ might seem to require detailed knowledge of the
low-$\w$ behavior of the $\Sigma_{A\sigma}^{I}(\w,0)$. However, provided only that  
$\Sigma_{A\sigma}^{I}(\w,0)$ vanishes as $\w \rightarrow 0$ no less slowly than the hybridization 
(i.e.\ $\Sigma_{A\sigma}^{I}(\w,0) \overset{\w \rightarrow 0}{\sim} |\w|^{\delta}$ with $\delta \geq r$)
-- which we show in sec.\ \ref{section:SSElowomegaLW} to be self-consistent in the Luttinger 
sense~\cite{Luttingerf1961} -- then it is merely a matter of algebra to show directly from   eq.\ \ref{eq:SSE_TSE} that the leading low-$\w$ behavior of $\Sigma^{I}(\w)$ is~\cite{fnLMA}
\begin{equation}
\label{eq:ssewto0lmasymm}
\frac{1}{\Gamma_{0}\w_{*}^{r}}~ \Sigma^{I}(\w) ~ \propto ~ 
\left(
\frac{|\w|}{\w_{*}}
\right)^{r}~
\end{equation}
(with coefficients ${\cal{O}}(1)$). $\Sigma^{I}(\w)$ thus vanishes with precisely the same power-law as the hybridization $\Gamma^{I}(\w)$, indicative of the NFL character of the LM phase
(the counterpart for the $r=0$ metallic case would be a constant $\Sigma^{I}(\w =0)$). 
Notice also that, as anticipated above, the resultant (dimensionless) $\Sigma^{I}(\w)/\Gamma_{0}\w_{*}^{r}$ 
indeed exhibits scaling in terms of $\tilde{\w} =\w/\w_{*}$.

The leading low-$\w$ behavior of $\Sigma^{R}(\w)$ can of course be obtained in the same way. Alternatively, 
we can deduce it immediately from the low-$\w$ behavior of $\Sigma^{I}(\w)$ in
eq.\ \ref{eq:ssewto0lmasymm}. From $F(\w)=F^{R}(\w) -iF^{I}(\w)$ with $F^{R}(\w)$ and $F^{I}(\w)$
related by Hilbert transformation, then if $F^{I}(\w) \overset{\w \rightarrow 0}{\sim} |\w|^{\lambda}$ with 
$-1 <\lambda <1$, the $\w \rightarrow 0$ behavior of $F^{R}(\w)$ is readily shown to follow as
\begin{equation}
\label{eq:HTgenresult}
F^{R}(\w)~-~F^{R}(0) ~ \overset{\w \rightarrow 0}{\sim}~-\mathrm{sgn}(\w)~\beta(\lambda)~ F^{I}(\w)
\end{equation}
where $\beta(\lambda) =\mathrm{tan}(\tfrac{\pi}{2}\lambda)$.~\cite{fn10a} Hence, since $r<1$,
eq.\ \ref{eq:ssewto0lmasymm} gives:
\begin{equation}
\label{eq:REssewto0lmasymm}
\Sigma^{R}(\w)-\Sigma^{R}(0) ~=~ -\mathrm{sgn}(\w)~\beta(r)~\Sigma^{I}(\w)
\end{equation}
Writing eq. \ref{eq:ssewto0lmasymm} as $\frac{1}{\Gamma_{0}\w_{*}^{r}}\Sigma^{I}(\w) = a_{a} |\tilde{\w}|^{r}$
(with $a_{a}\sim {\cal{O}}(1)$), the leading low-frequency behavior of the local propagator 
$\Gs(\w, 0) \equiv G(\w) = [ \w^{+}-\epsilon -\Gamma(\w,0)-\Sigma(\w) ]^{-1}$
follows from eqs.\ (\ref{eq:estargamma},\ref{eq:ssewto0lmasymm},\ref{eq:REssewto0lmasymm}) as
\begin{equation}
\label{eq:qpformLMasymm}
\begin{split}
&\Gamma_{0}\w_{*}^{r} ~G(\w)~\overset{|\tilde{\w}| \ll1}{\sim}
\\
~&\left[ \frac{\w_{*}^{1-r}}{\Gamma_{0}}\tilde{\w}^{+} -(\gamma -\frac{1}{\gamma}) +\left[\mathrm{sgn}(\tilde{\w}) \beta(r) +i\right](1+a_{a}) |\tilde{\w}|^{r}
\right]^{-1} 
\end{split}
\end{equation}
where $\tilde{\w}=\w/\w_{*}$.
This is the `quasiparticle form' for the propagator in the zero-field LM phase.
(The scaling regime, arising close to the QCP, corresponds to finite $\tilde{\w}$ in the 
limit that $\w_{*}\rightarrow 0$, and in that regime the $(\w_{*}^{1-r}/\Gamma_{0})~\tilde{\w}^{+}$ 
contribution to eq.\ \ref{eq:qpformLMasymm} may of course  be dropped.) Eq.\ \ref{eq:qpformLMasymm} 
yields the asymptotic scaling spectrum
\begin{equation}
\label{eq:speclowfreqLMasymm}
\pi \Gamma_{0}\w_{*}^{r}~ D(\w) 
~\overset{\tilde{\w}\rightarrow 0}{\sim} ~ \frac{(1+a_{a})}{(\gamma-\frac{1}{\gamma})^{2}}~|\tilde{\w}|^{r}
\end{equation}
which vanishes $\propto |\tilde{\w}|^{r}$ on approaching the Fermi level, as known e.g.\ from NRG 
calculations.~\cite{Vojta_Bulla2001} We emphasize again that it is in terms of the low-energy scale 
$\w_{*}$ of eq.\ \ref{eq:w*def} that $\Gamma_{0}\omega_{*}^{r}D(\w)$ scales universally, which is 
why $\w_{*}$ was thus defined.


\subsubsection{Particle-hole symmetric case}
\label{subsection:symmSSElowomega}

Here again, provided only that  $\Sigma_{A\sigma}^{I}(\w,0)$ vanishes as $\w \rightarrow 0$ no less 
slowly than the hybridization, it is a matter of algebra to show directly from   eq.\ \ref{eq:SSE_TSE} 
that the leading low-$\w$ behavior of $\Sigma^{I}(\w)$ is~\cite{fnLMA}
\begin{equation}
\label{eq:ssewto0lmsymm}
\frac{1}{\Gamma_{0}\w_{*}^{r}}~ \Sigma^{I}(\w) ~ \propto ~ 
\left(
\frac{|\w|}{\w_{*}}
\right)^{-r}
\end{equation}
(arising from the second term on the right of eq.\ \ref{eq:SSE_TSE}). The single self-energy thus diverges as 
$\w \rightarrow 0$, again symptomatic of the NFL nature of the LM phase.
The corresponding real part follows from eq.\ \ref{eq:HTgenresult}, 
so $\Sigma(\w)\overset{\w \rightarrow 0}{\sim} \Sigma^{R}(0) -[\mathrm{sgn}(\w)\beta(-r)+i]\Sigma^{I}(\w)$.
At p-h symmetry, $\epsilon +\Sigma^{R}(0) =0$ by symmetry. Hence, writing eq.\ \ref{eq:ssewto0lmsymm} as
$\frac{1}{\Gamma_{0}\w_{*}^{r}}\Sigma^{I}(\w) = a_{s} |\tilde{\w}|^{-r}$
(with $a_{s}\sim {\cal{O}}(1)$), the leading low-frequency behavior of the propagator $\Gs(\w, 0)\equiv G(\w)$ follows as 
$\Gamma_{0}\w_{*}^{r} G(\w) \sim [(-\mathrm{sgn}(\tilde{\w}) \beta(r) +i) ~a_{s}|\tilde{\w}|^{-r} ]^{-1}$.
The asymptotic scaling spectrum is thus
\begin{equation}
\label{eq:speclowfreqLMsymm}
\pi \Gamma_{0}\w_{*}^{r}~ D(\w) ~\overset{|\tilde{\w}| \rightarrow 0}{\sim}~\frac{1}{a_{s}}\frac{1}{1+\beta^{2}(r)} ~|\tilde{\w}|^{r},
\end{equation}
and likewise vanishes $\propto |\tilde{\w}|^{r}$.~\cite{nrglmacomp}

Note further that eq.\ \ref{eq:speclowfreqLMsymm} may be recast as
\begin{equation}
\label{eq:specsumruleLMsymm}
\underset{\w \rightarrow 0\pm}{\mathrm{lim}} \left[ \pi  (1+\beta^{2}(r)) ~\Sigma^{I}(\w)~ D(\w)\right] =1
~~~~:~ U>U_{c}
\end{equation}
(since $\Sigma^{I}(\w)\overset{\tilde{\w}\rightarrow 0}{\sim}\Gamma_{0}\w_{*}^{r}a_{s}|\tilde{\w}|^{-r}$). This  is the counterpart, in the LM phase, of the well known `pinning condition' on the single-particle spectrum in the zero-field SC phase of the symmetric PAIM,~\cite{MTGEPJB,MTG_SPAIM,nrglmacomp}
\emph{viz}
\begin{equation}
\label{eq: SC pinningagain}
\underset{\w \rightarrow 0\pm}{\mathrm{lim}}\left[\pi (1+\beta^{2}(r)) ~\Gamma^{I}(\w)~ D(\w)\right] =1
~~~~:~ U<U_{c}
\end{equation}
In the latter case, the local spectrum $D(\w)$ diverges $\propto |\w|^{-r}$ and the hybridization vanishes 
$\propto |\w|^{r}$; while for the LM phase by contrast (eq.\ \ref{eq:specsumruleLMsymm}), 
it is the self-energy which diverges $\propto |\w|^{-r}$ and the spectrum which vanishes as $|\w|^{r}$.
Note that the pinning condition in \emph{each} case is a particular example of eq.\ \ref{eq:dosps1a} 
for the case where $\delta(\w)$ is discontinuous across $\w =0$; with a discontinuity $\Delta = \tfrac{\pi}{2}r$ and $\delta \equiv \delta(\w=0) = \tfrac{\pi}{2}$.

Finally, the condition eq.\ \ref{eq:specsumruleLMsymm} holds of course throughout the LM phase at p-h symmetry,
and is exact. We have further confirmed that it is satisfied in NRG calculations of the single-particle spectrum.


\section{Luttinger self-consistency}
\label{section:SSElowomegaLW}

We  now consider briefly the low-$\w$ behavior of the self-energies $\Sigma_{A\sigma}(\w, 0)$ or $\Sigma_{\sigma}(\w, 0)$, as obtained self-consistently by considering the skeleton expansion for the 
self-energies, order-by-order in the interaction. This is done by adapting the original
analysis of Luttinger.~\cite{Luttingerf1961} That it can be done, even for the non-Fermi liquid LM phase, 
reflects of course the fact that the  self-energies $\Sigma_{A\sigma}(\w, 0)$ relevant to $U>U_{c}$  are
also expressible in skeleton form as functionals of the $G_{A\sigma}(\w, 0)$.
In the following we consider explicitly the self-energies $\Sigma_{A\sigma}(\w, 0)$
(as usual all results hold equally for the single self-energy $\Sigma_{\sigma}(\w, 0)$ 
appropriate to the SC phase, simply by dropping the $A$-labels). For brevity, explicit reference 
to the field (in)dependence  will be temporarily suppressed . 

 To proceed, one focuses on time-ordered Goldstone diagrams for the skeleton expansion (as readily obtained 
from any corresponding $n$-th order Feynman diagram). One begins by considering the second-order skeleton 
diagram, i.e.\ self-consistent second-order perturbation theory.
Due to the $\delta$-function constraints reflecting frequency conservation at any diagram vertex, 
when considering $\Sigma_{A\sigma}^{I}(\w)$ as $\w \rightarrow 0$ precisely the same (`phase space') constraints arise on interior frequency integrations as in Luttinger's original work.~\cite{Luttingerf1961} In consequence, only a knowledge the asymptotic low-frequency behavior of the spectrum $D_{A\sigma}(\w)$ is required. We assume it to be of form  $D_{A\sigma}(\w) \propto |\w|^{\lambda_{\sigma}}$ with the exponent $\lambda_{\sigma}$ to be determined self-consistently (and $\lambda_{\uparrow}$, $\lambda_{\downarrow}$ allowed in principle to be distinct~\cite{fn13}); with $\lambda_{\sigma} >-1$ necessarily, since $D_{A\sigma}(\w)$  must be integrable.
With this, the asymptotic behavior of the imaginary part of the self-energy is readily shown to be
\begin{equation}
\label{eq:seconsis1}
\Sigma_{A\sigma}^{I}(\w)~\overset{\omega \rightarrow 0}{\propto}~ |\w|^{2+\lambda_{\sigma}+2\lambda_{-\sigma}}
\end{equation}
(where $\lambda_{\sigma}+2\lambda_{-\sigma}$ reflects the fact that the second-order skeleton diagram 
contains one $\sigma$-spin and two $-\sigma$-spin propagators).  Eq.\ \ref{eq:seconsis1} reduces to 
conventional $\propto |\w|^{2}$ behavior if the $\lambda_{\sigma} =0$, as arises for a metallic Fermi liquid. 

To establish the self-consistent $\lambda_{\sigma}$, consider the single-particle spectrum
$D_{A\sigma}(\w)=-\tfrac{1}{\pi}\mathrm{Im}G_{A\sigma}(\w)$ expressed as
\begin{widetext}
\begin{equation}
\label{eq:seconspec}
\pi D_{A\sigma}(\w)~=~
\frac{\left[
\Gamma_{0}|\w|^{r} + \Sigma_{A\sigma}^{I}(\w)\right]}{
\left[\w -\epsilon_{A\sigma}^{*} +\mathrm{sgn}(\w)\beta(r)\Gamma_{0}|\w|^{r}
- \left(
\Sigma^{R}_{A\sigma}(\w) -\Sigma^{R}_{A\sigma}(0)
\right)
\right]^{2}
+\left[
\Gamma_{0}|\w|^{r} + \Sigma_{A\sigma}^{I}(\w)
\right]^{2} }
\end{equation}
\end{widetext}
with $\epsilon_{A\sigma}^{*}$ the renormalized level.
The low-$\w$ behavior of $D_{A\sigma}(\w)$ is controlled by whether $\epsilon_{A\sigma}^{*}\neq 0$, or 
vanishes. The generic case is of course $\epsilon_{A\sigma}^{*}\neq 0$ (for $\sigma =\uparrow$ and 
$\downarrow$), so we consider it first.
If $2+\lambda_{\sigma}+2\lambda_{-\sigma} >r$ -- i.e.\ $\Sigma_{A\sigma}^{I}(\w)$ in 
eq.\ \ref{eq:seconsis1} vanishes as  $\w \rightarrow 0$ more rapidly that the hybridization -- then the low-$\w$ behavior of $D_{A\sigma}(\w)$ is controlled by the hybridization and given from eq.\ \ref{eq:seconspec} as 
$D_{A\sigma}(\w) \propto |\w|^{r} \equiv |\w|^{\lambda_{\sigma}}$, whence $\lambda_{\sigma} =r$.
If by contrast $2+\lambda_{\sigma}+2\lambda_{-\sigma} <r$ were to arise,  then the
low-$\w$ behavior of the $\sigma$-spin spectrum would be controlled by the self-energy, 
$D_{A\sigma}(\w) \propto |\w|^{2+\lambda_{\sigma}+2\lambda_{-\sigma}} \equiv |\w|^{\lambda_{\sigma}}$, giving 
$\lambda_{-\sigma}=-1$; which is incompatible with the condition $\lambda_{-\sigma}>-1$ for an integrable spectrum, and hence not self-consistently possible. The sole self-consistent solution is thus
$2+\lambda_{\sigma}+2\lambda_{-\sigma} > r$ for $\sigma =\uparrow$ and $\downarrow$;
yielding $\lambda_{\uparrow} = r =\lambda_{\downarrow}$, and with $2+\lambda_{\sigma}+2\lambda_{-\sigma} > r$ requiring $r> -1$, as it is by construction. Hence, 
$\Sigma_{A\sigma}^{I}(\w)\propto |\w|^{2+3r}$ as $|\w| \rightarrow 0$.

The case $\epsilon_{A\sigma}^{*}= 0$ may be analyzed in the same fashion. 
However in contrast to $\epsilon_{A\sigma}^{*}\neq 0$ -- which arises throughout the LM phase (see 
sec.\ \ref{subsection:PAIMh=0lm}) and is equally generic in the SC phase (where 
$\epsilon_{A\sigma}^{*} \equiv \epsilon_{\sigma}^{*}$) -- the case $\epsilon_{A\sigma}^{*}= 0$ is quite 
specific at zero-field: it applies only to the SC phase at p-h symmetry, where 
$\epsilon_{\sigma}^{*} \equiv \epsilon^{*}(0)=0$ is guaranteed by symmetry. In this case of course, $\lambda_{\sigma} \equiv \lambda$ independently of $\sigma$, whence (eq.\ \ref{eq:seconsis1}) 
$\Sigma^{I}(\w) \equiv \Ss^{I}(\w) \propto |\w|^{2+3\lambda}$. If $2+3\lambda >r$, then the self-energy 
is again subsidiary to the hybridization as $|\w|\rightarrow 0$.  Eq.\ \ref{eq:seconspec} then gives 
$D(\w) \equiv D_{\sigma}(\w) \propto |\w|^{-r}$ by virtue of the vanishing renormalized level; whence 
$\lambda =-r$, and $\Sigma_{\sigma}^{I}(\w)\propto |\w|^{2-3r}$ is thus self-consistent for $2-3r > r$, 
i.e.\ provided $r<\tfrac{1}{2}$. This moreover is the only self-consistent solution, as the SC phase is perturbatively connected to the non-interacting limit, and in consequence~\cite{MTGEPJB} 
$\Ss^{I}(\w)$ ($\propto |\w|^{2+3\lambda}$) must vanish as $|\w| \rightarrow 0$ more rapidly than the 
hybridization $\propto |\w|^{r}$ (which requirement is familiar for the usual metallic model, $r=0$, where 
it amounts to the fact that $\Sigma_{\sigma}^{I}(\w =0)$ must vanish for the SC Fermi liquid). 

While the results above arise from explicit consideration of the second-order skeleton diagram, the contribution 
to the low-$\w$ asymptotic behavior of $\Sigma_{A\sigma}^{I}(\w)$ arising from arbitrary $n$-th order 
diagrams may also be analyzed, following directly Luttinger's original analysis.~\cite{Luttingerf1961}
And the same key result arises, namely that all $n$-th order diagrams contribute to the leading low-$\w$ dependence of the self-energy, the asymptotic behavior of which is precisely that deduced at second-order level. We can thus summarize the results obtained above,~\cite{fn14} which hold order-by-order in self-consistent perturbation theory in the interaction (and remembering that we are interested in $r \in [0,1)$, although eq.\ \ref{eq:selfconsesumm} encompasses $r \in (-1, +1)$):
\begin{subequations}
\label{eq:selfconsesumm}
\begin{align}
\Sigma_{A\sigma}^{I}(\w)~ &\overset{\w \rightarrow 0}{\propto} ~ |\w|^{2+3r} ~~~:~\epsilon_{A\sigma}^{*} \neq 0 ~ \mathrm{and}~
0 \leq r < 1
\\
\Sigma_{\sigma}^{I}(\w)~ &\overset{\w \rightarrow 0}{\propto} ~|\w|^{2-3r} ~~~:~\epsilon_{\sigma}^{*} = 0 ~ \mathrm{and}~
0 \leq r < \tfrac{1}{2}
\end{align}
\end{subequations}
Notice that (a) in all cases the imaginary part of the appropriate self-energy
vanishes at the Fermi level, $\w =0$, as asserted and used hitherto (eq.\ \ref{eq:ses=0} \emph{ff}); 
and (b) eq.\ \ref{eq:selfconsesumm}a for the LM phase indeed conforms to the condition used in
sec.\ \ref{subsection:SSElowomega} for analysis of the low-$\w$ behavior of the single self-energy,
\emph{viz} that $\Sigma_{A\sigma}^{I}(\w)$ vanishes no less slowly that the hybridization.


\subsection{Low-frequency behavior of $\mathbf{G(\w)}$: SC Phase}
\label{subsection:LWSCphase}

We now consider the implications of eq.\ \ref{eq:selfconsesumm}, mainly for the SC phase, 
beginning with the p-h asymmetric model (for which  eq.\ \ref{eq:selfconsesumm}a encompasses both phases). 
In the SC phase, $U<U_{c}$, the renormalized level $\epsilon^{*}_{\sigma} \equiv \epsilon^{*}(0)$ is 
non-vanishing; and from eq. \ref{eq:selfconsesumm}a the (spin-independent) single self-energy 
$\Ss^{I}(\w) =\Ss^{I}(\w, h=0) \propto |\w|^{2+3r}$, while its leading real part as $\w \rightarrow 0$ follows directly from Hilbert transformation and is linear in $\w$, \emph{viz}
$\Ss^{R}(\w)-\Ss^{R}(0) \sim -(\frac{1}{Z}-1)\w$, with  $Z=[ 1- (\partial \Ss^{R}(\w)/\partial \w)_{\w =0}]^{-1}$ 
the usual quasiparticle weight. The leading low-$\w$ quasiparticle form for the zero-field propagator can 
thus be obtained from eq.\ \ref{eq:Gs} for $G(\w) \equiv \Gs(\w, h=0)$. Defining the low-energy Kondo scale 
$\w_{*} \equiv \w_{K}$ in the SC phase by
\begin{equation}
\label{eq:defKscaleSC}
\w_{*} ~=~ \left[ \Gamma_{0} Z \right]^{\frac{1}{1-r}}
\end{equation}
(as familiar for the metallic model $r=0$, where $\w_{K} \propto \Gamma_{0}Z$), gives 
\begin{equation}
\label{eq:qpformSCasym}
\Gamma_{0}\w_{*}^{r}~G(\w)~\overset{|\tilde{\w}| \ll 1}{\sim}~\left[
\tilde{\w}^{+} - \frac{\epsilon^{*}(0)}{\Gamma_{0}\w_{*}^{r}} +\left(\mathrm{sgn}(\tilde{\w})\beta(r)+i
\right) |\tilde{\w}|^{r}
\right]^{-1}  
\end{equation}
($\tilde{\w} =\w/\w_{*}$);
where the quasiparticle damping embodied in $\Ss^{I}(\w)$ is asymptotically neglectable, as it vanishes more rapidly than both the hybridization and $\tilde{\w}$. Eq.\ \ref{eq:qpformSCasym} is the counterpart, in the SC phase, of the quasiparticle form for the zero-field LM phase given by eq.\ \ref{eq:qpformLMasymm}. 
The latter is of course also consistent with eq.\ \ref{eq:selfconsesumm}a for $\Sigma_{A\sigma}^{I}(\w)$, 
as detailed in sec.\ \ref{subsection:SSElowomega} where it leads to a conventional single self-energy 
$\Sigma^{I}(\w) \propto |\tilde{\w}|^{r}$ (eq.\ \ref{eq:ssewto0lmasymm}) that vanishes with 
precisely the same power as the hybridization.


\subsubsection{Particle-hole symmetry, and $U_{c}(r)$ as $r\rightarrow \tfrac{1}{2}$}
\label{subsubsection:LWSCphase_phsymm}

The p-h symmetric limit may obviously be handled similarly, now with
$\epsilon^{*}_{\sigma} \equiv \epsilon^{*}(0)=0$ for the zero-field SC phase, and
the single self-energy given by eq. \ref{eq:selfconsesumm}b. Importantly, note first in this case that 
eq.\ \ref{eq:selfconsesumm}b shows a symmetric SC phase to be self-consistently possible only for $r<\tfrac{1}{2}$. This explains the fact known from NRG studies~\cite{GB-Ingersent} that the critical $U_{c}(r)\rightarrow 0$ as
$r \rightarrow \tfrac{1}{2}-$, such that for $r>\tfrac{1}{2}$ a LM phase alone arises for any non-zero $U$.

For $r<\tfrac{1}{3}$ ($2-3r>1$), $\Sigma_{\sigma}^{R}(\w)$ again follows by Hilbert transformation of 
eq.\ \ref{eq:selfconsesumm}b as $\Ss^{R}(\w)-\Ss^{R}(0) \sim -(\frac{1}{Z}-1)~\w$. The resultant 
quasiparticle form is then given by eq. \ref{eq:qpformSCasym} but with $\epsilon^{*}(0)=0$. For
$\tfrac{1}{3}<r<\tfrac{1}{2}$ by contrast, $\Ss^{R}(\w)$ has the same leading low-$\w$ behavior as
$\Ss^{I}(\w)$ (see eq.\ \ref{eq:HTgenresult}), and the low-$\w$ behavior of $G(\w)$ in the SC phase is then
$\Gamma_{0}\w_{*}^{r}~G(\w) \sim~[
\frac{\w_{*}^{1-r}}{\Gamma_{0}}\tilde{\w}^{+} +(\mathrm{sgn}(\tilde{\w})\beta(r)+i) ~|\tilde{\w}|^{r}]^{-1}$.
In either case, of course, the ultimate  low-$\w$ behavior of the propagator is that of the non-interacting limit, 
such that the known~\cite{MTGEPJB}  `pinning condition' eq.\ \ref{eq: SC pinningagain} is recovered, 
reflecting the adiabatic continuity to the non-interacting limit that is inherent to the SC phase. 

The quasiparticle form in the LM phase at p-h symmetry is also naturally consistent with 
eq.\ \ref{eq:selfconsesumm}a for $\Sigma_{A\sigma}^{I}(\w)$, leading (see sec.\ \ref{subsection:symmSSElowomega})
to a conventional single self-energy  $\Sigma^{I}(\w) \propto |\tilde{\w}|^{-r}$ (eq.\ \ref{eq:ssewto0lmsymm}) whose low-energy divergence is indicative of the NFL nature of the  LM phase.


\section{Scaling and the Quantum critical point.}
\label{section:QCP}
We now consider further the scaling behavior of the zero-field propagator, and what can be deduced 
generally from it regarding the QCP itself.~\cite{fn17}

As discussed above, in both the LM and SC phases the problem is characterized by a low-energy scale  $\w_{*}$, 
eqs.\ (\ref{eq:w*def},\ref{eq:defKscaleSC}), that vanishes as $U\rightarrow U_{c}$ from either phase. 
A simple argument then gives the general form for the zero-field propagator $G(\w)$ $\equiv \Gs(\w, h=0)$
in the scaling regime; for as the transition is approached, $u = |1-\frac{U}{U_{c}}| \rightarrow 0$ and the low-energy scale vanishes,  as $\w_{*} \sim u^{a}$ with some power $a$. $G(\w)$ can then be expressed in the general scaling form $G(\w) \sim u^{-ab}\Psi_{\alpha}(\w/u^{a})$ in terms of two exponents $a$ and $b$, and with $\alpha =$ SC or LM denoting the phase; i.e.\ $\w_{*}^{b}~G(\w) \sim \Psi_{\alpha}(\w/\w_{*})$ [or, to be dimensionally precise, $[\Gamma_{0}]^{\frac{1-b}{1-r}}~ \w_{*}^{b} ~G(\w)=\Psi_{\alpha}(\w/\w_{*})$,
with $\Psi_{\alpha}$ dimensionless]. Note that it is $\w_{*}^{b}~G(\w)$ -- where the scale $\w_{*}$ is 
vanishing as the transition is approached -- and not e.g.\ $G(\w)$ itself, which exhibits universality as a function of $\tilde{\w}=\w/\w_{*}$. This equation embodies the scaling of the propagator close to the transition, and as such holds for \emph{any} finite $\tilde{\w} =\w/\w_{*}$ in the limit $\w_{*} \rightarrow 0$. 

However the exponent $b$ can be deduced simply and generally, solely from the low-$\tilde{\w}$ behavior of 
the propagator (the quasiparticle forms). The latter has already been obtained, for both asymmetric and p-h symmetric cases, and for both the LM phase 
(sec.\ \ref{subsection:SSElowomega}, eqs.\ (\ref{eq:speclowfreqLMasymm},\ref{eq:speclowfreqLMsymm})) 
and the SC phase (sec.\ \ref{subsection:LWSCphase}, eq.\ \ref{eq:qpformSCasym}). From this it follows directly 
that $b=r$ in \emph{all} cases. The general scaling form is thus
$\Gamma_{0} \w_{*}^{r} ~G(\w)=\Psi_{\alpha}(\tilde{\w})$; or equivalently 
\begin{equation}
\label{eq:scal1}
\pi \Gamma_{0}~ \w_{*}^{r} ~D(\w)~=~ 
\Psi_{\alpha}^{I}(\tilde{\w})~~~~~~:~\tilde{\w} =\w/\w_{*}
\end{equation}
for the local spectrum, where 
$\Psi_{\alpha}(\tilde{\w})=\Psi_{\alpha}^{R}(\tilde{\w})-i\Psi_{\alpha}^{I}(\tilde{\w})$.

As shown in secs.\ \ref{subsection:SSElowomega}, \ref{section:SSElowomegaLW}, on the
lowest energy scales $|\tilde{\w}| \ll 1$, $\Psi_{LM}^{I}(\tilde{\w}) \propto |\tilde{\w}|^{r}$ 
in the LM phase (whether ph-symmetric or asymmetric); and likewise 
$\Psi_{SC}^{I}(\tilde{\w}) \propto |\tilde{\w}|^{r}$ in the SC phase for the 
asymmetric model, but with $\Psi_{SC}^{I}(\tilde{\w}) \propto |\tilde{\w}|^{-r}$ at ph-symmetry.
This behavior has also been shown numerically by NRG, for the p-h symmetric~\cite{BPH1997,nrglmacomp} and asymmetric~\cite{Vojta_Bulla2001} models (it is also that arising within the LMA~\cite{MTG_SPAIM,MTG_APAIM,MTGSPAIMepl}). From NRG,~\cite{Vojta_Bulla2001} 
the coefficients of the leading low-$\tilde{\w}$ power-laws in $\Psi_{SC}^{I}(\tilde{\w})$ and $\Psi_{LM}^{I}(\tilde{\w})$ are further found to be equal for 
$\w \gtrless 0$, regardless of whether the model is p-h asymmetric or symmetric.
For the LM phase this is natural, given that in RG terms the p-h asymmetry flows to zero at the 
LM FP.~\cite{GB-Ingersent,Vojta_Bulla2001} For the asymmetric SC (ASC) phase by contrast, the reasons 
for this behavior are immediately clear from eq.\ \ref{eq:qpformSCasym}: the fact that the 
self-energy vanishes more rapidly than the hybridization as  $|\tilde{\w}|\rightarrow 0$, means that the low-$|\tilde{\w}|$ behavior of $\Psi_{SC}^{I}(\tilde{\w})$ is controlled exclusively by the hybridization,
which is symmetric in frequency by construction.

\begin{figure}
\includegraphics[height=8cm]{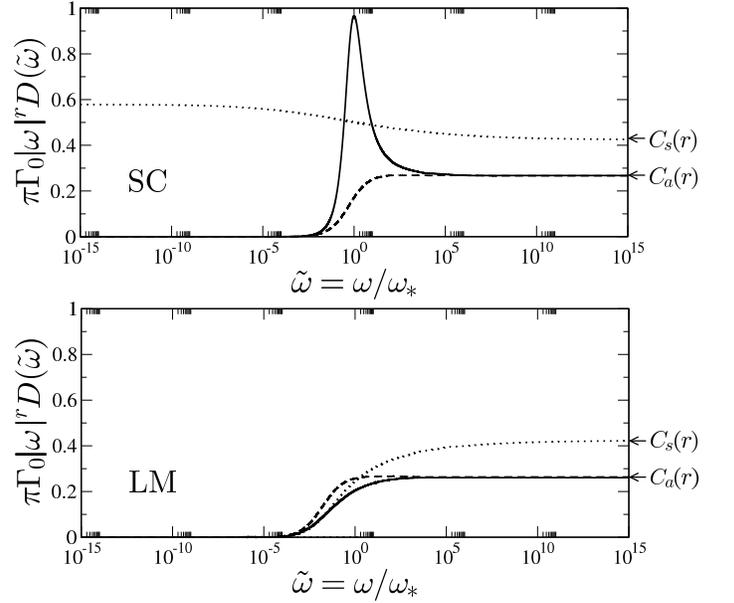}
\caption{\label{fig:scalspecr=0.45} For $r=0.45$, FDM-NRG determined scaling spectra shown as
$|\tilde{\w}|^{r}\Psi_{\alpha}^{I}(\tilde{\w}) \equiv \pi \Gamma_{0} |\w|^{r}D(\w)$ \emph{vs.}
$\tilde{\w}=\w/\w_{*}$ on a log-scale. \emph{Upper panel:} asymmetric SC phase (with $x=\epsilon +\tfrac{U}{2}<0$),
for both $|\tilde{\w}| <0$ (solid line) and $|\tilde{\w}| >0$ (dashed line); and symmetric SC phase ($x=0$, dotted line).
\emph{Lower panel:} corresponding LM phase results for both asymmetric and symmetric cases. For the asymmetric cases, results shown are independent of $|x|$. 
For both SC and LM phases, coefficients of proportionality in $\w_{*}$ have been (freely) chosen such that  spectral maxima occur at $|\tilde{\w}| =1$. The coefficients $C_{s}(r)$ and $C_{a}(r)$ relevant to the QCP as discussed in sec.\ \ref{subsection:QCPperse}, are indicated by arrows.
}
\end{figure}

We have also calculated full scaling spectra $\Psi_{\alpha}^{I}(\tilde{\w})$ using NRG,
for a representative range of $r\in (0,1)$, and varying the p-h asymmetry parameter 
$x=\epsilon +\tfrac{U}{2}$. Labelling temporarily their $x$-dependence, it is readily shown from a p-h transformation that $\Psi_{\alpha}^{I}(\tilde{\w}; x) = \Psi_{\alpha}^{I}(-\tilde{\w}; -x)$,
so that only either $x \leq 0$ or $x\geq 0$ need be considered. In fact, however,
subject only to fixed $\mathrm{sgn}(x)$, we find the full scaling spectra for the asymmetric model to be 
\emph{independent} of the asymmetry $|x|\neq 0$ (a point to which we shall return below).
Representative NRG scaling spectra are given in Fig.\ \ref{fig:scalspecr=0.45} for $r=0.45$, shown specifically in the form $|\tilde{\w}|^{r}\Psi_{\alpha}^{I}(\tilde{\w}) \equiv \pi \Gamma_{0} |\w|^{r}D(\w)$
(eq.\ \ref{eq:scal1}). The upper panel shows both the asymmetric SC (ASC) phase (with $x<0$) and the symmetric SC (SSC) phase, while the lower panel gives the corresponding LM phase spectra; and we note that for the asymmetric model, $\Psi_{\alpha}^{I}(\tilde{\w})\neq \Psi_{\alpha}^{I}(-\tilde{\w})$, i.e.\ the scaling spectra are not fully p-h symmetric.

\subsection{Quantum critical point}
\label{subsection:QCPperse}

Eq.\ \ref{eq:scal1} also yields very simply the exact behavior precisely \emph{at} the QCP, where $\w_{*} =0$: since the QCP must be scale-free (i.e.\ independent of $\w_{*}$), the asymptotic behavior of $\Psi_{\alpha}^{I}(\tilde{\w})$ for large-$\tilde{\w}$ follows immediately as 
$\Psi_{\alpha}^{I}(\tilde{\w}) \propto |\tilde{\w}|^{-r}$.
Hence the leading low-$\w$ dependence of the QCP spectrum is
\begin{equation}
\label{eq:qcpspec}
\pi \Gamma_{0}~D(\w)~\propto ~|\w|^{-r}
\end{equation}
(which we emphasize holds at the QCP for both the p-h symmetric \emph{and} asymmetric models, as indeed found  numerically using NRG~\cite{Vojta_Bulla2001,nrglmacomp}).
It is in otherwords the \emph{high}-$\tilde{\w}$ `tails' of the scaling spectra $\Psi_{\alpha}^{I}(\tilde{\w})$
which determine the leading low-$\w$ behavior of the QCP spectrum itself. From this one can obtain the 
asymptotic $\tilde{\w}$-dependence of the self-energies as the QCP is approached, and hence their leading $\w$-dependence at the QCP itself, as now shown. 

Consider first the approach to the QCP from the SC phase. From $\pi D(\w) = -\mathrm{Im} G(\w)$ in the 
SC phase (with zero-field propagator $G(\w) = [\w^{+} -\epsilon -\Gamma(\w) -\Sigma(\w)]^{-1}$),
the scaling spectrum $\pi\Gamma_{0} \w_{*}^{r}D(\w) = \Psi_{SC}^{I}(\tilde{\w})$ follows as
\begin{equation}
\nonumber
\pi\Gamma_{0} \w_{*}^{r}~D(\w) = -\mathrm{Im}
\frac{1}{-\tilde{\epsilon}^{*}(0) +|\tilde{\w}|^{r} \left[\mathrm{sgn}(\w) \beta (r)+i \right] -
\delta{\tilde{\Sigma}}(\tilde{\w})} 
\nonumber
\end{equation}
where 
$-|\tilde{\w}|^{r}[\mathrm{sgn}(\tilde{\w}) \beta (r)+i] = \Gamma(\w)/(\Gamma_{0}\w_{*}^{r})$,
with $\tilde{\w} =\w/\w_{*}$ and  $\w_{*} =\w_{K}$ the Kondo scale; and where
$\delta{\tilde{\Sigma}}(\tilde{\w})= \delta\Sigma(\w)/(\Gamma_{0}\w_{*}^{r})$
with $\delta\Sigma(\w)=\delta\Sigma^{R}(\w) -i\Sigma^{I}(\w)$
and $\delta\Sigma^{R}(\w) =\Sigma^{R}(\w)-\Sigma^{R}(0)$.
Likewise, $\tilde{\epsilon}^{*}(0)=\epsilon^{*}(0)/(\Gamma_{0}\w_{*}^{r})$, with
$\epsilon^{*}(0) =\epsilon +\Sigma^{R}(0)$ the renormalized level; and we assume
$\tilde{\epsilon}^{*}(0)$ to be bounded as the transition is approached and 
$\w_{*}$ vanishes (\emph{cf} the situation shown to arise in the LM phase, eq.\ \ref{eq:estargamma}).
Hence for large $|\tilde{\w}| \gg1 $,
\begin{equation}
\label{eq:qcpscside}
\begin{split}
& \pi \Gamma_{0} \w_{*}^{r}~D(\w) ~\sim ~
\\
&\frac{\left[|\tilde{\w}|^{r} +\tilde{\Sigma}^{I}(\tilde{\w})\right]}
{\left[ \mathrm{sgn}(\w)\beta(r)|\tilde{\w}|^{r}
-\delta\tilde{\Sigma}^{R}(\tilde{\w})\right]^{2}
+\left[|\tilde{\w}|^{r} +\tilde{\Sigma}^{I}(\tilde{\w})\right]^{2}
}~.
\end{split}
\end{equation}
But as shown above,
$\pi\Gamma_{0} \w_{*}^{r}~D(\w)=\Psi_{SC}^{I}(\tilde{\w}) \propto |\tilde{\w}|^{-r}$ for $|\tilde{\w}| \gg 1$; 
whence from eq.\ \ref{eq:qcpscside}, $\tilde{\Sigma}^{I}(\tilde{\w}) \propto |\tilde{\w}|^{\phi}$ with 
$\phi \leq r$ necessarily. If $\phi < r$, then for $|\tilde{\w}| \gg 1$ the self-energies would be irrelevant compared to the hybridization, and the QCP would be trivially non-interacting. Instead one naturally expects an
interacting QCP (as NRG calculations confirm), for which $\phi =r$ is thus required; i.e.\ 
the $|\tilde{\w}| \gg 1$ behavior of the self-energy must be of form
\begin{equation}
\label{eq:seqcpim0}
\frac{1}{\Gamma_{0}\w_{*}^{r}}\Sigma^{I}(\w)~ = ~
\tilde{\Sigma}^{I}(\tilde{\w}) \sim~ |\tilde{\w}|^{r}
\left[\alpha_{sc}^{+}\theta(\tilde{\w})+\alpha_{sc}^{-}\theta(-\tilde{\w})
\right]
\end{equation}
or equivalently
$\Sigma^{I}(\w)\sim ~\Gamma_{0}|\w|^{r}[\alpha_{sc}^{+}\theta(\w)+\alpha_{sc}^{-}\theta(-\w)]$.
Eq.\ \ref{eq:seqcpim0} gives the large-$|\tilde{\w}|$ `tails' of the scaling self-energy.
But precisely at the QCP, where $\w_{*} =0$, this behavior holds right down to $\w =0$.  And from Hilbert transformation, using only the asymptotic behavior in eq.\ \ref{eq:seqcpim0}, it can be shown that 
$\delta\Sigma^{R}(\w) \sim \Gamma_{0}|\w|^{r}(
-\mathrm{sgn}(\w)\beta(r) \tfrac{1}{2}[\alpha_{sc}^{+} +\alpha_{sc}^{-}]+
\tfrac{1}{\beta(r)}\tfrac{1}{2}[\alpha_{sc}^{+} -\alpha_{sc}^{-}])$,
likewise $\propto |\w|^{r}$.

The low-frequency QCP behavior $D(\w) \propto |\w|^{-r}$ is of course also that occurring at p-h symmetry in the SC phase (see e.g.\ eq.\ \ref{eq: SC pinningagain}). But whether or not the QCP spectrum itself is p-h symmetric away from the p-h symmetric point $\epsilon= -U/2$, is reflected in the coefficient of the leading $|\w|^{-r}$
divergence. The general form for the spectrum is clearly 
\begin{equation}
\label{eq:qcpspecgen}
\pi \Gamma_{0}~D(\w)~\sim ~|\w|^{-r}~\left[
C_{+} ~\theta (\w)~+~ C_{-}~\theta (-\w)
\right]
\end{equation}
with (dimensionless) coefficients $C_{\pm}$. Only if $C_{+} =C_{-}$ will the QCP spectrum be asymptotically p-h symmetric. The coefficients $C_{\pm}$ obtained from eqs.\ (\ref{eq:seqcpim0},\ref{eq:qcpscside}) are 
\begin{equation}
\label{eq:qcpspecC}
\begin{split}
&C_{\pm}~=~ 
\\
&\frac{1+\alpha_{sc}^{\pm}}
{
\left(
\beta(r)\left[1+\tfrac{1}{2}(\alpha_{sc}^{+}+\alpha_{sc}^{-})\right]
\mp
\frac{1}{\beta(r)}\tfrac{1}{2}
(\alpha_{sc}^{+}-\alpha_{sc}^{-})
\right)^{2}+
\left(
1+\alpha_{sc}^{\pm}
\right)^{2}
}
\end{split}
\end{equation}
from which $C_{+}=C_{-}$ only if $\alpha_{sc}^{+}=\alpha_{sc}^{-}$. The latter is of course guaranteed 
at p-h symmetry (where $\Sigma^{I}(\w) = \Sigma^{I}(-\w)$ $\forall ~\w$). But from  direct calculation 
using NRG we find that $\alpha_{sc}^{+}=\alpha_{sc}^{-} \equiv \alpha_{sc}$ arises \emph{regardless} of 
whether the model is p-h symmetric or asymmetric; such that (eq.\ \ref{eq:seqcpim0})
$\tilde{\Sigma}^{I}(\tilde{\w}) \sim \alpha_{sc} |\tilde{\w}|^{r}$ acquires an emergent p-h symmetry as 
the QCP is approached, and hence
\begin{equation}
\label{eq:qcpspecCSQCPsc}
C_{+}~=~C_{-}~=~C~=~ \mathrm{cos}^{2}(\tfrac{\pi}{2}r)~\frac{1}{[1+\alpha_{sc}(r)]}
\end{equation}
(using $[1+\beta^{2}(r)]^{-1} = \mathrm{cos}^{2}(\frac{\pi}{2}r) $, and with the $r$-dependence of
$\alpha_{sc}$ explicit). The asymptotic QCP spectrum is thus always p-h symmetric, as also reported in previous NRG studies.~\cite{Vojta_Bulla2001} As a corollary, note that since the QCP spectrum arises from the 
$|\tilde{\w}| \gg 1$ `tails' of the scaling spectrum $\Psi_{SC}^{I}(\tilde{\w})$, it follows that the ASC 
scaling spectrum $\Psi_{SC}^{I}(\tilde{\w})$ is effectively p-h symmetric for $|\tilde{\w}| \gg 1$ as well 
as for $|\tilde{\w}| \ll 1$ (as indeed seen clearly in Fig.\ \ref{fig:scalspecr=0.45}).

\begin{figure}
\includegraphics[height=5cm, width=8cm]{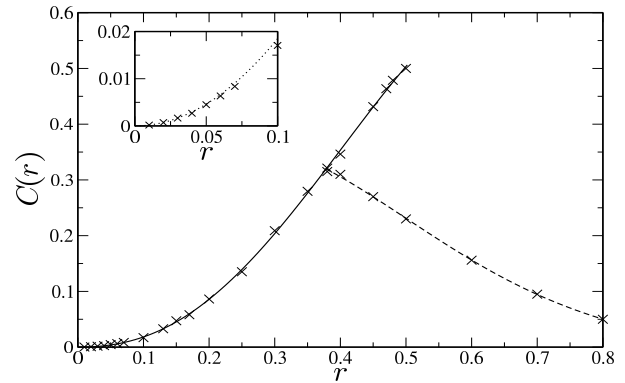}
\caption{\label{fig:QCPC(r)} $r$-dependence of 
coefficients $C\equiv C(r)$ (eq.\ \ref{eq:qcpspecCSQCPsc}) describing the QCP spectrum, 
$\pi\Gamma_{0}D(\w) = C|\w|^{-r}$. Calculated from FDM-NRG, showing both $C\equiv C_{s}(r)$
characteristic of the symmetric QCP for $r\leq \tfrac{1}{2}$ (points, with solid line as guide to eye); 
and $C\equiv C_{a}(r)$ (points with dashed line) characteristic of the asymmetric QCP for 
$r>r^{*} \simeq 0.375$. As $r\rightarrow 0$ the asymptotic behavior is found to be 
$C_{s}(r) =3\pi^{2}r^{2}/16$ (see inset, where dotted line shows $3\pi^{2}r^{2}/16$); while 
$C_{s}(r) = \tfrac{1}{2}$ as $r\rightarrow \tfrac{1}{2}$, as explained in text.
}
\end{figure}

The ubiquity of p-h symmetric behavior in the QCP spectrum may at first sight seem slightly counterintuitive, 
since from extensive NRG studies (notably\ [\onlinecite{GB-Ingersent}]) it is well known that distinct 
symmetric and asymmetric critical fixed points exist: the symmetric QCP (SQCP) is the critical point for both the p-h symmetric model, where it occurs for the entire $r$-range $0<r<\tfrac{1}{2}$ where the transition exists; and also for the p-h asymmetric model where it occurs for $0<r<r^{*}$, with  $r^{*} \simeq 0.375$ determined numerically.~\cite{GB-Ingersent} By contrast, for $r>r^{*}$ in the p-h asymmetric model, the  critical point is the asymmetric QCP (AQCP).~\cite{GB-Ingersent}  As illustrated in Fig.\ \ref{fig:QCPC(r)}, however, we find using NRG
that the distinction between the SQCP and the AQCP resides in the coefficients $C \equiv C(r)$. 
For $0<r<r^{*}$, $C \equiv C_{s}(r)$ is found to be independent of whether the model is p-h symmetric or 
asymmetric (as embodied in $x=\epsilon + \tfrac{U}{2}$); consistent in otherwords with a single SQCP in this $r$-range. For $r^{*} < r$ by contrast, the p-h asymmetric model (and hence the AQCP) has a $C \equiv C_{a}(r)$ 
which differs from the $C \equiv C_{s}(r)$ arising for the p-h symmetric model (and hence SQCP) in the interval
$r^{*} < r <\tfrac{1}{2}$, see both Fig.\ \ref{fig:QCPC(r)} and Fig.\ \ref{fig:scalspecr=0.45}.
For $r^{*} < r$, moreover, we find $C_{a}(r)$ to be \emph{independent} of the degree of p-h asymmetry embodied in $x \neq 0$ (as in fact follows from the $|x|$-independence of the full ASC scaling spectrum $\Psi_{SC}^{I}(\tilde{\w})$ mentioned above).
This in turn implies the occurrence of a single AQCP (as opposed to a line of critical fixed points parametrised 
by p-h asymmetry), as indeed inferred from perturbative RG study~\cite{Fritz_Vojta2004} of the maximally asymmetric model ($U=\infty$, $\epsilon$ finite) for $r$ values close to $\tfrac{1}{2}$ and $1$.

The approach to the QCP has been considered above from the SC phase, $U<U_{c}(r)$. Equally, 
one can of course approach it from the LM side, focusing as such on
$\pi D(\w) = -\frac{1}{2}\mathrm{Im} \sum_{\sigma} G_{A\sigma}(\w)$ and hence the self-energies
$\Sigma_{A\sigma}(\w)$; but otherwise proceeding in direct parallel to the above. For
$|\tilde{\w}| \gg1 $, $\Psi_{LM}^{I}(\tilde{\w})=\pi\Gamma_{0} \w_{*}^{r}D(\w)$ is given by
(\emph{cf} eq.\ \ref{eq:qcpscside})
\begin{equation}
\begin{split}
\label{eq:qcplmside}
&\pi\Gamma_{0} \w_{*}^{r}~D(\w) ~\sim ~ 
\\
&\tfrac{1}{2} \sum_{\sigma}
\frac{\left(|\tilde{\w}|^{r} +\tilde{\Sigma}_{A\sigma}^{I}(\tilde{\w})\right)}
{\left( \mathrm{sgn}(\w)\beta(r)|\tilde{\w}|^{r}
-\delta\tilde{\Sigma}_{A\sigma}^{R}(\tilde{\w})\right)^{2}
+\left(|\tilde{\w}|^{r} +\tilde{\Sigma}_{A\sigma}^{I}(\tilde{\w})\right)^{2}
}
\end{split}
\end{equation}
and likewise (\emph{cf} eq.\ \ref{eq:seqcpim0})
\begin{equation}
\label{eq:seqcpimLM}   
\Sigma_{A\sigma}^{I}(\w)~\sim ~ \Gamma_{0}|\w|^{r}~
\left[\alpha_{\sigma}^{+}\theta(\w)+\alpha_{\sigma}^{-}\theta(-\w)
\right]
\end{equation}
holding at the QCP down to $\w=0$. Using this, $\pi\Gamma_{0}D(\w)$ at the QCP is given by
eq.\ \ref{eq:qcpspecgen}, with: 
\begin{widetext}
\begin{equation}
\label{eq:qcpspecCLM} 
C_{\pm}~=~
\tfrac{1}{2} \sum_{\sigma} \frac{1+\alpha_{\sigma}^{\pm}}
{
\left(
\beta(r)\left[1+\tfrac{1}{2}(\alpha_{\sigma}^{+}+\alpha_{\sigma}^{-})\right]
\mp
\frac{1}{\beta(r)}\tfrac{1}{2}(\alpha_{\sigma}^{+}-\alpha_{\sigma}^{-})
\right)^{2}+
\left(
1+\alpha_{\sigma}^{\pm}
\right)^{2}
}
\end{equation}
\end{widetext}
From this it follows that $C_{+}=C_{-} \equiv C$ arises if either of two conditions is 
satisfied: $\alpha_{\sigma}^{+} = \alpha_{-\sigma}^{-}$ or $\alpha_{\sigma}^{+} = \alpha_{\sigma}^{-}$, 
with the former guaranteed by symmetry at the p-h symmetric point (where 
$\Sigma_{A\sigma}^{I}(\w) =\Sigma_{A-\sigma}^{I}(-\w)$ $\forall ~\w$).
From NRG calculations we find in practice that \emph{both} conditions are satisfied at the QCP, 
regardless of whether the model is p-h symmetric or asymmetric. All four coefficients $\alpha_{\sigma}^{\pm}$ 
thus coincide, with $\alpha_{\sigma}^{\pm}\equiv \alpha_{sc}(r)$ (and $C$ given by eq.\ \ref{eq:qcpspecCSQCPsc}). Hence, as the QCP is approached from either the LM or SC sides, all self-energies $\Sigma_{A\uparrow}(\w)$, $\Sigma_{A\downarrow}(\w)$ and $\Sigma(\w)$ are asymptotically coincident; and the QCP is thus (naturally) independent of the phase from which one accesses it (as also seen directly in Fig.\ \ref{fig:scalspecr=0.45}). 

Finally, we comment on the $r$-dependence of the $C_{s}(r)$ characteristic of the SQCP, obtained from NRG as
in Fig.\ \ref{fig:QCPC(r)}. First, note that as $r \rightarrow 0$, $C_{s}(r)$ is found to vanish $\propto r^{2}$; and hence from eq.\ \ref{eq:qcpspecCSQCPsc} ($\alpha_{\sigma}^{\pm} \equiv$)
$\alpha_{sc}(r) \propto  1/r^{2}$ diverges at low-$r$.
Specifically, the numerics give $C_{s}(r) = 3\pi^{2}r^{2}/16$ as $r \rightarrow 0$ (Fig.\ \ref{fig:QCPC(r)}, inset).
Remarkably, this result also arises from an LMA description~\cite{MTGSPAIMepl} of the QCP at p-h symmetry.~\cite{fn31}

 Second, it is seen from Fig.\ \ref{fig:QCPC(r)} that $C_{s}(r) = \tfrac{1}{2}$ as $r\rightarrow \tfrac{1}{2}$.
The reasons for this follow from the fact (shown in sec.\ \ref{subsubsection:LWSCphase_phsymm}) that as 
$r \rightarrow \tfrac{1}{2}-$ the critical $U_{c}(r)$ for the transition vanishes. The latter means the SQCP becomes non-interacting at $r=\tfrac{1}{2}$, whence $\alpha_{sc}(r)$ must vanish as $r\rightarrow \tfrac{1}{2}-$; from eq.\ \ref{eq:qcpspecCSQCPsc} it then follows directly that $C_{s}(r=\tfrac{1}{2})=\mathrm{cos}^{2}(\tfrac{\pi}{4}) = \tfrac{1}{2}$, as indeed found numerically.

\section{Finite field}
\label{section:finitefield}

Our focus above has been the zero-field case, and we now turn to finite fields,~\cite{Zitzler2002,Fritz_Vojta2004} 
in particular to what may be deduced using the Luttinger self-consistency arguments sketched in  sec.\ \ref{section:SSElowomegaLW}. The situation arising if the field is applied globally is quite simple. There the 
$\w =0$ hybridization $\Gamma^{I}_{\sigma}(0, h) = \Gamma_{0}|h|^{r}$ is non-zero, hence so too is the
single-particle spectrum at the Fermi level, and Luttinger self-consistency in this case gives
 $\Ss^{I}(\w, h\neq 0) \propto |\w|^{2}$ as $\w \rightarrow 0$.
The asymptotic low-energy physics is thus that of the normal metallic AIM.~\cite{Zitzler2002} \\

Importantly, however, the results given in eq.\ \ref{eq:selfconsesumm} hold equally at finite-$h$, for a field  
applied locally to the impurity; reflecting the fact that the hybridization is independent of $h$ 
(whence $D_{A\sigma}(\w, |h|)$ is again given by eq.\ \ref{eq:seconspec}, now in terms of $\Sigma_{A\sigma}(\w, |h|)$).
Only eq.\ \ref{eq:selfconsesumm}a is relevant for $h \neq 0$
(eq.\ \ref{eq:selfconsesumm}b applies solely to the $h=0$ p-h symmetric limit). Further, as discussed in 
sec.\ \ref{subsubsection:s-estructure}, for any non-zero field (say $h>0$), $\Gs(\w, h) \equiv G_{A\sigma}(\w, |h|)$ 
and $\Ss (\w, h) \equiv \Sigma_{A\sigma}(\w, |h|)$; i.e.\ only a single self-energy description 
arises, with the same low-$\w$ behavior $\Sigma_{A\sigma}^{I}(\w, |h|)\propto |\w|^{2+3r}$ 
(eq.\ \ref{eq:selfconsesumm}a) occurring for both $U\gtrless U_{c}$ (provided 
$\epsilon_{A\uparrow}^{*} \neq 0 \neq \epsilon_{A\downarrow}^{*}$).
While our convention at finite-$h$ is to retain the A-label for $U>U_{c}$ (for reasons explained
in sec.\ \ref{subsubsection:s-estructure}), we can drop it in the following because the essential results
below will be seen to be relevant only to $U<U_{c}$. And only $h \geq 0$ need be considered in the following, 
since the renormalized levels satisfy $\epsilon_{\sigma}^{*}(h)=\epsilon_{-\sigma}^{*}(-h)$.

Eq.\ \ref{eq:selfconsesumm}a does not however encompass all relevant self-consistent possibilities arising for 
non-zero local field, as now explained.  The renormalized levels are given by eq.\ \ref{eq:rlsc}
(or eq.\ \ref{eq:rllm}), \emph{viz} $\epsilon_{\sigma}^{*}(h)=\epsilon -\sigma h+\Sigma_{\sigma}^{R}(0,h)$. 
For general values of the field $h$, both $\epsilon_{\uparrow}^{*}(h)$ and $\epsilon_{\downarrow}^{*}(h)$ are 
non-vanishing, with Luttinger self-consistency giving
$\Sigma_{\sigma}^{I}(\w, h) \propto |\w|^{2+3r}$ (eq.\ \ref{eq:selfconsesumm}a).
However on tuning the field to a particular value, call it $h^{\prime}$,
one of the renormalized levels may vanish (say $\epsilon_{\sigma}^{*}(h^{\prime})$),
while the other,  $\epsilon_{-\sigma}^{*}(h^{\prime})$, remains non-zero.
This situation arises generally in the p-h asymmetric PAIM only, i.e.\ for
$x=\epsilon +\tfrac{U}{2} \neq 0$ (being precluded at p-h symmetry because there 
$\epsilon_{\sigma}^{*}(h) = - \epsilon_{-\sigma}^{*}(h)$ for \emph{any} field, 
as follows using the transformation eq.\ \ref{eq:phst}).
That it does so is obvious in the trivial non-interacting limit. Here
$\epsilon_{\sigma}^{*}(h)\equiv \epsilon -\sigma h$ vanishes at 
$h^{\prime} =\sigma \epsilon$ --- i.e.\ $\epsilon_{\downarrow}^{*}(h^{\prime})$ 
vanishes if $\epsilon <0$, and  $\epsilon_{\uparrow}^{*}(h^{\prime})$ if $\epsilon >0$ ---
while $\epsilon_{-\sigma}^{*}(h^{\prime}) = 2\epsilon \neq0$. In consequence the $-\sigma$-spin spectrum
$D_{-\sigma}(\w, h^{\prime}) \propto |\w|^{r}$ as $|\w| \rightarrow 0$. But since 
$\epsilon_{\sigma}^{*}(h^{\prime})$ vanishes at the field $h^{\prime}$, the $\sigma$-spin spectrum effectively acquires particle-hole symmetry at low-energies, and satisfies the `pinning condition' eq.\ \ref{eq: SC pinningagain}  also characteristic of the symmetric SC phase at zero-field:
\begin{equation}
\label{eq:pinninghprime}
\pi \Gamma_{0} D_{\sigma}(\w, h^{\prime})
\overset{|\w|\rightarrow 0}{\sim}
\mathrm{cos}^{2}(\tfrac{\pi}{2}r)~|\w|^{-r} ~:\epsilon_{\sigma}^{*}(h^{\prime}) =0
\end{equation}

The situation above is naturally not confined to the non-interacting limit; but with interactions present one must 
establish the self-consistent low-$\w$ behavior of the self-energies $\Sigma_{\sigma^{\prime}}^{I}(\w, h^{\prime})$ 
following the procedure outlined in  sec.\ \ref{section:SSElowomegaLW}.
With $D_{\sigma^{\prime}}(\w, h^{\prime}) \propto |\w|^{\lambda_{\sigma^{\prime}}}$ as $|\w|\rightarrow 0$, the 
self-energies 
$\Sigma_{\sigma^{\prime}}^{I}(\w, h^{\prime}) \propto |\w|^{2+\lambda_{\sigma^{\prime}}+2\lambda_{-\sigma^{\prime}}}$ 
(eq.\ \ref{eq:seconsis1}); and the spectrum $D_{\sigma^{\prime}}(\w, h^{\prime})$ is given by eq.\ \ref{eq:seconspec} 
in terms of $\Sigma_{\sigma^{\prime}}(\w, h^{\prime})$.
There are thus four possibilities to consider, \emph{viz} 
$2+\lambda_{\sigma^{\prime}}+2\lambda_{-\sigma^{\prime}} > r$ and $<r$,  for each of 
$\sigma^{\prime} =\sigma$ and $-\sigma$. As is readily checked, only one of them is Luttinger self-consistent, 
namely $2+\lambda_{\sigma^{\prime}}+2\lambda_{-\sigma^{\prime}} > r$ for both $\sigma^{\prime}$s 
(the others are ruled out by the requirement that the spectum be integrable, requiring $\lambda_{\sigma^{\prime}} >-1$).
Since $\epsilon_{-\sigma}^{*}(h^{\prime}) \neq 0$, the low-$\w$ behavior 
$D_{-\sigma}(\w, h^{\prime}) \propto |\w|^{r} \equiv |\w|^{\lambda_{-\sigma}}$ then follows using
eq.\ \ref{eq:seconspec}, i.e.\ $\lambda_{-\sigma}=r$; likewise, since $\epsilon_{\sigma}^{*}(h^{\prime}) =0$,
$D_{\sigma}(\w, h^{\prime}) \propto |\w|^{-r} \equiv |\w|^{\lambda_{\sigma}}$, i.e.\
$\lambda_{\sigma}=-r$. And the conditions $2+\lambda_{\sigma^{\prime}}+2\lambda_{-\sigma^{\prime}} > r$ require
$r<1$, as is so by construction. The self-energies thus have the asymptotic low-$\w$ behavior:
\begin{subequations}
\label{eq:seshnonzero}
\begin{align}
\Sigma_{\sigma}^{I}(\w, h^{\prime})&~\overset{\omega \rightarrow 0}{\propto}~|\w|^{2+r} ~~~~~~
:~ \epsilon_{\sigma}^{*}(h^{\prime}) =0
\\
\Sigma_{-\sigma}^{I}(\w, h^{\prime})&~\overset{\omega \rightarrow 0}{\propto}~|\w|^{2-r} ~~~~~~
:~\epsilon_{-\sigma}^{*}(h^{\prime}) \neq 0
\end{align}
\end{subequations}
The corresponding real parts $\Sigma_{\sigma^{\prime}}^{R}(\w, h^{\prime})$ follow from Hilbert transformation, and 
are necessarily linear in $\w$ as $|\w|\rightarrow 0$. The self-energies $\Sigma_{\sigma^{\prime}}(\w, h^{\prime})$ 
for either spin thus vanish more rapidly that the hybridization ($\propto |\w|^{r}$), and are  therefore irrelevant 
on the lowest energy scales. Hence the leading low-$\w$ dependence of the spectra is 
precisely that occurring in the \emph{non-interacting} limit; \emph{viz} $D_{-\sigma}(\w, h^{\prime}) \propto |\w|^{r}$ 
for the $-\sigma$-spin spectum, with the $\sigma$-spin spectrum again diverging as 
$D_{\sigma}(\w, h^{\prime}) \propto |\w|^{-r}$ and satisfying the condition
eq.\ \ref{eq:pinninghprime} that is characteristic of the non-interacting p-h symmetric 
model at zero-field. 

\begin{figure}
\includegraphics[height=10.0cm, width=8.0cm]{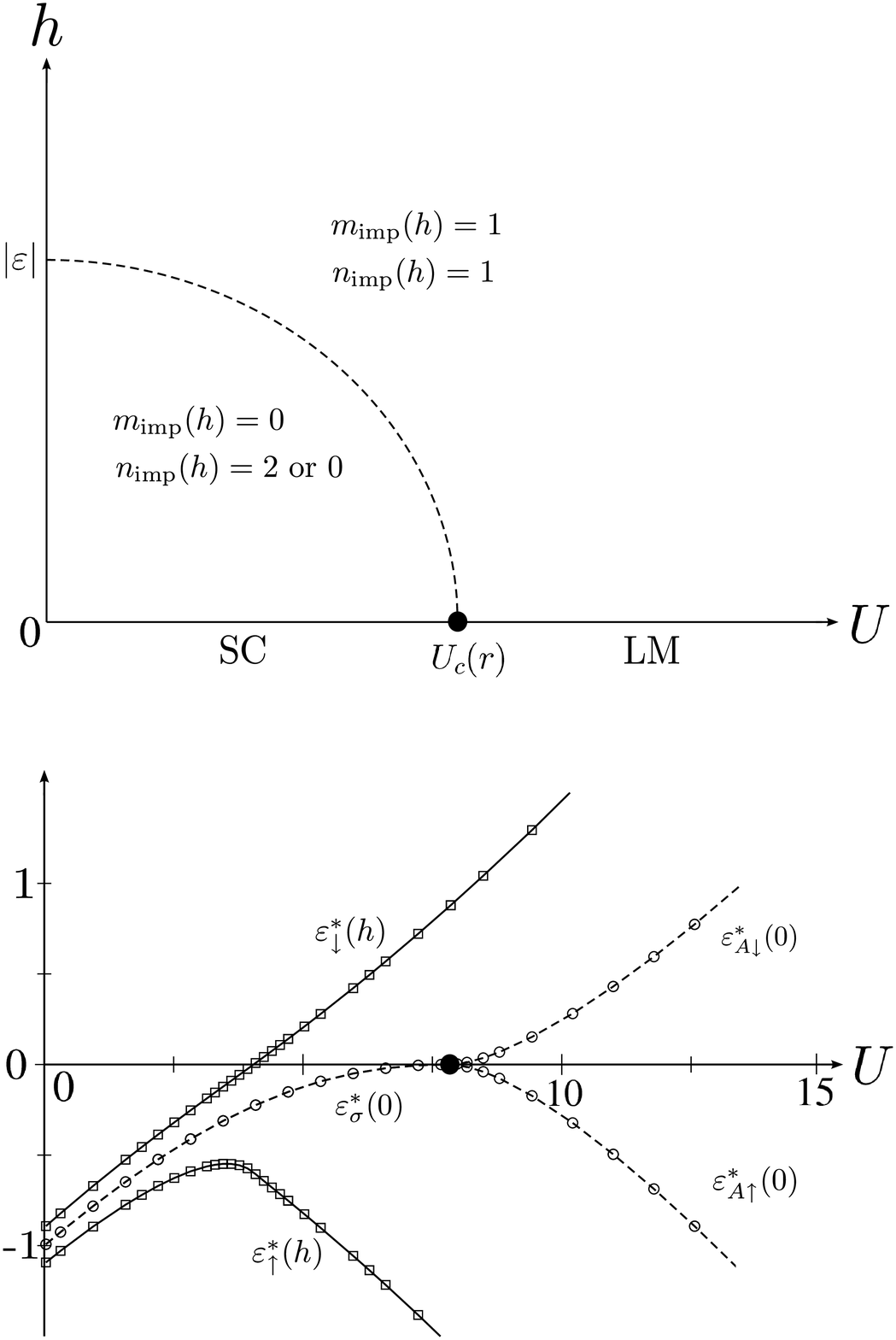}
\caption{\label{fig:fig6}  \emph{Upper}: schematic in the $(U, h)$-plane for the asymmetric PAIM 
($x=\epsilon +\tfrac{U}{2} \neq 0$), and a typical $r \in (0,1)$.
Dashed line shows points for which $\epsilon_{\sigma}^{*}(h) =0$ with
$\epsilon_{-\sigma}^{*}(h) \neq 0$ (terminating at $h=|\epsilon|$ in the non-interacting limit). The zero-field 
QCP at $U=U_{c}(r)$ is indicated, and is the only local QCP in the $(U, h)$-plane (see text).
\emph{Lower}: Renormalized levels $\epsilon_{\sigma}^{*}(h)$ \emph{vs} $U$, for $h=0.1$ 
(squares, with solid line as guide to eye) and $h=0$ (circles and dashed line);
shown for $r=0.35$ and fixed $x=\epsilon +\tfrac{U}{2} = -1$, and calculated from FDM-NRG.
For $h =0.1$, $\epsilon_{\downarrow}^{*}(h)$ changes sign at $U \simeq 4.1$ ($<U_{c}(r) \simeq 7.8$), while 
$\epsilon_{\uparrow}^{*}(h)<0$ remains. $[\Gamma_{0}]^{1/(1-r)} \equiv 1$ is taken as the energy unit.
}
\end{figure}

 We have numerically verified all preceding results using NRG calculations; in particular that 
eq.\ \ref{eq:pinninghprime} is satisfied at the field $h=h^{\prime}$, showing directly that the self-energies 
$\Sigma_{\sigma^{\prime}}(\w, h^{\prime})$ vanish more rapidly than the hybridization as
$|\w| \rightarrow 0$. For a typical $r \in (0,1)$, the situation in the $(U, h)$-plane (for some fixed 
$x\neq 0$) is summarized schematically in  Fig.\ \ref{fig:fig6} (upper),
with the dashed line showing the locus of points for which
$\epsilon_{\sigma}^{*}(h^{\prime}) =0$ with $\epsilon_{-\sigma}^{*}(h^{\prime}) \neq 0$. In practice 
the latter is found to arise (at a single field) for any $U <U_{c}(r)$, i.e.\ in the SC phase only;
the line terminating at $h^{\prime} =|\epsilon|$ in the non-interacting limit, as noted above, and with 
$h^{\prime} \rightarrow 0$ as $U\rightarrow U_{c}(r)-$.
The  zero-field QCP at $U=U_{c}(r)$ is also indicated in Fig.\ \ref{fig:fig6}. 
As explained below, it is the only local quantum critical point in the $(U, h)$-plane, reflecting
the fact that the local quantum phase transition is strictly destroyed for any non-zero field; 
while for any finite field the dashed line on which $\epsilon_{\sigma}^{*}(h^{\prime}) =0$
represents a simple bulk level-crossing. NRG results for the $U$-dependence of the renormalized levels 
$\epsilon_{\sigma}^{*}(h)$ for a fixed $h>0$ are shown in Fig.\ \ref{fig:fig6} (lower); from which
$\epsilon_{\downarrow}^{*}(h)$ is indeed seen to change sign at a certain $U<U_{c}(r)$, while
$\epsilon_{\uparrow}^{*}(h)$ remains sign-definite.
~\\

To elucidate the situation physically, recall as shown in sec.\ \ref{subsection:Luttintgen} that a Friedel sum 
rule holds at finite-field, with $n_{\mathrm{imp}, \sigma}(h)$ given by
eq.\ \ref{eq:nimpsboth} for any field. From this (remembering that $\Gamma_{\sigma}^{I}(0,h) =0$ for the 
local field), it follows that $n_{\mathrm{imp}, \sigma}(h) =1$ if 
$\epsilon_{\sigma}^{*}(h) <0$ and $n_{\mathrm{imp}, \sigma}(h) =0$ if
$\epsilon_{\sigma}^{*}(h) >0$.
With reference to Fig.\ \ref{fig:fig6} (upper), consider then the situation arising for any $U<U_{c}(r)$ 
upon increasing the field $h$ from zero towards and through  $h=h^{\prime}$.
For concreteness consider the case of $x<0$  where at zero-field (Fig.\ \ref{fig:phasediagram})
the ($\sigma$-independent) renormalized levels $\epsilon_{\sigma}^{*}(h=0) <0$ lie below the Fermi level;
 so that the zero-field excess charge $\nimp (h=0) =n_{\mathrm{imp}, \uparrow}+n_{\mathrm{imp}, \downarrow} =2$ 
while the excess magnetization $\mimp (h=0)=n_{\mathrm{imp}, \uparrow}-n_{\mathrm{imp}, \downarrow}$ naturally 
vanishes. On switching on the field $h$, both renormalized levels $\epsilon_{\sigma}^{*}(h)$ remain $<0$
for $h < h^{\prime}$. Hence eq.\ \ref{eq:nimpsboth} gives $\nimp (h)=2$ and $\mimp (h)=0$ -- just as at zero-field
(and we have verified numerically by NRG this striking result of vanishing magnetization for
$h \neq 0$, using eqs.\ \ref{eq:nimpsig}-\ref{eq:mimpnimp}).

But on crossing the field $h =h^{\prime}$, $\epsilon_{\uparrow}^{*}(h)<0$ remains, while 
$\epsilon_{\downarrow}^{*}(h)$ changes sign; so from eq.\ \ref{eq:nimpsboth}, $n_{\mathrm{imp}, \downarrow}(h)$ 
thus drops abruptly from $1$ to $0$ at $h=h^{\prime}$. Hence on increasing the field through the 
point $h =h^{\prime}$, $\nimp (h)$ drops discontinuously from $2$ to $1$  and $\mimp (h)$ increases 
abruptly from $0$ to $1$ (see Fig.\ \ref{fig:fig6} (upper)). By contrast, the \emph{local} impurity magnetization 
$m(h)$ behaves quite differently: it is non-vanishing for \emph{any} non-zero field (as is physically obvious, and 
confirmed directly by NRG calculations), and evolves continuously with increasing $h$.

To obtain a physical understanding of these results, remember that $\mimp (h)$ is the `excess' magnetization, i.e.\ is the magnetization of the entire system with the impurity present, minus that with the impurity absent. But since 
the field is purely local, the magnetization in the absence of the impurity vanishes (trivially).
$\mimp (h)$ is thus equivalently the magnetization of the entire system, including the impurity; and 
as such can be separated as
\begin{equation}
\label{eq:mimpsplit}
\mimp (h)~=~ m(h)~+~ M_{CB}(h)
\end{equation}
with $M_{CB}(h)$ the magnetization of the conduction band. As shown above, for all non-zero fields 
$h<h^{\prime}$ (and $U<U_{c}(r)$), $\mimp (h) =0$ while $m(h) \neq 0$.
Hence $M_{CB}(h) = -m(h)$, i.e.\ the field induces a magnetization in the conduction band that is 
equal and opposite to the local impurity magnetization. For fields $h>h^{\prime}$ by contrast, $\mimp (h) =1$, 
and hence $M_{CB}(h)$ changes discontinuously across $h=h^{\prime}$, to $M_{CB}(h) = -m(h)+1$.
But the local magnetization $m(h)$ evolves continuously as $h = h^{\prime}$ is crossed. Hence, on crossing 
the field $h = h^{\prime}$, the resultant increase in magnetization resides \emph{entirely} in the conduction band, 
with no weight whatever on the impurity.

 Analogous reasoning applies to the excess charge, $\nimp (h)$. With the impurity present, the charge of the entire 
system is $n(h)+ N_{CB}(h)$, with $n(h)$ the local impurity charge and  $N_{CB}(h)$ the charge of the conduction band. 
In the absence of the impurity, the charge of the system (now the free conduction band) is denoted $N_{CB}^{0}$, and 
is of course independent of the purely local $h$. Hence $\nimp (h) = n(h)+ N_{CB}(h) -N_{CB}^{0}$. 
But the local charge $n(h)$ evolves continuously as the field $h=h^{\prime}$ is crossed; while, as above, 
$\nimp (h)$ decreases by unity, i.e.\ $\nimp (h^{\prime}+)-\nimp (h^{\prime}-) = -1$.
Hence $\nimp (h^{\prime}+)-\nimp (h^{\prime}-)=N_{CB}(h^{\prime}+)- N_{CB}(h^{\prime}-) = -1$,
i.e.\ the single electron is lost exclusively from the conduction band on crossing $h=h^{\prime}$.

The physical picture is thus clear: on crossing $h = h^{\prime}$ \emph{the single electron
of definite spin that is lost by the entire system has no weight on the impurity}.
The associated discontinuous change in $\mimp (h)$ and $\nimp(h)$ reflects a simple level-crossing 
`transition' whereby the ground state charge of the entire system changes by unity as a single electron 
is lost from the bulk.~\cite{fn23a} As such, it has no bearing on the \emph{local} quantum critical behavior 
associated with the transition between SC and LM phases (and does not correspond to a `field-induced Kondo transition' 
as asserted in Ref.\ \onlinecite{Zitzler2002}). This indeed is obvious since the local magnetization -- 
the order parameter for the zero-field transition -- is always non-vanishing for any finite field.
The situation on crossing $U=U_{c}$ at zero-field, depicted in Fig.\ \ref{fig:fig2}, is just the $h=0$ limit of 
that shown in Fig.\ \ref{fig:fig6}; but with the crucial addition that at $U_{c}$ 
a permanent \emph{local} moment $|\tm| =m(0+)$ is on the verge of condensing on
the impurity (Fig.\ \ref{fig:fig2}) -- which is the characteristic signature of 
the local quantum critical point.


\subsection{Comment on Luttinger self-consistency}
\label{subsection:LWcomment}

One final, general remark about `Luttinger self-consistency' should be made.
The essential  results deduced above (eq.\ \ref{eq:seshnonzero}) and in sec.\ \ref{section:SSElowomegaLW} 
(eq.\ \ref{eq:selfconsesumm}) have of course been obtained from the skeleton expansion, order-by-order. In that 
sense they are perturbative (albeit that, by virtue of the underlying self-consistency, each order in the skeleton 
expansion corresponds to an infinite-order summation in $U$). This need not however be the only sense in which 
Luttinger self-consistency can arise. If for example some infinite-order subset of skeleton digrams for the 
self-energy generates an integrable singularity in an `internal' frequency (one integrated over), which is
not present in an order-by-order expansion, then a  self-consistent solution to the low-$\w$ behavior of the 
self-energy can still exist, but be different from that one would deduce on an order-by-order basis. 
From direct comparison with NRG calculations, we find that eqs.\ (\ref{eq:selfconsesumm},\ref{eq:seshnonzero}) 
indeed appear to be correct everywhere in the $(U,h)$-plane -- indicating the correctness of order-by-order 
self-consistency --  except precisely at zero-field and for $U>U_{c}$ in the LM phase. 
In this case, NRG calculations indicate that the self-energies in fact have the leading $|\w| \rightarrow 0$ 
behavior $\Sigma_{A\sigma}^{I}(\w, h=0) \propto |\w|^{r}$. We do not pursue this further here, although we 
have identified an infinite-order subset of skeleton diagrams that self-consistently generates
such behavior (arising from the interplay between single-particle dynamics and the emergence of an $\w =0$ 
pole in the dynamical local spin susceptibility as $h \rightarrow 0$ in the LM phase).
We would however emphasize that \emph{none} of our previous conclusions are affected by this consideration; 
since in vanishing with the same power as the hybridization, $\Sigma_{A\sigma}^{I}(\w, h=0) \propto |\w|^{r}$ 
indeed satisfies the requirement that it vanishes no less slowly than the hybridization, which was used in 
sec.\ \ref{subsection:SSElowomega} to deduce the low-$\w$ behavior of the conventional zero-field self-energy 
$\Sigma (\w)$ in the LM phase.


\section{Spin susceptibilities}
\label{section:susceptibilities}

 We consider now a range of relevant static spin susceptibilities, each of which probes -- in 
different but distinctive ways -- the underlying transition between SC and LM phases.
Note that the following considerations  are quite general, and not specific e.g.\ to the pseudogap AIM.

\subsection{T=0 local susceptibility as h$\rightarrow$0}
\label{section:localchi}

We turn first to the $T=0$ static local spin susceptibility in response to a local field,
defined for general $h$ by
\begin{equation}
\label{eq:chi1}
\chi_{s}(h) ~=~ \frac{\partial \langle \hat{s}_{z}\rangle}{\partial h}~=~
\frac{1}{2}\frac{\partial m(h)}{\partial h}
\end{equation}
with $\hat{s}_{z} =\tfrac{1}{2}\sum_{\sigma} \sigma d_{\sigma}^{\dagger}d_{\sigma}^{\phantom\dagger}$
the impurity spin $z$-component; and with $\chi_{s}(h)=\chi_{s}(-h)$ since $m(h)=-m(-h)$ is odd in $h$.
Our natural focus is the low-field susceptibility.

Consider first the LM phase, for which $m(h)$ is shown schematically in Fig.\ \ref{fig:fig1} (upper).
From this the low-field behavior of $\chi_{s}(h)$ follows as
\begin{equation}
\label{eq:chi2}
\chi_{s}(h\rightarrow 0) ~=~ |\tm |~ \delta (h)~+~\chi_{s}(h=0\pm),
\end{equation}
where the $\delta$-function piece reflects the existence of the permanent local moment $|\tm|$
characteristic of the LM phase. We omit this (known) contribution from further consideration and focus
on the non-trivial part of the susceptibility, denoted simply by $\chi_{s}$ and given by
\begin{equation}
\label{eq:chi3}
\chi_{s} ~=~
\frac{1}{2}\left(\frac{\partial m(h)}{\partial h}\right)_{h=0+}.
\end{equation}
The SC phase is likewise encompassed by the above, merely by setting $|\tm| =0$.

Our aim here is simply to obtain a general result for the form of $\chi_{s}$ in either phase 
(eq.\ \ref{eq:chipi} below); and in consequence (sec.\ \ref{subsubsection:PB}) a condition for the QPT itself. \\

Both phases can be handled on a common footing in the following (e.g.\ for the LM phase, 
$G_{\sigma}(\w, h) \equiv G_{A\sigma}(\w, |h|)$, 
$\Ss(\w, h) \equiv \Sigma_{A\sigma}(\w, |h|)$, $n_{\sigma}(h) \equiv n_{A\sigma}(|h|)$ and so on).
It also proves helpful to use $T=0$ Matsubara propagators; with 
$G_{\sigma}(i\w, h)$  given by (\emph{cf} eq.\ \ref{eq:Gs})
\begin{equation}
\label{eq:GsX}
G_{\sigma}(i\w, h)~=~ \left[i\w -\epsilon  +\sigma h  -\Gamma(i\w) -\Sigma_{\sigma}(i\w, h) \right]^{-1}
\end{equation}
and the hybridization $\Gamma(i\w) = \sum_{\mathbf{k}}V_{\mathbf{k}}^{2}[i\w -\epsilon_{\mathbf{k}}]^{-1}$ 
independent of the purely local $h$. In terms of this propagator, $m(h)$ is given by 
(\emph{cf} eqs.\ (\ref{eq:mn},\ref{eq:nsig}))
\begin{equation}
\label{eq:nsig2}
m~=~\sum_{\sigma}\sigma\int^{+\infty}_{-\infty}\frac{d\w}{2\pi} ~e^{i\w 0^{+}}~ G_{\sigma}(i\w)
\end{equation}
where for brevity we drop explicit reference to the $h$-dependence from here on; and
where $\partial /\partial h$ in the following implicitly means evaluated at zero field.
From eq.\ \ref{eq:GsX}, 
\begin{equation}
\label{eq:dGdh}
\frac{\partial G_{\sigma}(i\w)}{\partial h} ~=~ -G^{2}_{\sigma}(i\w)\left[\sigma - \frac{\partial\Sigma_{\sigma}(i\w)}{\partial h}\right],
\end{equation}
whence eqs.\ (\ref{eq:chi3},\ref{eq:nsig2}) give (recall $\sigma^{2} =1$)
\begin{equation}
\label{eq:chisone}
\chi_{s}^{\pd}=-\tfrac{1}{2}\sum_{\sigma}\int^{+\infty}_{-\infty} \frac{d\w}{2\pi}~G_{\sigma}^{2}(i\w) \left[1~-~\sigma \frac{\partial \Sigma_{\sigma}(i\w)}{\partial h} \right]
\end{equation}
(where the $e^{i\w 0^{+}}$ convergence factor can be dropped).

 The essential trick is now to separate the skeleton expansion for the self-energy as
\begin{equation}
\label{eq:sedecomp}
\Sigma_{\sigma}(i\w)=\Sigma_{\sigma}^{0}+\tilde{\Sigma}_{\sigma}(i\w)
= \tfrac{1}{2} U [ n - \sigma m]+\tilde{\Sigma}_{\sigma}(i\w)
\end{equation}
into the first order, purely static ($\w$-independent) contribution, $\Sigma_{\sigma}^{0} = Un_{-\sigma}$,
and the remainder $\tilde{\Sigma}_{\sigma}(i\w)$ arising from all `dynamical' skeleton diagrams.
From eq.\ \ref{eq:mn}, $n_{\sigma}=\tfrac{1}{2}[n+\sigma m]$ in terms of the local charge
and magnetization, whence $\Ss^{0}= \tfrac{1}{2}U[n-\sigma m]$. Consistent with $n(h)=n(-h)$ 
(sec.\ \ref{section:LMgeneral}), we consider the case where
$(\partial n/\partial h)_{h=0}=0$
whence eqs.\ (\ref{eq:chi3},\ref{eq:sedecomp}) give
\begin{equation}
\label{eq:dSdhone}
\sigma\frac{\partial\Sigma_{\sigma}(i\w)}{\partial h}~=~ 
-U\chi_{s} +\sigma \frac{\partial\tilde{\Sigma}_{\sigma}(i\w)}{\partial h} .
\end{equation}
Eq.\ \ref{eq:chisone} thus yields:
\begin{equation}
\label{eq:chistwo}
\frac{\chi_{s}^{\pd}}{1+U\chi_{s}^{\pd}}=-\tfrac{1}{2}\sum_{\sigma}\int \frac{d\w}{2\pi}~G_{\sigma}^{2}(i\w) \left[1-\frac{1}{1+U\chi_{s}^{\pd}}\sigma \frac{\partial \tilde{\Sigma}_{\sigma}(i\w)}{\partial h} \right]
\end{equation} 

Now focus on $\partial \tilde{\Sigma}_{\sigma}(i\w)/\partial h$. Since 
$\tilde{\Sigma}_{\sigma}$ is a  functional of the propagators,
\begin{equation}
\label{eq:dSdh}
\frac{\partial \tilde{\Sigma}_{\sigma}(i\w)}{\partial h}=\sum_{\sigma^{\prime}} \int \frac{d\w^{\prime}}{2\pi}~ \frac{\delta \tilde{\Sigma}_{\sigma}(i\w)}{\delta G_{\sigma^{\prime}}(i\w^{\prime})}
~\frac{\partial G_{\sigma^{\prime}}(i\w^{\prime})}{\partial h}.
\end{equation}
We define an irreducible two-particle vertex function as a functional derivative of $\tilde{\Sigma}_{\sigma}(i\w)$
(which itself excludes the static first-order contribution), \emph{viz}
\begin{equation}
\label{eq:Iprime}
\tilde{I}_{\sigma\sigma^{\prime}}(i\w,i\w^{\prime})~=~
\frac{\delta \tilde{\Sigma}_{\sigma}(i\w)}{\delta G_{\sigma^{\prime}}(i\w^{\prime})};
\end{equation}
and which satisfies
$\tilde{I}_{\sigma\sigma^{\prime}}(i\w,i\w^{\prime})=\tilde{I}_{\sigma^{\prime}\sigma}(i\w^{\prime},i\w)$
by virtue of the fact that 
$\tilde{\Sigma}_{\sigma}(i\w) =\delta \tilde{\Phi}_{LW}/\delta G_{\sigma}(i\w)$,
where $\tilde{\Phi}_{LW}$ is the usual Luttinger-Ward functional excluding its first-order contribution.
Using eqs.\ (\ref{eq:dGdh},\ref{eq:dSdhone}), eq.\ref{eq:dSdh} gives
\begin{widetext}
\begin{equation}
\label{eq:noddy1again}
\frac{-1}{1+U\chi_{s}^{\pd}}~\sigma \frac{\partial \tilde{\Sigma}_{\sigma} (i\w)}{\partial h}~=
~\sum_{\sigma^{\prime}}\int \frac{d\w^{\prime}}{2\pi}~\tilde{I}_{\sigma\sigma^{\prime}}(i\w,i\w^{\prime})~G_{\sigma^{\prime}}^{2}(i\w^{\prime})~\sigma\sigma^{\prime}~\left[1~-~\frac{1}{1+U\chi_{s}^{\pd}}~\sigma^{\prime}\frac{\partial \tilde{\Sigma}_{\sigma^{\prime}} (i\w^{\prime})}{\partial h} \right],
\end{equation}
and hence by iteration
\begin{equation}
\label{eq:bigears}
\frac{-1}{1+U\chi_{s}^{\pd}}~\sigma \frac{\partial \tilde{\Sigma}_{\sigma} (i\w)}{\partial h}~=~\sum_{\sigma^{\prime}}~\sigma\sigma^{\prime}\int \frac{d\w^{\prime}}{2\pi}~\tilde{\Gamma}_{\sigma\sigma^{\prime}}(i\w,i\w^{\prime})~G_{\sigma^{\prime}}^{2}(i\w^{\prime})
\end{equation}
with $\tilde{\Gamma}_{\sigma\sigma^{\prime}}(i\w,i\w^{\prime})$ the reducible vertex defined from 
$\tilde{I}_{\sigma\sigma^{\prime}}(i\w,i\w^{\prime})$ via the Bethe-Salpeter equation
\begin{equation}
\label{eq:BS}
\tilde{\Gamma}_{\sigma\sigma^{\prime}}(i\w,i\w^{\prime})~=~\tilde{I}_{\sigma\sigma^{\prime}}(i\w,i\w^{\prime})+~~
\sum_{\sigma^{\prime\prime}} \int \frac{d\w^{\prime\prime}}{2\pi}~\tilde{I}_{\sigma\sigma^{\prime\prime}}(i\w,i\w^{\prime\prime})~G^{2}_{\sigma^{\prime\prime}}(i\w^{\prime\prime})~\tilde{\Gamma}_{\sigma^{\prime\prime}\sigma^{\prime}}(i\w^{\prime\prime},i\w^{\prime}).
\end{equation}
Eq.\ref{eq:chistwo} thus gives
\begin{equation}
\label{eq:burble}
\frac{\chi_{s}^{\pd}}{1+U\chi_{s}^{\pd}} ~=~-\tfrac{1}{2}\sum_{\sigma}\int^{+\infty}_{-\infty} \frac{d\w}{2\pi}~G_{\sigma}^{2}(i\w) \left[1~+~\sum_{\sigma^{\prime}}\sigma\sigma^{\prime}\int \frac{d\w^{\prime}}{2\pi}~\tilde{\Gamma}_{\sigma\sigma^{\prime}}(i\w,i\w^{\prime})G_{\sigma^{\prime}}^{2}(i\w^{\prime})\right]   ~~~
\equiv ~ ^{0}\tilde{\Pi}
\end{equation}
\end{widetext}
with $^{0}\tilde{\Pi}={^{0}\tilde{\Pi}}(U)$ thereby defined, and hence:
\begin{equation}
\label{eq:chipi}
\chi^{\pd}_{s}~=~\frac{^{0}\tilde{\Pi}(U)}{1~-~{U}~^{0}\tilde{\Pi}(U)}
\end{equation}
This is the essential result for the static spin susceptibility, and is exact. Strikingly, moreover, 
note that it has just the `RPA-like' form that would arise at the approximate level 
of `mean-field plus fluctuations' (wherein the vertex $\tilde{\Gamma}_{\sigma \sigma^{\prime}} \equiv 0$ in 
eq.\ \ref{eq:burble}, and the propagators are non-interacting rather than fully self-consistent).

The key to obtaining eq.\ \ref{eq:chipi} has been decomposition of the self-energy
as in eq.\ \ref{eq:sedecomp},~\cite{fnlocsusc} and recognition that the exact static contribution
$\Ss^{0}$ contains the magnetization $m$ again. This is physically natural, since the symmetry that is broken in 
the zero-field LM phase ($m(h=0+) =|\tm| \neq 0$) is explicitly contained in the static contribution to the self-energy.
That general strategy is not moreover particular to a single-level AIM, since for typical impurity problems one expects
the symmetry broken in the NFL phase to be directly apparent in the static $\Ss^{0}$.


\subsubsection{Condition for quantum phase transition}
\label{subsubsection:PB}

To obtain a condition for the critical $U_{c}$ at which the quantum phase
transition occurs (eq.\ \ref{eq:cond7} below), we consider specifically the
zero-field LM phase (whence $G_{\sigma}(i\w)\equiv G_{A\sigma}(i\w, |h|=0)$ in the following); and 
the approach to the transition $U\rightarrow U_{c}+$, where the
local moment $|\tm| = m(h=0+)$ vanishes (continuously by assumption).
It is convenient to define
\begin{equation}
\label{eq:y}
y~=~\tfrac{1}{2}U|\tm|,
\end{equation}
which likewise vanishes at $U=U_{c}$. Separating the self-energy as in eq.\ \ref{eq:sedecomp}, the static part may be
written as $\Sigma_{\sigma}^{0} =\tfrac{1}{2}Un -\sigma y$, and hence the propagator
\begin{equation}
\label{eq:cond1}
G_{\sigma}(i\w)=\left[i\w -\epsilon -(\tfrac{U}{2}n -\sigma y) -\Gamma(i\w) -\tilde{\Sigma}_{\sigma}(i\w)
\right]^{-1}.
\end{equation}
Eq.\ \ref{eq:nsig2} for ($m(0+) =$) $|\tm| = 2y/U$ then gives
\begin{equation}
\label{eq:cond2}
\frac{2}{U}y= \sum_{\sigma} \sigma \int^{+\infty}_{-\infty}\frac{d\w}{2\pi}~ e^{i\w 0^{+}}~G_{\sigma}(i\w)
~~~ \equiv f(y;U)
\end{equation}
where $f(y;U)$ is thus defined. $f(y;U)$  is clearly odd in $y$ (since from eq.\ \ref{eq:cond1},
$G_{\sigma}(i\w, y) = G_{-\sigma}(i\w, -y)$ with the $y$-dependence temporarily explicit).
On the natural assumption that $f(y;U)$ vanishes linearly in $y$ as
$U\rightarrow U_{c}+$, the condition for the transition is thus
$\tfrac{2}{U_{c}} =(\partial f(y;U_{c})/\partial y)_{y=0}$, i.e.\
\begin{equation}
\label{eq:cond3}
1~=~\tfrac{1}{2}U_{c} \left( \frac{\partial f(y;U_{c})}{\partial y}\right)_{y=0}.
\end{equation}
Since $G_{\sigma}(i\w, y) = G_{-\sigma}(i\w, -y)$ as above, the charge $n \equiv n(y)$
satisfies $n(y)=n(-y)$; consistent with which we consider the case where $(\partial n/\partial y)_{y=0} =0$.
Hence from eqs.\ (\ref{eq:cond2},\ref{eq:cond1}),
\begin{equation}
\label{eq:cond4}
\frac{\partial f(y;U_{c})}{\partial y}= -\sum_{\sigma}\int \frac{d\w}{2\pi} ~\Gs^{2}(i\w)\left[
1-\sigma \frac{\partial\tilde{\Sigma}_{\sigma}(i\w)}{\partial y} \right]
\end{equation}
(where $\partial/\partial y$ from here on implicitly means evaluated at $y=0$, and hence $U=U_{c}$).

 Since $\tilde{\Sigma}_{\sigma}(i\w)$ is a functional of the $\{G_{\sigma}(i\w)\}$, 
\begin{equation}
\label{eq:cond5}
\frac{\partial \tilde{\Sigma}_{\sigma}(i\w)}{\partial y}=\sum_{\sigma^{\prime}} \int \frac{d\w^{\prime}}{2\pi}~ \frac{\delta \tilde{\Sigma}_{\sigma}(i\w)}{\delta G_{\sigma^{\prime}}(i\w^{\prime})}
~\frac{\partial G_{\sigma^{\prime}}(i\w^{\prime})}{\partial y}
\nonumber
\end{equation}
(\emph{cf} eq.\ \ref{eq:dSdh}). Proceeding in direct parallel to eqs.\ \ref{eq:dSdh} \emph{ff} then leads to
\begin{equation}
\label{eq:cond6}
-\sigma \frac{\partial\tilde{\Sigma}_{\sigma}(i\w)}{\partial y} =
\sum_{\sigma^{\prime}}\sigma\sigma^{\prime}
\int \frac{d\w^{\prime}}{2\pi}
\tilde{\Gamma}_{\sigma\sigma^{\prime}}(i\w,i\w^{\prime})G_{\sigma^{\prime}}^{2}(i\w^{\prime})
\end{equation}
(\emph{cf} eq.\ \ref{eq:bigears}), with $\tilde{\Gamma}_{\sigma\sigma^{\prime}}(i\w,i\w^{\prime})$
the reducible vertex given by the Bethe-Salpeter eq.\ \ref{eq:BS}.
From eqs.\ (\ref{eq:cond6},\ref{eq:cond4}), on comparison to eq.\ \ref{eq:burble} defining
$^{0}\tilde{\Pi}$, the condition eq.\ \ref{eq:cond3} for the transition is thus:
\begin{equation}
\label{eq:cond7}
1~=~U_{c}~  ^{0}\tilde{\Pi}(U_{c})
\end{equation}

From eq.\ \ref{eq:chipi} the local static spin susceptibility $\chi_{s}$ thus diverges as $U=U_{c}$ is 
approached and the local moment $|\tm|$ vanishes (as expected physically, since $\chi_{s}$ is the susceptibility 
corresponding to the local magnetization). More generally, since $\chi_{s} >0$ for all $U$, eq.\ \ref{eq:chipi} implies 
\begin{equation}
\label{eq:cond8}
0 ~\leq ~U {^{0}\tilde{\Pi}}(U) ~\leq ~ 1
\end{equation}
for all $U$, where the upper limit corresponds to the transition at $U_{c}$; and with 
$\chi_{s}$ otherwise finite throughout both the LM and SC phases.

For the pseudogap AIM, we have numerically confirmed these results via NRG. 
$\chi_{s}$ (eqs.\ (\ref{eq:chi3},\ref{eq:chipi})) is indeed found to diverge as
the QCP is approached -- from either phase -- but remains finite throughout
the SC and LM phases. This is in fact consistent with the NRG calculations of
[\onlinecite{Pixley2012}], who find a divergent $\chi_{s}$ as the transition is approached 
from the SC phase, but a divergent spin susceptibility throughout the LM phase. The latter 
reflects the `trivial' $|\tm|\delta(h)$ contribution to $\chi_{s}(h)$ (eq.\ \ref{eq:chi2});
while $\chi_{s} \equiv \chi_{s}(h=0+)$ (eq.\ \ref{eq:chi3}) is itself finite throughout the phase, 
diverging only as $U\rightarrow U_{c}+$ (and with the same exponent as the approach from the SC side).


\subsection{h=0 local susceptibility as T$\rightarrow$0}
\label{subsection:localsuscfiniteT}

The transition between SC and LM phases is of course pristine only for $T=0=h$. The local 
$\chi_{s}$ considered above corresponds first to taking the limit $T=0$, and then considering $h\rightarrow 0$. 
Equally, one can consider the reverse order: first $h=0$, then $T\rightarrow 0$. The order of limits do not 
commute in a degenerate LM phase, and we denote the latter susceptibility by 
$\tilde{\chi}_{s}(T)$, \emph{viz}
\begin{equation}
\label{eq:chitil1}
\tilde{\chi}_{s}(T)~=~\left(
\frac{\partial\langle\hat{s}_{z}\rangle}{\partial h}
\right)_{h=0}.
\end{equation}
Since $H=H^{0} -2h\hat{s}_{z}$ (with $H^{0}$ the zero-field Hamiltonian), but $[H^{0},\hat{s}_{z}]\neq 0$, 
it follows that
$\tilde{\chi}_{s}(T)$ $=2\int_{0}^{\frac{1}{T}}d\tau\langle\hat{s}_{z}(\tau)\hat{s}_{z}\rangle$ 
(with $k_{B}\equiv 1$). Direct analysis of the Lehmann representation of $\tilde{\chi}_{s}(T)$ then gives 
the leading $T\to 0$ Curie-like behavior
\begin{equation}
\label{eq:chitil2}
\tilde{\chi}_{s}(T)~\overset{T\rightarrow 0}{\sim}~ \frac{2}{T} \times \frac{1}{g_{0}}\sum_{\alpha,\beta}|\langle\alpha|\hat{s}_{z}|\beta\rangle|^{2}
\end{equation}
with $\{|\alpha\rangle\}$ the $g_{0}$-fold degenerate ground states. 

We consider specifically $\tilde{\chi}_{s}(T)$ in the LM phase, where the essential feature of the $T=0$ ground 
state is that it is globally $g_{0}=2$-fold spin-degenerate, with eigenstates denoted (for obvious reasons) by 
$|A\rangle$ and $|B\rangle$ which are eigenfunctions of the \emph{total} spin $z$-component 
$\hat{S}_{z}$ ($ = \hat{s}_{z}+\tfrac{1}{2}\sum_{\mathbf{k},\sigma}\sigma \hat{n}_{\mathbf{k}\sigma}^{\pd}$); \emph{viz}
$|\alpha\rangle =|A\rangle \equiv |S_{z}=+\tfrac{1}{2}\rangle$ and
$|B\rangle \equiv |S_{z}=-\tfrac{1}{2}\rangle$, such that $2\langle A|\hat{s}_{z}|A\rangle = +|\tm|$ and
$2\langle B|\hat{s}_{z}|B\rangle = -|\tm|$ with $|\tm|$ the usual permanent local moment.
From eq.\ \ref{eq:chitil2}, 
$T\tilde{\chi}(T)\sim \sum_{\alpha,\beta \in \{ A,B\}}|\langle\alpha|\hat{s}_{z}|\beta\rangle|^{2}$.
But since $\hat{S}_{z}$ and $\hat{s}_{z}$ commute, 
$\langle A| \hat{S}_{z}\hat{s}_{z}|B\rangle =\langle A| \hat{s}_{z}\hat{S}_{z}|B\rangle$
gives $\langle A|\hat{s}_{z}|B\rangle = - \langle A|\hat{s}_{z}|B\rangle =0$, and hence
\begin{equation}
\label{eq:chitil4}
\underset{T\rightarrow 0}{\mathrm{lim}} ~ T\tilde{\chi}_{s}(T) ~=~ \tfrac{1}{2}|\tm|^{2}
\end{equation}
The SC phase by contrast has a non-degenerate ground state ($|\alpha\rangle \equiv|0\rangle$), whence
the right side of eq.\ \ref{eq:chitil2} vanishes, since the impurity spin is quenched by 
the Kondo effect ($\langle 0|\hat{s}_{z}|0\rangle =0$). In this case $\tilde{\chi}_{s}(T\rightarrow 0)$ is
naturally finite, so $\mathrm{lim}_{T\to 0}~T\tilde{\chi}(T)$ vanishes. From eq.\ \ref{eq:chitil4},
$\mathrm{lim}_{T\to 0}~T\tilde{\chi}(T)$ is thus finite in the LM phase, but vanishes as the transition 
$U\to U_{c}+$ is approached; with an exponent of twice that for the vanishing of the local moment $|\tm|$
(as can also be shown from hyperscaling arguments based on a scaling 
ansatz~\cite{Ingersent-Si2002,Kircan_Vojta2003,Fritz_Vojta2004}). For the pseudogap AIM we have numerically 
confirmed eq.\ \ref{eq:chitil4} in full from NRG, and the same behavior is also found from NRG calculations 
for the Bose-Fermi Kondo model.~\cite{MTGKIBFKPRB2007}


\subsection{Uniform field susceptibilities}
\label{subsection:impsusceps}

The susceptibilities considered above refer to a field $h$ applied locally to the impurity. We now consider 
the spin susceptibilities in response to a globally applied uniform field, $h\equiv h_{u}$, with a natural
focus on the LM phase. 

The \emph{local} susceptibility in response to the global field, here denoted $\tilde{\chi}_{u,\mathrm{loc}}(T)$, is 
given (\emph{cf} eq.\ \ref{eq:chitil1}) by
\begin{equation}
\label{eq:chitil5}
\tilde{\chi}_{u,\mathrm{loc}}(T)~=~\left(
\frac{\partial\langle\hat{s}_{z}\rangle}{\partial h}
\right)_{h=0} ~~~~~~:~ h=h_{u}.
\end{equation}
In this case $H=H^{0} -2h\hat{S}_{z}$ (with $\hat{S}_{z}$ the total spin $z$-component), and
since $[H^{0},\hat{S}_{z}]=0$ it follows trivially that
\begin{equation}
\label{eq:chitil6}
\tilde{\chi}_{u,\mathrm{loc}}(T)~\overset{T\rightarrow 0}{\sim}~ \frac{2}{T} \times \frac{1}{2} \sum_{\alpha \in \{A,B\}}\langle\alpha|\hat{s}_{z}\hat{S}_{z}|\alpha\rangle .
\end{equation}
Hence (using sec.\ \ref{subsection:localsuscfiniteT}),
\begin{equation}
\label{eq:chitil7}
\underset{T\rightarrow 0}{\mathrm{lim}} ~ T\tilde{\chi}_{u,\mathrm{loc}}(T) ~=~ \tfrac{1}{2}|\tm|
\end{equation}
exhibits characteristic Curie-like behavior in the LM phase; but $\propto |\tm|$, and hence vanishing as 
$U\rightarrow U_{c}+$ with the same exponent as the local moment (in contrast to the $\propto|\tm|^{2}$ 
behavior for the local susceptibility eq.\ \ref{eq:chitil4} in response to a local field).
For the pseudogap AIM, we have again confirmed this result numerically by NRG.

The global uniform spin susceptibility is correspondingly given by
(\emph{cf} eq.\ \ref{eq:chitil5})
\begin{equation}
\label{eq:chitil8}
\tilde{\chi}_{u}(T)~=~\left(
\frac{\partial\langle\hat{S}_{z}\rangle}{\partial h}
\right)_{h=0} ~~~~~~:~ h=h_{u}.
\end{equation}
Hence 
$T\tilde{\chi}_{u,\mathrm{loc}}(T)\overset{T\rightarrow 0}{\sim} \sum_{\alpha \in \{A,B\}}\langle\alpha|\hat{S}_{z}^{2}|\alpha\rangle$,
and thus
\begin{equation}
\label{eq:chitil9}
\underset{T\rightarrow 0}{\mathrm{lim}} ~ T\tilde{\chi}_{\mathrm{imp}}(T)~\equiv~
\underset{T\rightarrow 0}{\mathrm{lim}} ~ T\tilde{\chi}_{u}(T) ~=~ 
\tfrac{1}{2}
\end{equation}
exhibits `full' free spin-$\tfrac{1}{2}$ Curie behavior.~\cite{fn32} 
As indicated, eq.\ \ref{eq:chitil9} applies also to the `excess' susceptibility
$\tilde{\chi}_{\mathrm{imp}} =\tilde{\chi}_{u}-\tilde{\chi}_{u}^{0}$ (where $\tilde{\chi}_{u}^{0}$ 
refers to the absence of the impurity), on recognizing that $\tilde{\chi}_{u}^{0}(T=0)$ is simply a constant.
Since $\tilde{\chi}_{u}(0)$ is finite in the SC phase, note then that in this case
$\mathrm{lim}_{T\to 0} ~T\tilde{\chi}_{u}(T) = \mathrm{lim}_{T\to 0} ~T\tilde{\chi}_{\mathrm{imp}}(T)$
vanishes abruptly as the LM$\rightarrow$SC transition is crossed.\\

Finally, we reiterate that the essential results above for the Curie-like form of
$\tilde{\chi}_{s}$, $\tilde{\chi}_{u,\mathrm{loc}}$ and $\tilde{\chi}_{u}$
or $\tilde{\chi}_{\mathrm{imp}}$ -- eqs.\ (\ref{eq:chitil4},\ref{eq:chitil7},\ref{eq:chitil9}) --
all reflect and arise from the global degeneracy of the zero-field LM ground state, embodied in 
the states $|A\rangle$, $|B\rangle$ (sec.\ \ref{subsection:localsuscfiniteT}). 
Precisely at zero-field,
$\langle \hat{S}_{z}\rangle_{T=0}= \tfrac{1}{2}(\langle A|\hat{S}_{z}|A\rangle +
\langle B|\hat{S}_{z}|B\rangle) =0$ naturally vanishes. 
For any non-zero field the strict degeneracy is of course lifted; whence e.g.\ for $h=0+$, 
$\langle \hat{S}_{z}\rangle_{T=0}=\langle A|\hat{S}_{z}|A\rangle = \tfrac{1}{2}$
(and $\langle \hat{S}_{z}\rangle_{T=0}=\langle B|\hat{S}_{z}|B\rangle = -\tfrac{1}{2}$ for $h=0-$), 
or equivalently $\mimp (h=0\pm) = \pm 1$ (as the impurity-free contribution to $\mimp$,
sec.\ \ref{subsection:excessprops}, vanishes as $h\rightarrow 0$).
$\mimp(0\pm)=\pm 1$ is thus quite generally characteristic of a LM phase and, together 
e.g.\ with the full free spin-$\tfrac{1}{2}$ Curie behavior of $\tilde{\chi}_{\mathrm{imp}}$,
is equally symptomatic of the global spin-degeneracy of the LM ground state.


\section{Multilevel impurities}
\label{section:Luttintgen}

While our principal focus has been on single-level quantum impurity models, LM phases in fact abound in
multilevel problems (sec.\ \ref{sec:intro}). As a relevant exemplar, we touch briefly on rich and much 
studied~\cite{fnreftwo-levelgen,CJW2009,CJWfield} Anderson-like models in which a two-level impurity 
(or quantum dot) is coupled to metallic leads in 1-channel fashion. Our aim here is simply to derive and 
understand on general grounds some key results hitherto inferred numerically.~\cite{CJW2009}

The impurity has one-electron energies $\epsilon_{i}$ ($i=1,2$), with tunnel couplings $V_{i}$ to 
conduction band states. Local correlations enter via an on-site Coulomb repulsion (charging energies) for 
each level, an inter-level Coulomb repulsion, and an inter-orbital ferromagnetic coupling 
$-J_{H}\hat{\mathbf{s}}_{1}\cdot\hat{\mathbf{s}}_{2}$ (with $\hat{\mathbf{s}}_{i}$ the local spin operator). 
In accordance with Hund's first rule, the latter acts to favor a local triplet state in the two-electron sector of 
the free impurity. It is this which is ultimately responsible~\cite{fnreftwo-levelgen,CJW2009,CJWfield}
for the two distinct phases arising in the $(\epsilon_{1},\epsilon_{2})$-plane on coupling the impurity/dot 
to the conduction band, characterized by distinct FPs and separated generically by a closed
line of Kosterlitz-Thouless quantum phase transitions;~\cite{CJW2009}  \emph{viz} a SC
phase and an underscreened (USC) spin-1 phase.~\cite{nozblan} The former is a Fermi liquid, perturbatively 
connected to the non-interacting limit; while the USC ground state is a degenerate LM phase with the impurity 
spin only partially quenched (and despite the `spin' language, note that the USC/LM phase is not confined to
integral impurity valence, but encompasses generally mixed-valent behavior).
Full details may be found e.g.\ in [\onlinecite{CJW2009}].

For a two-level impurity, the local propagators, single-particle spectra, self-energies and one-electron hybridization,
are of course $2\times 2$ matrices: $G_{ij;\sigma}(\w)$,  
$D_{ij;\sigma}(\w)$ $(=-\tfrac{1}{\pi}\mathrm{Im}G_{ij;\sigma}(\w))$,
$\Sigma_{ij;\sigma}(\w)$ and $\Gamma_{ij}(\w)$ 
(such that $\Gamma^{I}_{ij}(\w)=\sum_{\mathbf{k}}V_{i}V_{j}\delta (\w -\epsilon_{\mathbf{k}})$, with non-vanishing  
$\Gamma^{I}_{ij}(\w =0)\equiv \Gamma_{ij}$ reflecting the metallic nature of the conduction band).
We focus on the central single-particle spectrum which determines the zero-bias conductance $G_{c}(T=0)$
across the dot. Here denoted by $D_{\sigma}(\w)$, it is given by~\cite{CJW2009}
$D_{\sigma}(\w)=\tfrac{1}{\Gamma_{11}+\Gamma_{22}}\sum_{i,j} \Gamma_{ij} D_{ij;\sigma}(\w)$,
such that $G_{c}(0) = \tfrac{2e^{2}}{h}\pi [\Gamma_{11}+\Gamma_{22}]D_{\sigma}(0)$.
For obvious reasons we consider explicitly the zero-field case (and suppress notational reference to 
it in the following), although finite-field is also easily handled.


\subsection{SC phase}
\label{subsection:2LDSC}

 Consider first the SC, Fermi liquid phase. Here, as shown in [\onlinecite{CJW2009}], 
\begin{subequations}
\label{eq:2LD1}
\begin{align}
\pi [\Gamma_{11}+\Gamma_{22}] D_{\sigma}(0)~=&~\mathrm{sin}^{2}(\delta_{\sigma})
\label{eq:2LD1a}
\\
=&~\mathrm{sin}^{2}(\pi \nimps+I_{L_{\sigma}})
\label{eq:2LD1b}
\end{align} 
\end{subequations}
where (\emph{cf} eq.\ \ref{eq:pssc}) the phase shift $\delta_{\sigma}$ is defined by
$\delta_{\sigma} = \mathrm{arg}[\mathrm{det}\mathbf{G}_{\sigma}(\w)]\big{|}_{\w =-\infty}^{\w = 0}$
(with $\mathbf{G}_{\sigma}$ the $2\times 2$ propagator matrix), and
in turn satisfies a Friedel-Luttinger sum rule (eq.\ \ref{eq:flsrsc}),
$\delta_{\sigma} = \pi \nimps+I_{L_{\sigma}}$ ($\equiv \tfrac{\pi}{2} \nimp+I_{L_{\sigma}}$ 
since $\nimps$ is independent of spin at zero field). The Luttinger integral 
in this case is given by
\begin{equation}
\label{eq:2LD2}
I_{L_{\sigma}}~=~ \mathrm{Im~Tr}\int_{-\infty}^{0}d\w ~\frac{\partial\mathbf{\Sigma}_{\sigma}(\w)}{\partial\w}
\mathbf{G}_{\sigma}(\w)
\end{equation}
and is an obvious matrix generalization of eq.\ \ref{eq:LISC}. But $I_{L_{\sigma}}=0$ in the SC phase,
by just the same argument given in sec.\ \ref{subsection:Luttintgen} (noting that the $t$-ordered
$\Sigma_{ij;\sigma}^{t}(\w)=\delta \Phi_{LW}/\delta G_{ji;\sigma}^{t}(\w)$, and $\Sigma_{ij;\sigma}^{I}(0) =0$).
Hence from eq.\ \ref{eq:2LD1b},~\cite{CJW2009}
\begin{equation}
\label{eq:2LD3}
\frac{G_{c}(0)}{(2e^{2}/h)} ~=~ \pi [\Gamma_{11}+\Gamma_{22}] D_{\sigma}(0)
~=~ \mathrm{sin}^{2}(\tfrac{\pi}{2}\nimp).
\end{equation}


\subsection{LM (USC) phase}
\label{subsection:2LDLM}

Now consider the LM/USC phase. As in secs.\ \ref{section:LMgeneral}-\ref{section:sumrules}, 
this is handled simply by focusing on the A-type state, where the
local moment $|\tm| = m_{A}(0)$ ($=\sum_{i=1,2}[n_{iA\uparrow}(0) - n_{iA\downarrow}(0)]$,
\emph{cf} eq.\ \ref{eq:noddy1}) is non-vanishing. In precise parallel to the SC phase it 
follows that 
\begin{subequations}
\label{eq:2LD4}
\begin{align}
\pi [\Gamma_{11}+\Gamma_{22}] D_{A\sigma}(0)~=&~\mathrm{sin}^{2}(\delta_{A\sigma})
\label{eq:2LD4a}
\\
=&~\mathrm{sin}^{2}(\pi \nimpas+I_{L_{A\sigma}}),
\label{eq:2LD4b}
\end{align} 
\end{subequations}
where the phase shift 
$\delta_{A\sigma} = \mathrm{arg}[\mathrm{det}\mathbf{G}_{A\sigma}(\w)]\big{|}_{\w =-\infty}^{\w = 0}$
likewise satsifies a Friedel-Luttinger sum rule, with Luttinger integral
$I_{L_{A\sigma}}= \mathrm{Im~Tr}\int_{-\infty}^{0}d\w ~(\partial\mathbf{\Sigma}_{A\sigma}(\w)/\partial\w)
\mathbf{G}_{A\sigma}(\w)$ (and self-energy matrix $\mathbf{\Sigma}_{A\sigma}(\w)$).
But once again, $I_{L_{A\sigma}}=0$, by the argument given in sec.\ \ref{subsection:Luttintgen};
reflecting the fact that the Luttinger-Ward functional in the LM phase is the same functional of
the $\{\mathbf{G}_{A\sigma}^{t}\}$ that it is of $\{\mathbf{G}_{\sigma}^{t}\}$ in the SC phase.
Writing $\nimpas =\tfrac{1}{2}(\nimp + \sigma \mimpa )$ 
with ($\mimpa \equiv $) $\mimpa (0) = \mimp(0+)$ (eq.\ \ref{eq:mimpLM}),
and recalling that $\mimp(0+) =1$ for a LM phase (sec.\ \ref{section:susceptibilities}), 
eq.\ \ref{eq:2LD4b} gives
\begin{equation}
\label{eq:2LD5}
\pi [\Gamma_{11}+\Gamma_{22}] D_{A\sigma}(0)~=~\mathrm{sin}^{2}(\tfrac{\pi}{2}[\nimp +\sigma]).
\end{equation}
But for zero field the ($\sigma$-independent) propagators $\mathbf{G}_{\sigma}(\w)$
are of course given by $\mathbf{G}_{\sigma}(\w)=\tfrac{1}{2}\sum_{\sigma} \mathbf{G}_{A\sigma}(\w)$, 
and hence $D_{\sigma}(\w)=\tfrac{1}{2}\sum_{\sigma} D_{A\sigma}(\w)$ (eq.\ \ref{eq:lincolncity}).
Eq.\ \ref{eq:2LD5} thus yields
\begin{equation}
\label{eq:2LD6}
\frac{G_{c}(0)}{(2e^{2}/h)}~=~\pi [\Gamma_{11}+\Gamma_{22}] D_{\sigma}(0)
~=~\mathrm{cos}^{2}(\tfrac{\pi}{2}\nimp).
\end{equation}
Eq.\ \ref{eq:2LD6} is the essential result for the conductance in the LM/USC phase, 
previously deduced numerically using NRG,~\cite{CJW2009} but here shown to arise as a consequence 
of a Luttinger theorem in terms of a two-self-energy description (i.e.\ $I_{L_{A\sigma}}=0$).
Since $\nimp$ varies continuously~\cite{CJW2009} on crossing the line of Kosterlitz-Thouless transitions
from the SC to the LM/USC phase, eqs.\ (\ref{eq:2LD3},\ref{eq:2LD6}) show that the zero-bias conductance jumps 
discontinuously on crossing the transition (although the Kondo scale itself vanishes continuously as the 
transition is approached from the SC phase); as detailed further in [\onlinecite{CJW2009}].

As discussed throughout, the zero-field LM phase can equally -- and more traditionally -- be described in 
terms of the conventional single self-energy $\mathbf{\Sigma}(\w)$, defined by the Dyson equation
$[\mathbf{G}_{\sigma}(\w)]^{-1} =[\mathbf{G}_{0}(\w)]^{-1} -\mathbf{\Sigma}(\w)$
with $\mathbf{G}_{0}(\w)$ the non-interacting propagator matrix. Accordingly (in parallel to 
sec.\ \ref{subsection:Luttinth=0}), one can repeat the same calculation that led to eq.\ \ref{eq:2LD1}, 
but now in terms of the single self-energy; leading rather obviously to
\begin{equation}
\label{eq:2LD7}
\pi [\Gamma_{11}+\Gamma_{22}] D_{\sigma}(0)~=~\mathrm{sin}^{2}(\tfrac{\pi}{2} \nimp+I_{L})
\end{equation}
but with a Luttinger integral now given (\emph{cf} eq.\ \ref{eq:LIsse}) by
\begin{equation}
\label{eq:2LD8}
I_{L}~=~ \mathrm{Im~Tr}\int_{-\infty}^{0}d\w ~\frac{\partial\mathbf{\Sigma}(\w)}{\partial\w}
\mathbf{G}_{\sigma}(\w)
\end{equation}
in terms of the single self-energy $\mathbf{\Sigma}(\w)$. This Luttinger integral cannot of course be argued to 
vanish, \emph{cf} secs.\ \ref{subsection:Luttinth=0}, \ref{subsection:ILprime}. But its magnitude 
follows directly from the equivalence of eqs.\ (\ref{eq:2LD6},\ref{eq:2LD7}), \emph{viz}
\begin{equation}
\label{eq:2LD9}
|I_{L}|~=~\frac{\pi}{2}
\end{equation}
~\\
(as indeed confirmed numerically~\cite{CJW2009} by NRG calculations of the $\w$-integral in 
eq.\ \ref{eq:2LD8}). As for the pseudogap and gapped Anderson models (sec.\ \ref{subsection:ILprime}), 
and the simple atomic limit (sec.\ \ref{section:atomlimh=0}), $|I_{L}|$ throughout the LM
phase thus has the characteristic universal value of $\tfrac{\pi}{2}$. And essentially similar arguments 
to those above also give this same result for the LM phases of e.g.\ triple quantum dot models.~\cite{Jarrold2013}


\section{Concluding remarks}
\label{sec:conclusion}

In this paper we have studied elements of the globally degenerate, broken symmetry local moment phases 
that arise in locally correlated quantum impurity models. Such phases occur commonly, without fine tuning 
of parameters, in a wide range of impurity models. They represent the typical `significant other' phase, a
non-Fermi liquid, that is separated from a Fermi liquid state by an interaction-driven quantum phase transition.
Our main focus has been on what can be shown exactly about local moment phases. We believe it fair to say 
that a diverse and rather rich range of results has been obtained (as already summarized in 
sec.\ \ref{subsec:overview}); which at heart is made possible by applying a field which removes the 
strict global degeneracy, and then switching it off. Equally, however, we have arguably only scratched 
the surface of the subject, and much clearly remains to be understood about the wealth of non-trivial physics 
characteristic of local moment and related phases.


\begin{acknowledgments}
Helpful discussions with H R Krishnamurthy and A K Mitchell are gratefully acknowledged.
We thank the EPSRC for financial support, under grant EP/I032487/1.
\end{acknowledgments}
~\\


\begin{thebibliography}{70}
\bibitem[{fnf()}]{fnfinetune}
\bibinfo{note}{Many important models do of course require fine-tuning to
  realize a NFL phase; e.g.\ the spin-$\tfrac{1}{2}$ 2-channel Kondo (2CK)
  model with exchange couplings $J_{KR}, J_{KL}$, where the NFL 2CK phase
  occurs only for $J_{KR}= J_{KL}$. We do not consider such models in this
  paper.}
\bibitem[{fn-()}]{fn-2}
\bibinfo{note}{Without loss of generality we consider a symmetric conduction
  band, $\epsilon_{\mathbf{k}}=-\epsilon_{-\mathbf{k}}$, and $V_{\mathbf{k}}
  =V_{-\mathbf{k}}$.}
\bibitem[{fnA()}]{fnABnotation}
\bibinfo{note}{The notation underscores that $h\geq0$ is by definition required
  for the A-type state, and $h\leq 0$ for the B-type state.}
\bibitem[{fn8()}]{fn8a}
\bibinfo{note}{The self-evident absence of higher order skeleton contributions
  can also be shown by diagrammatic arguments}.
\bibitem[{fn9()}]{fn9a}
\bibinfo{note}{This also shows why we currently exclude the p-h symmetric point
  $\epsilon =-U/2$. In that case the renormalized levels
  $\epsilon_{A\sigma}^{*} =-U/2 + \Sigma_{A\sigma}^{R}(0,0)$ satisfy
  $\epsilon_{A\uparrow}^{*}= -\epsilon_{A\downarrow}^{*}$ (and indeed do so for
  all $|h|$). The divergence of $\epsilon^{*}$ that would be implied by eq.\
  \ref{eq:levelitolevels} in this case (with the $\epsilon_{A\sigma}^{*}$
  themselves non-vanishing) is related to the resultant singular behavior of
  the self-energy $\Sigma(\w)$ as $\w \rightarrow 0$, as considered in sec.\
  \ref{subsection:SSElowomega}.}
\bibitem[{fnL()}]{fnLMA}
\bibinfo{note}{This arises also within the LMA,~\cite{MTG_APAIM,MTG_SPAIM} but
  as shown here is exact, and independent of approximations used in LMA.}
\bibitem[{fn1({\natexlab{a}})}]{fn10a}
\bibinfo{note}{For $\lambda \in (-1,0)$, $F^{R}(0)$ in eq. \ref{eq:HTgenresult}
  is to be understood as the ${\cal{O}}(1)$ correction to the leading low-$\w$
  divergence, rather than the value of $F^{R}(\w)$ at $\w=0$ itself.}
\bibitem[{fn1({\natexlab{c}})}]{fn14}
\bibinfo{note}{The derivation implicitly assumes that at low-$\w$,
  $D_{A\sigma}(\w) \propto |\w|^{\lambda_{\sigma}}$, independent of whether $\w
  \rightarrow 0+$ or $0-$. This can in fact be shown to be the only solution
  that is self-consistent.}
\bibitem[{fn1({\natexlab{d}})}]{fn17}
\bibinfo{note}{Parts of the argument below are given, incompletely, in Ref.
  \onlinecite{MTG_APAIM} (sec.\ 3). 
  The present argument is however complete.}
\bibitem[{fn3({\natexlab{a}})}]{fn31}
\bibinfo{note}{Immediately way from p-h symmetry however, the LMA
  yields~\cite{MTG_APAIM} $C_{+}\neq C_{-}$; in which regard it does not
  capture correctly the QCP for the asymmetric model.}
\bibitem[{fn3({\natexlab{b}})}]{fn32}
\bibinfo{note}{The factor of $\tfrac{1}{2}$ in eq.\ \ref{eq:chitil9}, rather
  than e.g. $(g\mu_{B})^{2}/4$, simply reflects our chosen definition of
  $\tilde{\chi}_{u}(T)$ (eq.\ \ref{eq:chitil8}), recalling that
  $h=\tfrac{1}{2}g\mu_{B}B$ with $B$ the applied field.}
\end{thebibliography}
\end{document}